\documentclass[prx,twocolumn,superscriptaddress,longbibliography,amsmath,amssymb]{revtex4-2}

\usepackage{graphicx,color}
\usepackage{braket,slashed}
\usepackage{hyperref}
\usepackage{bm}
\usepackage{tikz}
\usepackage{bbold}
\usetikzlibrary{decorations.pathmorphing}
\usepackage{tikz-network}
\usepackage{tikz-cd}
\usepackage[normalem]{ulem}

\newcommand{\Tr}{\mathrm{Tr}}
\newcommand{\Hom}{\mathrm{Hom}}

\newcommand{\Ob}{\mathrm{Ob}}
\newcommand{\id}{\mathrm{id}}
\newcommand{\ii}{\mathrm{i}}

\begin{document}
\title{
  Entanglement Spectrum  Resolved by Loop Symmetries
}

\author{Haruki Yagi}
\affiliation{Department of Applied Physics, The University of Tokyo, Tokyo 113-8656, Japan}
\author{Zongping Gong}
\affiliation{Department of Applied Physics, The University of Tokyo, Tokyo 113-8656, Japan}

\begin{abstract}

A rigorous analysis is presented for the entanglement spectrum of quantum many-body states possessing a higher-form group-representation symmetry generated by topological Wilson loops, which is generally non-invertible. 
A general framework based on elementary algebraic topology and category theory is developed to determine the block structure of reduced density matrices for arbitrary bipartite manifolds on which the states are defined. Within this framework, we scrutinize the impact of topology on the entanglement structure for low-dimensional manifolds, including especially the torus, the Klein bottle, and lens spaces.
By further incorporating gauge invariance, we refine our framework to determine the entanglement structure for topological gauge theories in arbitrary dimensions.
In particular, in two dimensions, it is shown for the Kitaev quantum double model that not only the topological entanglement entropy can be reproduced, but also the Li–Haldane conjecture concerning the full entanglement spectrum holds exactly. 
\end{abstract}

\maketitle

\section{Introduction}
Entanglement has become one of the central organizing principles of modern quantum many-body physics and quantum field theory \cite{LA08,NL16,DH16,nishiokaEntanglementEntropyHolography2018,EW18,casiniLecturesEntanglementQuantum2023}.
Measures such as the entanglement R\'{e}nyi entropy provides precise probes of quantum correlations between local degrees of freedom, revealing universal information about phases of matter and field-theoretic structures \cite{ryuHolographicDerivationEntanglement2006,JE10,DAA19}.
For instance, in critical one-dimensional (1D) quantum many-body states described by conformal field theory, the entanglement entropy exhibits universal scaling behavior and is proportional to the central charge of the theory \cite{holzheyGeometricRenormalizedEntropy1994,GV03,calabreseEntanglementEntropyQuantum2004}. Another example is 2D topologically ordered systems described by topological quantum field theories (TQFTs) \cite{XGW90}. The negative subleading correction to the area law of entanglement entropy exhibits a universal bound known as the topological entanglement entropy \cite{IHK23}, which serves as a definitive indicator of long-range entanglement \cite{kitaevTopologicalEntanglementEntropy2006,levinDetectingTopologicalOrder2006}.
This insight established the now standard view that topological order can be understood as a pattern of long-range entanglement \cite{XGW17}.

As a far-reaching generalization of various entanglement entropy measures, the full entanglement spectrum encodes much richer information about the underlying quantum many-body state \cite{liEntanglementSpectrumGeneralization2008}. 
For example, the degeneracy pattern of the entanglement spectrum, which usually cannot be detected by entropy measures, serves as a remarkable feature of symmetry-protected topological phases \cite{pollmannEntanglementSpectrumTopological2010}. 
Even for volume-law entangled states, spectral statistics of the reduced density matrix can be used to distinguish dynamically localized and thermal phases, 
whose qualitative differences are again invisible to entropy \cite{geraedtsManybodyLocalizationThermalization2016}.

On the other hand, symmetry has long played an even more fundamental 
role in physics, constraining dynamics, classifying phases, and governing anomalies and defects \cite{EPW64,DJG96,weinbergQuantumTheoryFields1995,weinbergQuantumTheoryFields1996}.
Conventional symmetries, which can be either global or gauge, are described by reversible group actions \cite{MH64}. 
Recently, a greatly extended framework has emerged to deal with \emph{generalized symmetries} \cite{gaiottoGeneralizedGlobalSymmetries2015,bhardwajLecturesGeneralizedSymmetries2023,schafer-namekiICTPLecturesNonInvertible2023,shaoWhatsDoneCannot2024,mcgreevyGeneralizedSymmetriesCondensed2023}.
The central idea is that every topological defect in a quantum field theory can be regarded as a symmetry operator acting on extended objects \cite{gaiottoGeneralizedGlobalSymmetries2015,kapustinCouplingQFTTQFT2014}.
Two prominent examples are higher-form symmetries, in which symmetry operators act on submanifolds of higher codimension, and non-invertible symmetries, in which the fusion of symmetry operators no longer forms a group but a richer categorical structure \cite{gaiottoGeneralizedGlobalSymmetries2015,thorngrenHigherSPTsGeneralization2015,bhardwajUnifyingConstructionsNonInvertible2023}.
This perspective not only opens up new avenues for exploring unprecedented phases of quantum matter and field theories \cite{jianPhysicsSymmetryProtected2021,qiFractonPhasesExotic2021,KI22,paceExactEmergentHigherform2023a,CZ24,seifnashriClusterStateNoninvertible2024,choiNoninvertibleHigherform2025}, 
but also sheds new light on well-established models. For example, the Kramers-Wannier duality of a critical quantum Ising chain is a non-invertible symmetry \cite{aasenTopologicalDefectsLattice2016}, while the toric code is a symmetry-broken phase with respect to the $1$-form symmetry \cite{gaiottoGeneralizedGlobalSymmetries2015,lakeHigherformSymmetriesSpontaneous2018}.

These developments raise a fundamental question:
How does a generalized symmetry constrain the structure of entanglement? 
For conventional $0$-form invertible symmetries, this has essentially been answered by the seminal works of Wigner and Dyson who initiated  systematic applications of group theory to quantum mechanics \cite{EW59,dysonThreefoldWayAlgebraic1962}. While their works concerned Hamiltonians, the results can be readily translated to reduced density matrices and lead to the notion of symmetry-resolved entanglement \cite{MG18,RB19,AN21}. The resolved symmetry sectors turn out to be labeled by irreducible representations (irreps), while the geometry and topology of (sub)system does not play a role \cite{yagiThreefoldWayTypical2025}. In contrast, as is already clear from previous studies on topological ordered phases, higher-form symmetry would greatly influence the entanglement structure in a topology-dependent way \cite{bondersonAnyonicEntanglementTopological2017}. Also, more sophisticated algebraic approaches are under development for dealing with the impact of non-invertible symmetries on entanglement \cite{saura-bastidaCategoricalsymmetryResolvedEntanglement2024}. Despite considerable progress made along this direction, it remains elusive to achieve a systematic understanding on the entanglement resolved by generalized symmetries.

A canonical example that is yet to be understood arises from the Kogut–Susskind formulation of lattice gauge theory \cite{kogutHamiltonianFormulationWilsons1975}.
When the gauge field configuration is flat, Wilson loops become topological defect operators.
Being 1D objects, they generate a loop symmetry \footnote{This is a higher-form symmetry what is called (d-2)-form symmetry in the field theory community under the convention of considering spacetime as d-dimensional. However, we determined that the emphasis should be on the fact that the symmetry operator is defined on a closed one-dimensional loop. Therefore, this paper actively employs the term ``loop symmetry.''}.
If the gauge group is non-Abelian, the operators associated with its irreducible representations lose invertibility, providing explicit realizations of non-invertible symmetries.
Unlike group symmetries, these non-invertible symmetry operators do not have a tensor-product decomposition when acting on spatially bipartite subsystems. In addition, the higher-form nature introduces another complexity arising from the underlying space topology, making the entanglement structure even more nontrivial.

In this work, we develop a general and rigorous framework for analyzing entanglement structures of quantum many-body states that respect such generalized symmetries.
Our approach applies to arbitrary space dimensions, general (sub)manifolds that can even be non-orientable, and all the finite-group-representation symmetries which are generally non-invertible. 
It naturally extends the concept of symmetry-resolved entanglement entropy and provides a systematic algorithm for evaluating entanglement structures under generalized symmetry actions. A key step is to make use of the Seifert-van Kampen theorem, a fundamental result in algebraic topology. This helps us to precisely encode the topological information into the entanglement structure.

Our framework, when supplemented with gauge invariance, further enables a systematic analysis of topological gauge theories in arbitrary dimensions.
Within this refined formulation, we derive explicit expressions for the reduced density matrices associated with general spatial bipartitions in non-Abelian topological gauge theories. One can then determine the topological entanglement entropy in a fully rigorous manner. 
In particular, in 2D space our results apply to the Kitaev quantum double model and verify the Li-Haldane conjecture \cite{liEntanglementSpectrumGeneralization2008}, the latter asserts that the entanglement structure of a topologically ordered phase corresponds precisely to the Hilbert space of a rational conformal field theory (RCFT).

\section{Preliminaries}\label{sec-setup}

\begin{figure}
\begin{center}
       \includegraphics[width=8.5cm, clip]{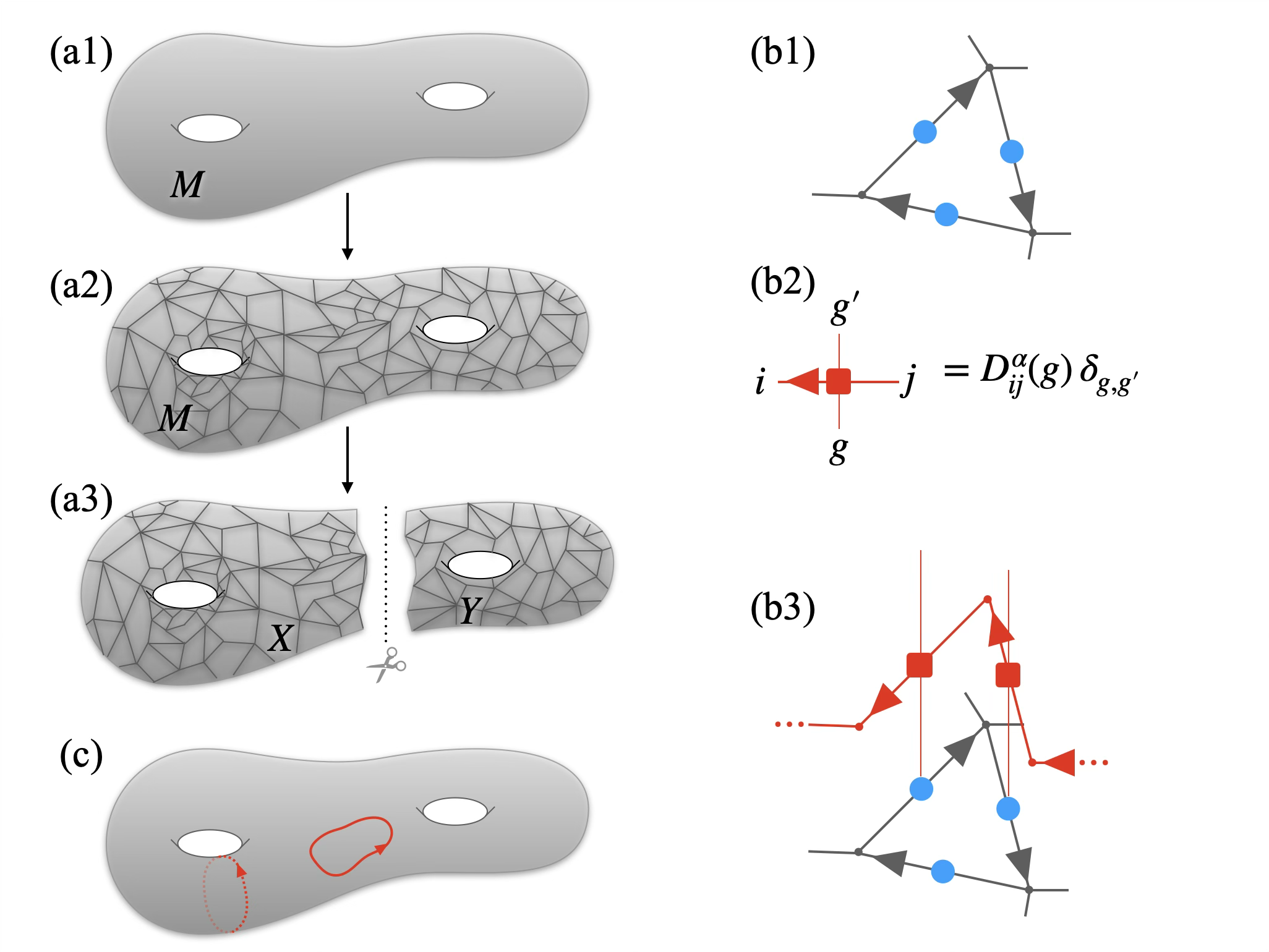}
       \end{center}
   \caption{
   (a) Procedure for defining a lattice and its bipartition. 
   (a1) Fix a manifold $M$, for example
   a genus-2 surface. 
   (a2) Fix a sufficiently fine-grained discretization of $M$. (a3) Fix a bipartition of $M$ into $X$ and $Y$ with no common edges. 
   (b) Definition of Hilbert space and the symmetry action on it.
   (b1) Fix a finite group $G$. For each edge, a local Hilbert space $\mathbb{C}^{|G|}$ spanned by $|g\rangle$ (blue dot) and an orientation (arrows) is assigned.
   (b2) Building block of the matrix product operator (MPO) of ${\rm Rep}(G)$ loop symmetry. Here $D^\alpha_{ij}(g)$ is the representation matrix of irrep $\alpha$.
   (b3) Action of MPO on the state. The orientation of an MPO is fixed. If its orientation aligns with (opposite to) that of the edge, we adopt the building block in (b2) (with $D^\alpha_{ij}(g)$ replaced by $D^\alpha_{ij}(g^{-1})$). 
   (c) Constraint on the many-body state from the loop symmetries. Holonomies along contractible loops (red solid loop) are enforced to be trivial, while those 
   along non-contractible loops (red dashed loop) may not be trivial.
   }
   \label{fig:setup}
\end{figure}

We start by providing a concise explanation of the setup,
as illustrated in Fig.~\ref{fig:setup}. Let $M$ be a $d$D ($d\ge2$) connected, compact, and closed manifold of space. We discretize $M$ into a lattice consisting of finite set of vertices $V$, oriented edges $E$, and plaquettes $P$.
Fix a finite group $G$ which is generally non-Abelian. Analogous to the toric code, each edge $e \in E$ is associated with a $G$-spin, whose local Hilbert space can be 
spanned by $|g\rangle$ ($g\in G$) which satisfies the orthonormal relation $\braket{g|h}=\delta_{g,h}$. Without further constraint, the entire Hilbert space is simply a tensor product of $|E|$ local Hilbert spaces $\mathbb{C}^{|G|}$, and any many-body state can be expressed as $|\Psi\rangle=\sum_{\{g_e\}_{e\in E}}\Psi_{\{g_e\}_{e\in E}}\bigotimes_{e\in E}|g_e\rangle$, or in short $|\Psi\rangle=\sum_{\{g\}_{E}}\Psi_{\{g\}_{E}}|\{g\}_{E}\rangle$.

We now introduce the higher-form group-representation symmetry closely related to \emph{Wilson loops}. Associated to a closed oriented loop $\ell \subset E$, the Wilson loop of $|\{g\}_{E}\rangle$ is defined as the conjugacy class of $\prod_{i=1}^{|\ell|} g_i$ provided the edge orientations all align with $\ell$. Here $\{g_i\}^{|\ell|}_{i=1}$ is an ordered projection of $\{g_e\}_{e\in E}$ onto $\ell$. 
This construction is ambiguous in the choice of starting point, so only the conjugacy class is well-defined for non-Abelian groups. More generally, when some edges align oppositely with $\ell$ (see Fig.~\ref{fig:setup}(b3)), we replace those $g_i$ with $g_i^{-1}$.
Following the duality between conjugacy classes and irreps (as basis of class functions), we can use the following matrix product operator (MPO) to extract the holonomic information of Wilson loop along $\ell$:
\begin{equation}
\begin{split}
  \Lambda_\ell^\alpha =\sum_{\{g\}_{\ell}} \Tr \left[D^\alpha(g_1^{s_1})\cdots D^\alpha(g_{|\ell|}^{s_{|\ell|}})\right] \ket{\{g\}_{\ell}}\bra{\{g\}_{\ell}}.
\end{split}
\end{equation}
Here $D^\alpha$ is the unitary representation matrix of irrep $\alpha$, and $s_i=1$ ($s_i=-1$) if the $i$th edge aligns (oppositely) with $\ell$. These MPOs labeled by 
irreps satisfy the \emph{fusion rule} of $\mathrm{Rep}(G)$, the representation category of group $G$:
\begin{equation}
    \Lambda_\ell^\alpha\times \Lambda_\ell^\beta = \sum_\gamma N^{\alpha \beta}_\gamma \Lambda_\ell^\gamma,
    \label{fr}
\end{equation}
where $N^{\alpha \beta}_\gamma \in \mathbb{Z}_{\geq 0}$ is the Clebsch-Gordan coefficient between irreps $\alpha,\beta$ and $\gamma$.
Whenever $G$ is non-Abelian, such an algebraic structure is not a group. 
In particular, since elements other than the identity $\Lambda^{1}_{\ell}$ ($1$: trivial irrep) also appear on the right-hand side (rhs) of Eq.~(\ref{fr}), they are generally non-invertible. 
One may obtain such $\mathrm{Rep}(G)$ loop symmetry by \emph{gauging} $0$-form $G$ symmetry \cite{schafer-namekiICTPLecturesNonInvertible2023}, though we do not care about the physical origin but rather claim this loop symmetry to be our starting point. Hereafter, we will abuse ${\rm Rep}(G)$ as the set of irreps, which are the simple objects in the representation category. 

Now we are ready to introduce the notion of loop-symmetric many-body states. By definition, such a state is symmetric with respect to $\Lambda_\ell^\alpha$ for any $\alpha$ and contractible $\ell$:
\begin{equation}
\Lambda_\ell^\alpha |\Psi\rangle=d_\alpha|\Psi\rangle,\;\;\;\;{}^\forall\alpha\in{\rm Rep}(G),\;\;[\ell]=[{\rm pt}]\in \pi_1(M).
\label{ls}
\end{equation}
Here $d_\alpha$ is the dimension of irrep $\alpha$, $[\ell]$ denotes the homotopy class of $\ell$, $\pi_1(M)$ is the fundamental group of $M$, and $[{\rm pt}]$ is the trivial element (unit) in it. As shown in Appendix~\ref{app:lsfc}, any loop-symmetric product state $|\{g\}_E\rangle$ is ``flat'' (and vice versa), meaning that all the contractible Wilson loops, or equivalently those around all the plaquettes, possess trivial holonomies \footnote{This setup is reminiscent of lattice gauge theory. However, we do not impose gauge invariance. In contrast, we require that the Wilson loops are topological. The theory can be interpreted as a topological gauge theory (in the Kogut-Susskind formalism) without gauge invariance.}. Our main focus is the structure of bipartite entanglement of these states. We denote the two subsystems as $X$ and $Y$, which are two subsets of $E$ complementary to each other. We assume that $X$ and $Y$ are regular enough as if they can be discretized from two complementary submanifolds of $M$ with a common boundary $\partial$ \footnote{Our framework is directly applicable to open manifolds as well, but this paper focuses on the case of closed manifolds. Therefore, it should not cause confusion to interpret $\partial$ as the common boundary of $X$ and $Y$ obtained as a partition of $M$. However, when a manifold follows $\partial$ in the following, $\partial$ should be interpreted as the boundary operator.}, which is again a manifold with one lower dimension. We will abuse $X$ and $Y$ to refer also to these underlying submanifolds for simplicity. Now consider a state of interest as follows:
\begin{equation}
  \ket{\Psi}
  =\sum_{\{g\}_X,\{g\}_Y} \Psi_{\{g\}_X,\{g\}_Y}\ket{{\{g\}_X}}\ket{\{g\}_Y}.
\end{equation}
The bipartite entanglement of this state $\ket{\Psi}$ can be completely characterized by a rectangular matrix $W$: 
\begin{equation}
  W:=\sum_{\{g\}_X,\{g\}_Y} \Psi_{\{g\}_X,\{g\}_Y}\ket{{\{g\}_X}}\bra{\{g\}_Y},
\end{equation}
whose singular value spectrum is the square root of the entanglement spectrum, as the reduced density matrix on $X$ ($Y$) is given by $\rho_X=WW^\dagger$ ($\rho_Y=W^\dagger W$). 
Our objective is to find the structure of $W$ constrained by the group-representation loop symmetry (\ref{ls}). 
We will provide a general algorithm to achieve this in the next section.

\section{Gauge transformation, Seifert–van Kampen theorem, and the General Algorithm}\label{sec-SvK}

The flatness condition arising from loop symmetries is reminiscent of the zero-divergence property for a solenoidal vector field. Indeed, the former can be regarded as a discretized, non-Abelian version of the latter. Here, we can generate a flat or loop-symmetric configuration by taking the ``gradient'' of a group-element valued function on the vertex set $V$. This is nothing but performing a gauge transformation from the trivial configuration $|\{1\}\rangle_M$ (i.e., $g_e=1$ ${}^\forall e\in E$). It is also known that a divergence-free vector field may not always be a gradient if the underlying manifold is nontrivial, and such a gap is fully characterized by the first de Rham cohomology. Likewise, starting from the representative configurations characterized by different non-Abelian cohomology classes, we can generate all the loop-symmetric states \footnote{The proof of this fact is also provided in the Supplemental Materials}. This statement applies equally to each subsystem. Having these ideas in mind, we are ready to develop a general algorithm for determining the block structure of the reduced density matrix.

For technical reasons that will become clear later, we prescribe a set $A\subset V$ of base points, which can be flexibly chosen on demand. We then obtain a few representative loop-symmetric states labeled by different elements $\Phi$ in ${\rm Hom}(\pi_1(M,A),G)$, where $\pi_1(M,A)$ is the fundamental groupoid (or group, if $A$ contains a single point) of $M$. We define gauge transformation explicitly for a function $g_v$: $V\backslash A\to G$. 
The exclusion of $A$ ensures that the non-Abelian cohomology class will not be changed. Given any product state $|\{g\}_{E}\rangle$, this function induces commuting local actions for each vertices $v \in V\backslash A$ as follows:
\begin{equation}
  \bigotimes_{e_v^+} \ket{g_{e_v^+}} \bigotimes_{e_v^-} \ket{g_{e_v^-}} \mapsto \bigotimes_{e_v^+} \ket{g_{e_v^+}g_v^{-1}} \bigotimes_{e_v^-} \ket{g_v g_{e_v^-}}
\end{equation}
where $e_v^+ (e_v^-)$ is an outgoing (incoming) edge from (to) the vertex $v$. The complete orthonormal basis of loop-symmetric states can then be denoted as $|\{g_v\}_{v\in V\backslash A}, \Phi\rangle$, or $|\{g\}_{V\backslash A}, \Phi\rangle$ in short, which is generated by a gauge transform $g_v$: $V\backslash A\to G$ on the representative with non-Abelian cohomology class $\Phi$.

Regarding a bipartition $M=X\cup Y$, it is natural to choose $A\subset \partial=X\cap Y$ such that the based points are common for $X,Y$ and $M$. More precisely, we pick up exactly one based point from each connected component in $\partial$ to avoid loss of topological information about how $X$ and $Y$ are glued into $M$. Following the same analysis above, we can express a loop-symmetric subsystem state on $X$ ($Y$) as $|\{g\}_{\overline V_X\backslash A},\phi_X\rangle$ ($|\{g\}_{\overline V_Y\backslash A},\phi_Y\rangle$) with $\phi_X\in {\rm Hom}(\pi_1(X,A),G)$ ($\phi_Y\in {\rm Hom}(\pi_1(Y,A),G)$). Here $V_{X,Y}$ denotes the set of vertices inside $X,Y$, while $\overline{V}_{X,Y}=V_{X,Y}\cup V_\partial$ is its closure with the vertices in $\partial$ included. Moreover, given $\Phi\in{\rm Hom}(\pi_1(M,A),G)$, $\phi_{X,Y}$ should be uniquely determined as $\Pi_{X,Y}(\Phi)$, where $\Pi_{X,Y}$ is induced by the inclusion map from $X,Y$ to $M$. Using the gauge-transform degree of freedom, one can always adjust the representatives on $M$ such that whenever $\Pi_{X,Y}(\Phi)$ coincides the reduced representatives on $X,Y$ coincide as well. Therefore, we have 
\begin{equation}
|\{g\}_{V\backslash A}, \Phi\rangle = |\{g\}_{\overline{V}_X\backslash A}, \Pi_X(\Phi)\rangle |\{g\}_{\overline{V}_Y\backslash A}, \Pi_Y(\Phi)\rangle,
\end{equation}
where the functions on the rhs (i.e., $g_v:\overline{V}_{X/Y}\backslash A \to G$) are simply restrictions of that on the lhs (i.e., $g_v:\overline{V}\backslash A \to G$), thus their further restrictions on $V_{\partial}\backslash A$ necessarily coincide. This feature implies that the $W$ matrix is fragmented into a direct sum of $|G|^{|V_\partial|-|A|}$ individual blocks, each of which has a tensor product structure between a geometric part of size $|G|^{|V_X|}\times|G|^{|V_Y|}$ and a topological part determined by $\Pi_{X,Y}$.

The remaining task is to determine the topological part in the block structure. It turns out that the set of non-Abelian cohomology classes on $M$ is a fiber product of those on $X$ and $Y$, as we demonstrate in the following. We first reinterpret the partition of $M$ as a commutative diagram in the category of manifolds: 
\begin{equation}
\begin{tikzcd}
M & X \arrow[l,hookrightarrow]  \\
Y \arrow[u,hookrightarrow]  & \partial \arrow[l,hookrightarrow] \arrow[u,hookrightarrow]
\end{tikzcd}
\end{equation}
Here the object of the diagram is a manifold, possibly with boundaries, and the morphism is the inclusion map. By employing the Seifert–van Kampen theorem, a fundamental result in algebraic topology, we transform this diagram into a diagram in the category of groupoids via the fundamental groupoid functor: 
\begin{equation}
\begin{tikzcd}
\pi_1(M,A)& \pi_1(X,A) \arrow[l, "p_X"]  \\
\pi_1(Y,A)\arrow[u, "p_Y"'] & \pi_1(\partial,A) \arrow[l, "i_Y"'] \arrow[u, "i_X"]
\end{tikzcd}
\label{pofg}
\end{equation}
Here the fundamental groupoid is the object, and the morphism is a homomorphism between them. We emphasize that the boundary may have multiple connected components in general, so this pushout diagram is preserved only for sufficiently many base points (at least one point in each connected component). This explains the reason why $A$ may contain more than one point. 
To figure out the relation between the entire and subsystem non-Abelian cohomology classes, we further take the $\Hom (*, G)$ functor. This functor is contravariant, and the limit/colimit is sent to the corresponding colimit/limit.
Therefore, the pushout (\ref{pofg}) is turned into a pullback in the category of sets:
\begin{equation}
\begin{tikzcd}
\Hom (\pi_1(M,A),G) \arrow[r, "\Pi_X"]\arrow[d, "\Pi_Y"']& \Hom(\pi_1(X,A),G) \arrow[d, "r_X"] \\
\Hom (\pi_1(Y,A),G) \arrow[r, "r_Y"'] & \Hom(\pi_1(\partial,A),G)   \\
\end{tikzcd}
\label{HomvK}
\end{equation}
Now it is clear that the topological degree of freedom takes the form of $\Phi=(\phi_X,\phi_Y)$ for those $r_X(\phi_X)=r_Y(\phi_Y)$, and $\Pi_{X,Y}(\Phi)=\phi_{X,Y}$ is simply a projection.
Subsequently, the matrix $W$ is an element of the vector space decomposed into the block structure as follows:
\begin{equation}
  \bigoplus_{
    \phi \in \mathrm{Im}
    }
  \bigoplus_{j=1}^{|G|^{|V_\partial|-|A|}}
  \mathbb{C}^{
    |r_X^{-1}(\phi)||G|^{|V_X|} \times |r_Y^{-1}(\phi)||G|^{|V_Y|}
  }
  \label{bs}
\end{equation}
where $\mathrm{Im} \subseteq \Hom (\pi_1(\partial,A),G)$ is the image of $r_X\circ \Pi_X=r_Y\circ \Pi_Y$ and $r_{X,Y}^{-1}$ denotes the preimage. 
To emphasize the separation between topological and geometric degrees of freedom, we may recast Eq.~(\ref{bs}) into
\begin{equation}
 \left(\bigoplus_{
    \phi \in \mathrm{Im}
    }\mathbb{C}^{|r_X^{-1}(\phi)|\times|r_Y^{-1}(\phi)|}\right)\otimes
    \left(\bigoplus^{|G|^{|V_\partial|-|A|}}_{
    j = 1
    }\mathbb{C}^{|G|^{V_X}\times|G|^{V_Y}}\right).
    \label{tg}
\end{equation}

We remark that such a block structure (\ref{bs}) is derived under the (subsystem) loop-symmetric many-body basis built from gauge transforming cohomology representatives. The number of total degrees of freedom reads $\sum_{\phi\in{\rm Im}}|r_X^{-1}(\phi)||r_Y^{-1}(\phi)||G|^{|V_\partial|-|A|+|V_X|+|V_Y|}=|{\rm Hom}(\pi_1(M,A),G)||G|^{|V|-|A|}$, consistent with the Hilbert-space dimension $|{\rm Hom}(\pi_1(M),G)||G|^{|V|-1}$ of loop-symmetric states on the entire lattice. In two or higher space dimensions, this is a quantity exponentially smaller (in terms of system size) than $|G|^{|E|}$, the Hilbert-space dimension of many-body states with no symmetry constraint. 
We also note that no degeneracy is enforced by ${\rm Rep}(G)$ loop symmetries, in contrast to non-Abelian $0$-form group symmetries. This is consistent with the observations in Ref.~\cite{xuEntanglementPropertiesGauge2025} for specific 2D models with $\mathbb{Z}_2$ 1-form symmetries. Moreover, while the block structure exhibits an ``area law'', i.e., the number of blocks is proportional to the boundary Hilbert-space dimension, the entanglement entropy may take volume-law values, as is the case in quantum thermalization \cite{OF23}.

We end the section by showcasing how to execute our general algorithm for a bipartite 2D sphere as a trivial example \footnote{We do not demonstrate the case of $S^1$ as it does not allow discretization with plaquettes, implying no symmetry is actually imposed.
Interestingly, even this no-symmetry result can be reproduced in our framework, as we demonstrate in Appendix~\ref{app:S1}.}. In this case we have $M=S^2$, $X=Y=D^2$ (2D disk) and $\partial=S^1$, and $A$ can be chosen to contain only a single point. As the fundamental group of $S^2$ or $D^2$ is trivial, there is no topological degree of freedom.
The matrix $W$ thus takes the form
\begin{equation}
  \bigoplus_{j=1}^{|G|^{|V_\partial|-1}}
  \mathbb{C}^{
    |G|^{|V_X|} \times |G|^{|V_Y|}
  }.
  \label{tri}
\end{equation}
In fact, provided $\partial$ is connected, this result holds true as long as the fundamental group of $M$ is trivial, even if those of $X$ and $Y$ are nontrivial (see Sec.~\ref{sec:HS} for a concrete example). This is the case for higher-dimensional spheres, and actually there are much more simply connected high-dimensional manifolds other than spheres. Another simple generalization of this trivial example is when $X$ (or $Y$) has a trivial fundamental group and connected boundary. In this case, $\Pi_X$ and $r_X$ in Eq.~(\ref{HomvK}) are trivial and the topological degrees of freedom in the entanglement block structure completely inherit from the entire system. The matrix $W$ thus takes the form
\begin{equation}
  \bigoplus_{j=1}^{|G|^{|V_\partial|-1}}
  \mathbb{C}^{
    |G|^{|V_X|} \times |{\rm Hom}(\pi_1(M),G)| |G|^{|V_Y|}
  }.
  \label{triX}
\end{equation}
In the following, we will present more nontrivial examples where both subsystems exhibit topological degrees of freedom.

\section{Minimal Examples}
\label{sec-examples}
While our framework applies to arbitrary dimensional manifolds, general manifolds with dimensions greater than $4$ are notorious for the intrinsic complexity of their topology. In fact, any finitely presented group is the fundamental group of some $4$D manifold, while it has been proved that no algorithm exists for determining whether a finite-dimensional representation exists 
\cite{MRB15}. This result may be roughly understood as a consequence of the undecidability of the word problem. To gain some analytical insights, we restrict ourselves to low-dimensional manifolds, which already exhibit nontrivial topological effects beyond Eqs.~(\ref{tri}) and (\ref{triX}). We also present an example of higher-dimensional torus, which is highly regular and does not suffer from the curse of complexity.

\begin{figure}
\begin{center}
       \includegraphics[width=8.5cm, clip]{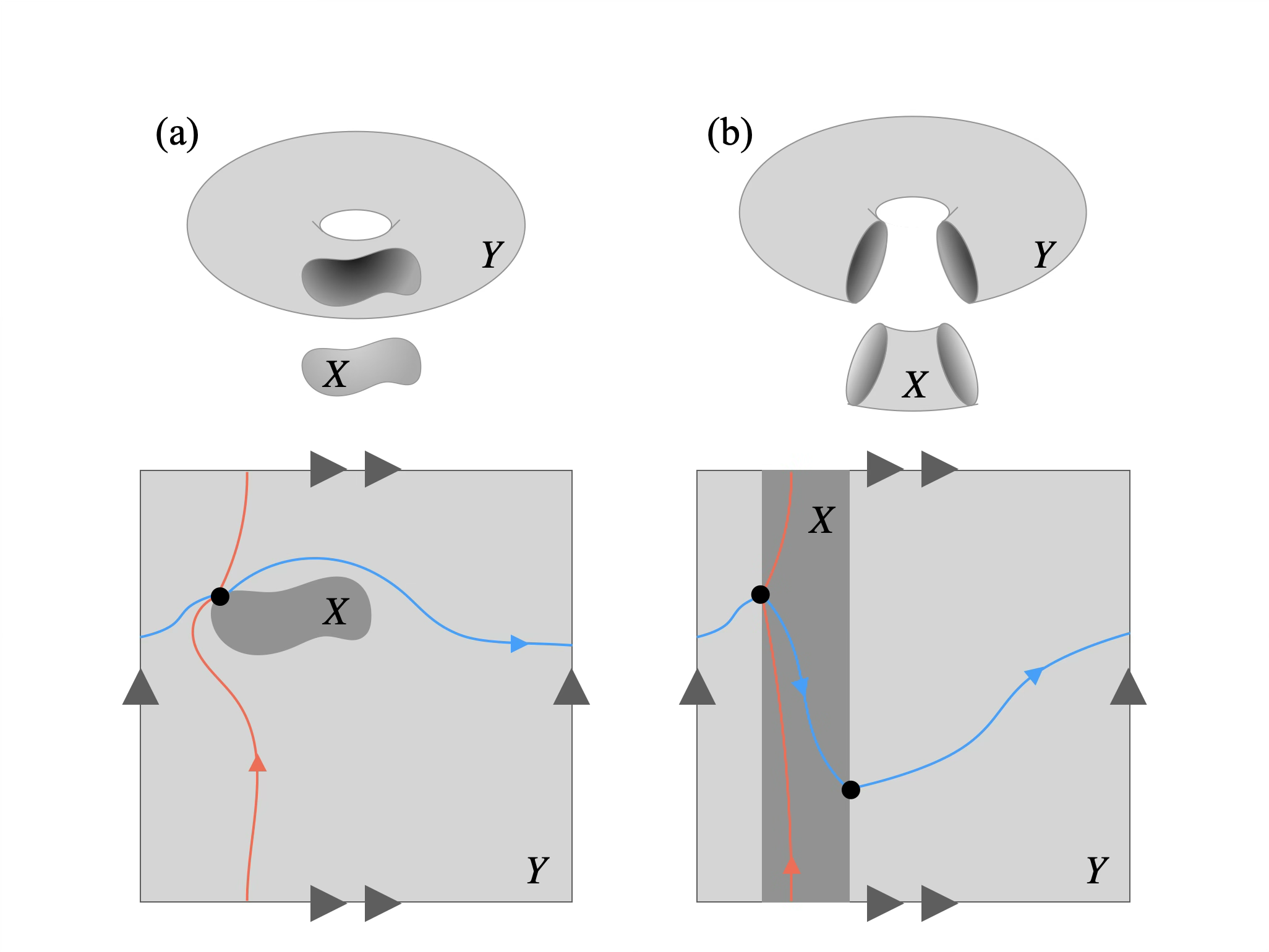}
       \end{center}
   \caption{
   Torus has essentially two topologically different ways of bipartition. (a) Disk + the other. Since $\partial$ is simply connected, only one base point is necessary. The holonomies along red and blue curves must commute. 
   (b) Two tubes. 
   Here $\partial$ has two connected components. 
   The non-contractible loop along the horizontal axis is inevitably segmented.
   } 
   \label{fig:torus-bipartitions}
\end{figure}

\subsection{Torus $\mathbb{T}^2$}

We consider torus $\mathbb{T}^2=S^1\times S^1$ as the first nontrivial example. 
There are essentially two different ways of decomposition as shown in
Fig.~\ref{fig:torus-bipartitions}. 
The trivial one as shown Fig.~\ref{fig:torus-bipartitions}(a)
takes the form of Eq.~(\ref{triX}):
\begin{equation}
  \bigoplus_{j=1}^{|G|^{|V_\partial|-1}}
  \mathbb{C}^{
    |G|^{|V_X|} \times |\mathrm{Rep}(G)||G|^{|V_Y|+1}
  },
  \label{torustri}
\end{equation}
where the cardinality of ${\rm Hom}(\pi_1(\mathbb{T}^2),G)=\{(g,h)\in G^{\times2}:gh=hg\}$ has been obtained to be $|G||{\rm Rep}(G)|$. This follows from a straightforward calculation using the stabilizer-orbit theorem as well as the duality between irreps and conjugacy classes.
In contrast, for the nontrivial decomposition shown in Fig.~\ref{fig:torus-bipartitions}(b) with two base points, we obtain
\begin{equation}
  \bigoplus_{c\in G}
  \bigoplus_{j=1}^{|[c]| |G|^{|V_\partial|-2}} \mathbb{C}^{|{\rm C}_c|
  |G|^{|V_X|}\times 
  |{\rm C}_c||G|^{|V_Y|}},
  \label{eq:blocksize-torus-nontrivial}
\end{equation}
where ${\rm C}_c=\{g\in G:gc=cg\}$ is the centralizer group of $c\in G$, and $[c]=\{gcg^{-1}:g\in G\}$ consists of the group elements in the same conjugacy class of $c$. According to the orbit-stabilizer theorem, we have $|[c]||{\rm C}_c|=|G|$, so the total degrees of freedom of Eqs.~(\ref{torustri}) and (\ref{eq:blocksize-torus-nontrivial}) can be checked to be the same. We mention that $|{\rm C}_c|$
can also be calculated from the data of irrep characters and the grand orthogonality as $|{\rm C}_c|=\sum_{\alpha\in{\rm Rep}(G)}|\chi^\alpha(c)|^2$, which will be used for consistency check later.

\begin{figure*}[t]
  \centering
  \includegraphics[width=\textwidth]{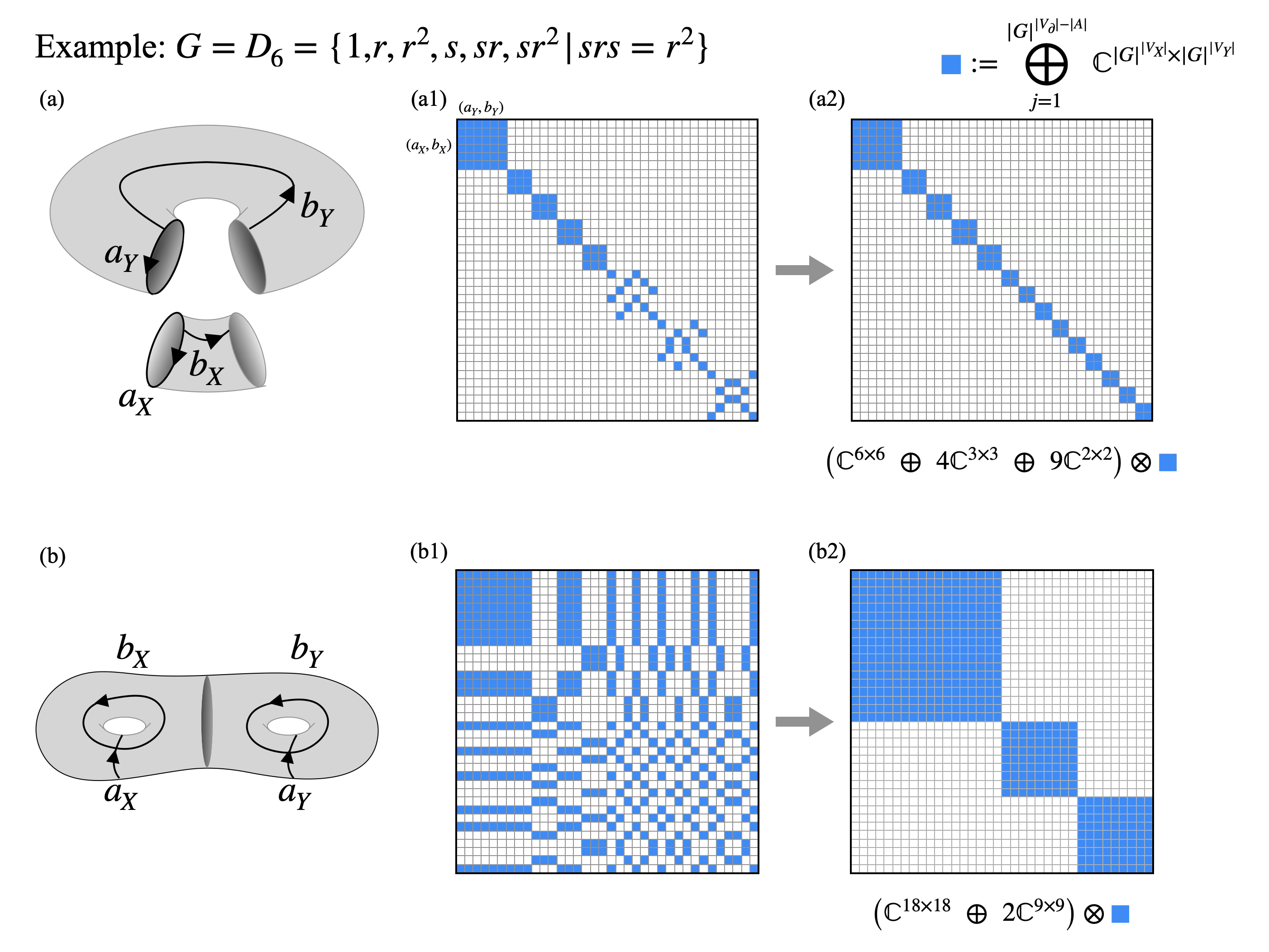}
  \caption{
      Nonzero elements in $W$ in two settings under fixing $G=D_6$. The blue square denotes the blocks generated by gauge transformation on $V \backslash A$, corresponding to the geometric part. 
      (a) Nontrivial bipartition of torus.
      (a1) A table showing where nonzero elements can be placed for nontrivial bipartition of torus. 
      The rows are labeled by $(a_X,b_X)$, and the columns are labeled $(a_Y,b_Y)$.
      The part where the blue square is placed is where nonzero elements are assigned when the basis is arranged in the lexicographical order $(1,1),(1,r),\dots,(1,sr^2),(r,1),\dots, (sr^2,sr^2)$, and the parts where it is not placed must be zero.
      (a2) Result of rearranging the basis vectors into blocks. The direct sum decomposition of the topological part is $\mathbb{C}^{6\times 6}\oplus 4 \mathbb{C}^{3\times 3}\oplus 9 \mathbb{C}^{2\times 2}$. 
      (b) Symmetric bipartition of genus-2 surface. 
      (b1) A table showing where nonzero elements can be placed for this bipartition. 
      The rows are labeled by $(a_X,b_X)$, and the columns are labeled $(a_Y,b_Y)$.
      The basis is arranged in the lexicographical order again. It appears to have a far more complex structure than the examples seen before, and it is not so obvious that it can be reduced to a block structure simply by rearranging the basis.
      (b2) The result of rearranging the basis vectors into blocks. The direct sum decomposition of the topological part is $\mathbb{C}^{18\times 18}\oplus 2 \mathbb{C}^{9\times 9}$.
  }
  \label{fig:example-torus-2torus-d6}
\end{figure*}

Let us provide some intuitions into Eq.~(\ref{eq:blocksize-torus-nontrivial}) without referring to the explicit calculations, which can be found in Appendix~\ref{app:torus}. Unlike the trivial case where $X$ is essentially a disk, now $X$ as well as $Y$ is a cylinder on which the azimuthal holonomy can be nontrivial. For one base point this value can be an arbitrary group element, corresponding to the first direct sum in Eq.~(\ref{eq:blocksize-torus-nontrivial}). As the azimuthal loop can be continuously deformed from one edges to the other, the holonomies around the two based points should be in the same conjugacy class. This explains the factor $|[c]|$ in the second direct sum. These two holonomies are related by a conjugation with respect to the group element accumulated along an inter-base point path. The number of solutions is exactly $|{\rm C}_c|$, which indeed gives the remaining topological degrees of freedom in Eq.~(\ref{eq:blocksize-torus-nontrivial}). In Fig.~\ref{fig:example-torus-2torus-d6}(a), we present an example for the minimal non-Abelian group $G=D_6$ to show how the topological part in the block structure looks like.

\subsection{Genus $\gamma$ Surface}
\label{sec:ggs}
Torus is among the simplest examples of connected, oriented, closed, and compact 2D manifolds, which are completely classified by the genus \cite{gallierGuideClassificationTheorem2013,francisConwaysZIPProof1999}.
While the entanglement structure on a torus is relatively comprehensible, we can hardly rely on intuitions for general genus $\gamma$ surfaces which admit $\mathcal{O}(\gamma^2)$ topologically different ways of decomposition. Nevertheless, their fundamental groups are of very low complexity, and thus the calculations within our general framework are fully tractable.

Let $\Sigma_{\gamma,n}$ be the surface with $\gamma$ genus and $n$ boundary circles, where $\gamma$ and $n$ can be arbitrary non-negative integers. A general bipartition of $\Sigma_{\gamma,0}$ is given by 
\begin{equation}
\begin{tikzcd}
M=\Sigma_{\gamma,0} & X=\Sigma_{\gamma_X,n} \arrow[l, hookrightarrow]  \\
Y=\Sigma_{\gamma_Y,n}  \arrow[u, hookrightarrow] & \partial = \bigsqcup^n S^1 \arrow[u, hookrightarrow] \arrow[l, hookrightarrow] 
\end{tikzcd}
\end{equation}
where $\gamma=\gamma_X+\gamma_Y+n-1$. Note that the bipartitions of torus considered previously correspond to $(\gamma_X,\gamma_Y,n)=(0,1,1)$ and $(0,0,2)$. In general, thanks to the existence of boundaries, the fundamental group of $X$ or $Y$ is simply a free group of $2\gamma_{X,Y}+n-1$ generators, making their non-Abelian cohomology extremely simple. The only thing that deserves some attention is to determine which element in the extended fundamental groupoid is projected into the boundary cohomology (holonomy) set via $r_{X,Y}$ in Eq.~(\ref{HomvK}). As explicitly derived in Appendix~\ref{app:ggs}, the block structure of reduced density matrix is identified as
\begin{equation}
  \bigoplus_{c\in G^{\times n}
  } \bigoplus_{j=1}^{|G|^{|V_\partial|-n}} \mathbb{C}^{R_{\gamma_X,n}(c)
  |G|^{|V_X|}\times R_{\gamma_Y,n}(c)
  |G|^{|V_Y|}},
  \label{ggsblock}
\end{equation}
where $c=(c_1,...,c_n)$ ($c_i\in G$) and
\begin{equation}
R_{\gamma,n}(c)
  =\sum_{\alpha \in{\rm Rep}(G)} \left(\frac{|G|}{d_\alpha}\right)^{2\gamma+n-2}\prod_{i=1}^n \chi^\alpha (c_i)
\label{eq:blocksize-genus-g-surface}
\end{equation}
is a non-negative class function on $G^{\times n}$. We remark that if $R_{\gamma_X,n}(c)$ or $R_{\gamma_Y,n}(c)$ vanishes, such a boundary holonomy $c$ will not contribute any block, implying that such a configuration is not compatible with the loop symmetry. In other words, ${\rm Im}$ in Eq.~(\ref{bs}) should be $\{c\in G^{\times n}:R_{\gamma_X,n}(c)\neq0,R_{\gamma_Y,n}(c)\neq0\}$ in this example.

Let us check that the general formula (\ref{ggsblock}) does reproduce the previous results for torus. Taking $(\gamma_X,\gamma_Y,n)=(0,1,1)$, we have $R_{0,1}(c)=\sum_{\alpha\in{\rm Rep}(G)}d_\alpha\chi^\alpha(c)/|G|=\delta_{c,1}$ corresponding to the character of regular representation. We thus only have compute $R_{1,1}(1)=\sum_{\alpha\in{\rm Rep}}(|G|/d_\alpha)\chi^\alpha(1)$, which gives $|G||{\rm Rep}(G)|$ since $\chi^\alpha(1)=d_\alpha$ and reproduces Eq.~(\ref{torustri}). Taking $(\gamma_X,\gamma_Y,n)=(0,0,2)$, we have $R_{0,2}(c)=\sum_{\alpha\in{\rm Rep}}\chi^\alpha(c_1)\chi^\alpha(c_2)=|{\rm C}_{c_1}|\delta_{[c_1],[c^{-1}_2]}$ using the orthogonality relation mentioned previously. We thus reproduce the block structure in Eq.~(\ref{eq:blocksize-torus-nontrivial}). In general, one can check that Eq.~(\ref{eq:blocksize-genus-g-surface}) satisfies $R_{\gamma,n}(c)=R_{\gamma,n}(c^{-1})$ ($c^{-1}=(c_1^{-1},...,c_n^{-1})$) and 
\begin{equation}
|G|^nR_{\gamma,0}=\sum_{c\in G^{\times n}}R_{\gamma_X,n}(c)R_{\gamma_Y,n}(c),
\label{RRR}
\end{equation}
where the lhs is the number of total topological degrees of freedom $|G|^n|{\rm Hom}(\pi_1(\Sigma_\gamma),G)|$ with $\gamma=\gamma_X+\gamma_Y+n-1$.

We end this subsection by providing a minimal example of higher genus. As shown in Fig.~\ref{fig:example-torus-2torus-d6}(b), we consider a genus 2 surface decomposed into two punctured tori, a situation corresponding to $(\gamma_X,\gamma_Y,n)=(1,1,1)$. The quantity $R_{1,1}(c)=\sum_{\alpha\in {\rm Rep}(G)} |G|\chi^\alpha(c)/d_\alpha$ actually counts how many times the commutator $[g,h]=ghg^{-1}h^{-1}$ takes value $c$ when $g,h$ runs over $G$. Note that $c$ is necessarily in the commutator subgroup $[G,G]$. For the minimal non-Abelian group $G=D_6$, we have $[D_6,D_6]\simeq\mathbb{Z}_3$ where each element can be generated from a single commutator. The corresponding block structure is shown in Fig.~\ref{fig:example-torus-2torus-d6}(b2).

\subsection{
Klein Bottle $\mathbb{K}$}
Aside from genus $\gamma$ surfaces, there are a large class of non-orientable 2D manifolds to which our general framework applies equally. The arguably most renowned example is the Klein bottle $\mathbb{K}$, which is the orientable counterpart of torus. As their fundamental groups are different, the entanglement block structures differ already for a trivial decomposition. In fact, one can show that the number of topological degrees of freedom in Eq.~(\ref{triX}) is $|G| |{\rm Rep}^*(G)|$ for the Klein bottle, where ${\rm Rep}^*(G)$ consists of real and quaternionic irreps. Recall that this quantity is $|G| |{\rm Rep}(G)|$ for torus, which is generally larger. Moreover, the Klein bottle has two different nontrivial decompositions, which are distinguished by the orientability of subsystems. We analyze both cases in the following.

\subsubsection{Tube + Tube}
\begin{figure}
\begin{center}
       \includegraphics[width=8.5cm, clip]{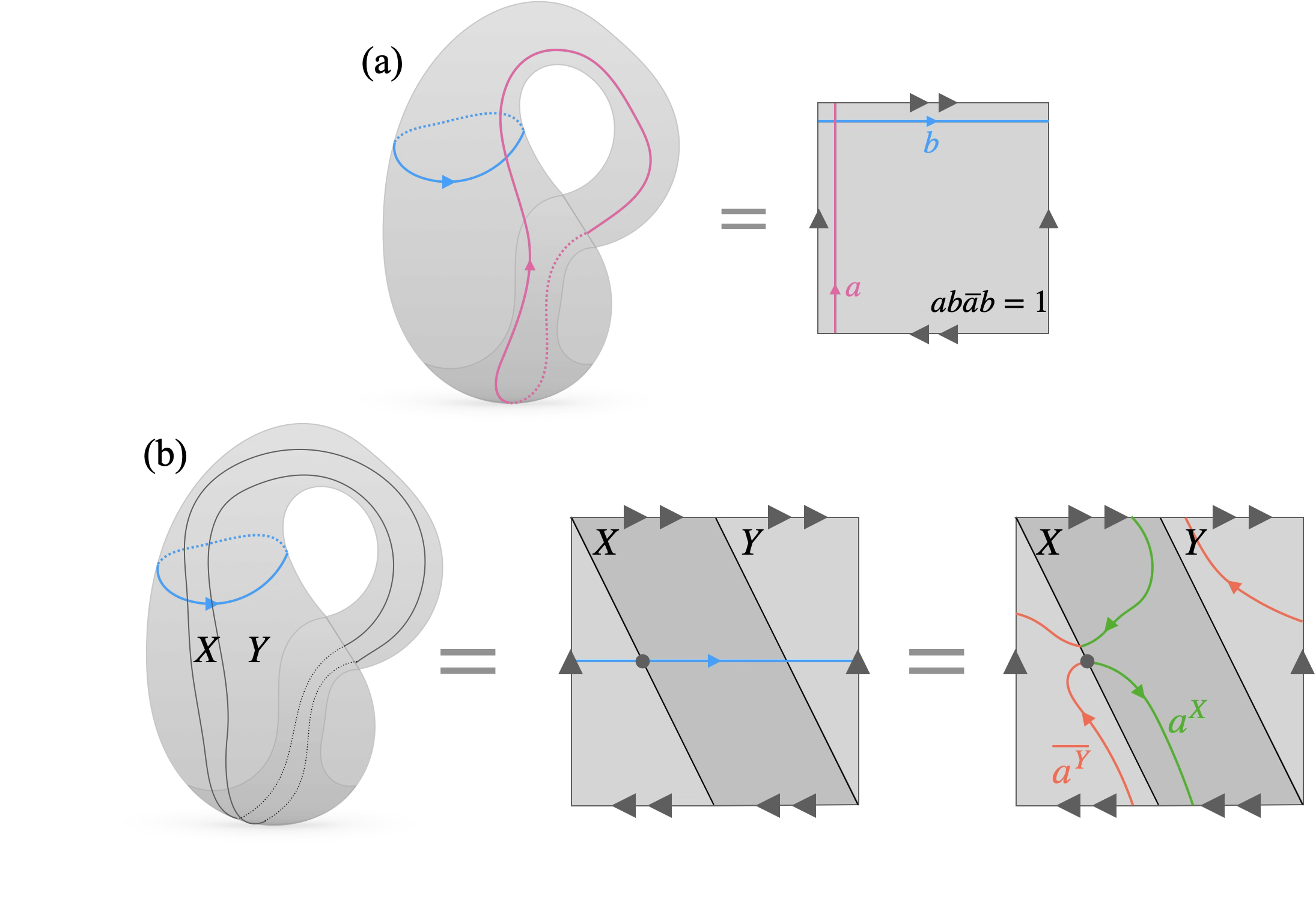}
       \end{center}
   \caption{
    (a) Constructing the Klein bottle from a square. The relation of holonomies is a bit different from the case of torus.
    (b) Bipartition of the Klein bottle to two Möbius bands. Since the boundary of the Möbius band is $S^1$, we only need one base point.
    The holonomy along the horizontal axis (blue line) can be understood as a product of two vertical holonomies (red and green curves) respectively associated to $X$ and $Y$.
   }
   \label{fig:klein-bottle-bipartition}
\end{figure}

We first consider orientable subsystems, which are actually tubes just like the case of torus. As shown in Fig.~\ref{fig:klein-bottle-bipartition}(a), the essential difference compared to torus (see Fig.~\ref{fig:torus-bipartitions}(b)) is that now one boundary circle is glued with antipodal flip, while the other boundary is not flipped. This antipodal flip leads to the non-orientable nature of $\mathbb{K}$. Algebraically, this means the holonomy around one based point in the two tubes should be identified, while that around the other based point should be the inverse of each other. Taking this effect into account, we find that the matrix $W$ takes the form
\begin{equation}
  \bigoplus_{c\in G_*}\bigoplus_{j=1}^{|[c]||G|^{|V_\partial|-1}}
  \mathbb{C}^{
    |{\rm C}_c||G|^{|V_X|} \times |{\rm C}_c||G|^{|V_Y|}
  },
  \label{ttblock}
\end{equation}
where $G_*=\{c\in G: [c]=[c^{-1}]\}$ is the set of group elements whose conjugacy classes are the same as those of their inverses. This result resembles Eq.~(\ref{eq:blocksize-torus-nontrivial}) a lot except for that only those inverse-closed conjugacy classes are involved. One can check that the number of total topological degrees of freedom is indeed $|G||{\rm Rep}^*(G)|$ following the duality between inverse-closed conjugacy classes and non-complex irreps, a refinement of the well-known duality between conjugacy classes and irreps.

\subsubsection{Möbius band + Möbius band}
In a qualitatively different bipartition, the subsystems can be non-orientable themselves. In the case of Klein bottle, the subsystems are nothing but M\"{o}bius bands, while the boundary is a single circle.
See Fig.~\ref{fig:klein-bottle-bipartition} for pictorial explanations. The fundamental group of a M\"{o}bius band is the same as the tube (as well as $S^1$), so the subsystem and boundary ${\rm Hom}$ sets in Eq.~(\ref{HomvK}) are all $G$. On the other hand, now $r_{X,Y}$ is not an isomorphism but rather sends $g$ to $g^2$, as the boundary loop is actually square of the generater. Using this fact, we find that the matrix $W$ takes the form
\begin{equation}
  \bigoplus_{c\in G:K(c)>0}\bigoplus_{j=1}^{|G|^{|V_\partial|-1}}
  \mathbb{C}^{K(c)|G|^{|V_X|} \times K(c)|G|^{|V_Y|}
  },
  \label{mmblock}
\end{equation}
where $K(c)=|\{g\in G:g^2=c\}|$ counts how many times $g^2$ gives $c$ when $g$ runs over $G$. In terms of irrep characters, this quantity can be rewritten as 
\begin{equation}
  K(c)=\sum_{\alpha \in{\rm Rep}(G)} \iota^\alpha \chi^\alpha(c), 
\label{K11}
\end{equation}
where $\iota^\alpha=|G|^{-1}\sum_{g\in G}\chi^\alpha(g^2)$ is the Frobenius-Schur indicator that distinguishes complex ($\iota_\alpha=0$), real ($\iota_\alpha=1$) and quaternionic ($\iota_\alpha=-1$) irreps.

Noting that $|{\rm Rep}^*(G)|$ in the previous subsection can also be expressed as $\sum_{\alpha\in{\rm Rep}(G)}\iota^2_\alpha$, we expect that the Frobenius-Schur indicator appears ubiquitously in the entanglement structure of non-orientable surfaces. In the next section, we consider general partitions of general non-orientable surfaces and show that this expectation is indeed correct. 
This may not be a big surprise since both orientability and the Frobenius-Schur indicator are some $\mathbb{Z}_2$ or involutionary properties. 
In particular, within the context of TQFT, time-reversal symmetry and non-orientable surfaces are inseparable concepts, and the Frobenius-Schur indicator appears \cite{Tachikawa_2017,Orii_2025,orii2025generalizationanomalyformulatime,barkeshliReflectionTimeReversal2020}.

\subsection{General Non-orientable Surface}
According to the classification theorem for 2-dimensional topological surfaces \cite{gallierGuideClassificationTheorem2013,francisConwaysZIPProof1999}, it suffices to consider biparition of surfaces with $k$ \emph{crosscaps} denoted as $N_k$, where $k$ can be any positive integers. They can be considered as the non-orientable counterparts of genus $\gamma$ surfaces $\Sigma_\gamma$, and $2\gamma$ turns out to be the appropriate analogue of $k$. Indeed, a local handle with genus $1$ can always be deformed into two crosscaps on a non-orientable surface. Note that the Klein bottle, the non-orientable counterpart of $\Sigma_1$, is topologically equivalent to $N_2$. Unlike the orientable case where subsystems are always orientable, here there are three qualitatively different ways of bipartition: both orientable, both non-orientable, or orientable plus non-orientable.

\subsubsection{Both Orientable}
We construct a closed non-orientable surface $N_k$ from two orientable surfaces $\Sigma_{\gamma_X,n}$ and $\Sigma_{\gamma_Y,n}$ with $n$ punctures.
The data of antipodal flip is specified as a sequence of signs $s=\{s_i=\pm 1\}^n_{i=1}$.
If all the signs coincide, the resulting manifold 
is orientable and has been considered in Sec.~\ref{sec:ggs}. Otherwise, the entire manifold is non-orientable and its crosscap number is given by 
$k=2\gamma_X+2\gamma_Y+2n-2$.
The diagram of manifold partition is as follows:
\begin{equation}
\begin{tikzcd}
M=N_{2\gamma_X+2\gamma_Y+2n-2} & X=\Sigma_{\gamma_X,n} \arrow[l, hookrightarrow]  \\
Y=\Sigma_{\gamma_Y,n}  \arrow[u, hookrightarrow] & \partial = \sqcup^n S^1 \arrow[u, hookrightarrow] \arrow[l, hookrightarrow] 
\end{tikzcd}
\end{equation}
Introducing the notation $c^s=(c_1^{s_1},...,c_n^{s_n})$ as an involution on $G^{\times n}$, we find that the matrix $W$ takes the form
\begin{equation}
\begin{split}
  \bigoplus_{c\in G^{\times n}}
  \bigoplus_{j=1}^{|G|^{|V_\partial|- n}}
  \mathbb{C}^{
    R_{\gamma_X,n}(c)|G|^{|V_X|} \times 
    R_{\gamma_Y,n}(c^s)
    |G|^{|V_Y|}
  },
\end{split}
\label{2oblock}
\end{equation}
where the class function $R_{\gamma,n}(c)$ on $G^{\times n}$ has already been given in Eq.~(\ref{eq:blocksize-genus-g-surface}). Just like Eq.~(\ref{ggsblock}), it is better to restrict the first sum range to ${\rm Im}=\{c\in G^{\times n}:R_{\gamma_X,n}(c)>0,R_{\gamma_Y,n}(c^s)>0\}$.

Let us check the consistency with the result (\ref{ttblock}) for the tube decomposition of the Klein bottle, which corresponds to $(\gamma_X,\gamma_Y,n)=(0,0,2)$ and $s_1=-s_2$. As we have encountered in Sec.~\ref{sec:ggs}, $R_{0,2}(c)=|{\rm C}_{c_1}|\delta_{[c_1],[c_2^{-1}]}$ and  $R_{0,2}(c^s)=|{\rm C}_{c_1}|\delta_{[c_1],[c_2]}$. Therefore, ${\rm Im}=\{(c_1,c_2)\in G^{\times 2}: [c_1]=[c_2]=[c_2^{-1}]\}$ and indeed $c_1$ (as well as $c_2$) should lie in $G_*$. From Eq.~(\ref{2oblock}), we can also figure out the total topological degrees of freedom to be $|G|^n\sum_{\alpha\in{\rm Rep}(G)}\iota_\alpha^k (|G|/d_\alpha)^{k-2}$, consistent with $|G|^{n-1}|{\rm Hom}(\pi_1(N_k),G)|$. While $k$ is always even here, this result holds also for odd $k$.

\subsubsection{Both Non-orientable}
We move on to consider the case that both subsystems are non-orientable, a situation unique to non-orientable manifolds. As a generalization of $N_k$, we use $N_{k,n}$ to denote a non-orientable surface with $k$ crosscaps and $n$ boundaries. Suppose $N_k=N_{k,0}$ can be decomposed into $N_{k_X,n}$ and $N_{k_Y,n}$, where $k_X,k_Y,n$ are all positive integers, we have $k=k_X+k_Y+2n-2$. The diagram of manifold partition is as follows:
\begin{equation}
\begin{tikzcd}
M=N_{k_X+k_Y+2n-2} & X=N_{k_X,n} \arrow[l, hookrightarrow]  \\
Y=N_{k_Y,n}  \arrow[u, hookrightarrow] & \partial = \sqcup^n S^1 \arrow[u, hookrightarrow] \arrow[l, hookrightarrow] 
\end{tikzcd}
\end{equation}
While formally one can consider antipodal flips of boundaries specified by $s\in\{\pm1\}^{\times n}$, this turns out to have no effect at all as a result of non-orientability. The matrix $W$ is found to take the form
\begin{equation}
\begin{split}
  \bigoplus_{ c\in G^{\times n}}
  \bigoplus_{j=1}^{|G|^{|V_\partial|- n}}
  \mathbb{C}^{K_{k_X,n}(c)|G|^{|V_X|} \times 
    K_{k_Y,n}(c)|G|^{|V_Y|}
  },
\end{split}
\label{2nblock}
\end{equation}
where $K_{k,n}(c)$ is a class function on $G^{\times n}$ given by
\begin{equation}
K_{k,n}(c)=\sum_{\alpha\in{\rm Rep}(G)} \iota_\alpha^k \left(\frac{|G|}{d_\alpha}\right)^{k+n-2} \prod^n_{i=1}\chi^\alpha(c_i).
\label{Kkn}
\end{equation}
This function can be regarded as the non-orientable counterpart of $R_{\gamma,n}(c)$ in Eq.~(\ref{eq:blocksize-genus-g-surface}). Again we find that $2\gamma$ is now replaced by $k$, and more crucially there appears Frobenius-Schur indicators that exclude complex irreps in the sum. In particular, unlike $R_{\gamma,n}(c)$, $K_{k,n}(c)=K_{k,n}(c^s)$ is invariant under the involution by any $s$, since $\chi^\alpha(c)=\chi^\alpha(c^{-1})$ for those non-complex irreps.

One can easily check the consistency with the result (\ref{mmblock}) for the M\"{o}bius-band decomposition of the Klein bottle, which corresponds to $(k_X,k_Y,n)=(1,1,1)$. Indeed, $K(c)$ in Eq.~(\ref{K11}) is nothing but $K_{1,1}(c)$ in Eq.~(\ref{Kkn}). More generally, one can confirm the relation
\begin{equation}
|G|^nK_{k,0}= \sum_{c\in G^{\times n}}K_{k_X,n}(c)K_{k_Y,n}(c),
\label{KKK}
\end{equation}
where the lhs is the number of total topological degrees of freedom $|G|^{n-1}|{\rm Hom}(\pi_1(N_k),G)|$ with $k=k_X+k_Y+2n-2$.

\subsubsection{Orientable Plus Non-orientable}
Finally, let us discuss the remaining case where only one of the subsystems is non-orientable. This way of bipartition is possible for any $N_k$ except for the Klein bottle $\mathbb{K}=N_2$. The reason is that suppose $N_k$ is decomposed into $\Sigma_{\gamma_X,n}$ and $N_{k_Y,n}$, we have $k=2\gamma_X+k_Y+2n-2$. Since $\gamma_X\ge0, k_Y\ge1$ and $n\ge1$, the only value that $k$ cannot take is $2$.
The diagram of manifold partition is as follows:
\begin{equation}
\begin{tikzcd}
M=N_{2\gamma_X+k_Y+2n-2} & X=\Sigma_{\gamma_X,n} \arrow[l, hookrightarrow]  \\
Y=N_{k_Y,n}  \arrow[u, hookrightarrow] & \partial = \sqcup^n S^1 \arrow[u, hookrightarrow] \arrow[l, hookrightarrow] 
\end{tikzcd}
\end{equation}
Similar to the both non-orientable case, antipodal flips do not make an essential difference. The matrix $W$ is found to take the form 
\begin{equation}
\begin{split}
  \bigoplus_{ c\in G^{\times n}}
  \bigoplus_{j=1}^{|G|^{|V_\partial|- n}}
  \mathbb{C}^{R_{\gamma_X,n}(c)|G|^{|V_X|} \times 
    K_{k_Y,n}(c)|G|^{|V_Y|}
  },
\end{split}
\label{onblock}
\end{equation}
which involves both orientable and non-orientable class functions $R_{\gamma,n}(c)$ and $K_{k,n}(c)$. While $R_{\gamma,n}(c)$ is generally not invariant under $c\to c^s$, the direct sum over $G^{\times n}$ and the involution invariance of $K_{k,n}(c)$ make the block structure independent of $s$.

We remark that the simplest non-orientable surface $N_1$, which is nothing but the real projective plane $\mathbb{RP}^2$, has a unique bipartition $(\gamma_X,k_Y,n)=(0,1,1)$ falling into the current class. Of course the block structure is relatively trivial in the sense of Eq.~(\ref{triX}), where $|{\rm Hom}(\pi_1(N_1),G)|=K_{1,1}(1)=\sum_{\alpha\in{\rm Rep}(G)}\iota_\alpha d_\alpha$ (recall that $R_{0,1}(c)=\delta_{c,1}$). In general, there is also a consistency relation for topological degrees of freedom analogous to Eqs.~(\ref{RRR}) and (\ref{KKK}):
\begin{equation}
|G|^nK_{k,0}= \sum_{c\in G^{\times n}}R_{\gamma_X,n}(c)K_{k_Y,n}(c),
\label{KRK}
\end{equation}
where $k=2\gamma_X+k_Y+2n-2$.

\subsection{Heegaard Splitting of 3D Manifolds}
\label{sec:HS}
So far we have provided results for general bipartitions of 2D manifolds, which can be either orientable or non-orientable. 
In contrast, 3D manifolds are of much higher complexity and the ultimate classification of them, including those non-orientable and/or with boundaries, contains many unsolved problems and is still an on-going project \cite{kirby1978problems,Rubinstein_2007}.
Indeed, the Thurston's geometrization theorem \cite{thurston1982three,perelman2002entropyformularicciflow,perelman2003ricciflowsurgerythreemanifolds,perelman2003finiteextinctiontimesolutions} remained a conjecture until this century, though the theorem is severely restricted to the orientable and closed subclass. As a thorough study on bipartite 3D manifolds would be a huge program that relies heavily on pure mathematics progress, we focus on an elementary class of bipartition called \emph{Heegaard spliting} \cite{heegaard1898forstudier,fomenko2013algorithmic,saveliev2011lectures}. In fact,
every 3D closed orientable manifold admits a Heegaard splitting into two genus $\gamma$ handlebodies $H_\gamma$. On the other hand, there can be several different Heegaard splittings for a given manifold, as will be exemplified below. 

\begin{figure}
\begin{center}
       \includegraphics[width=8cm, clip]{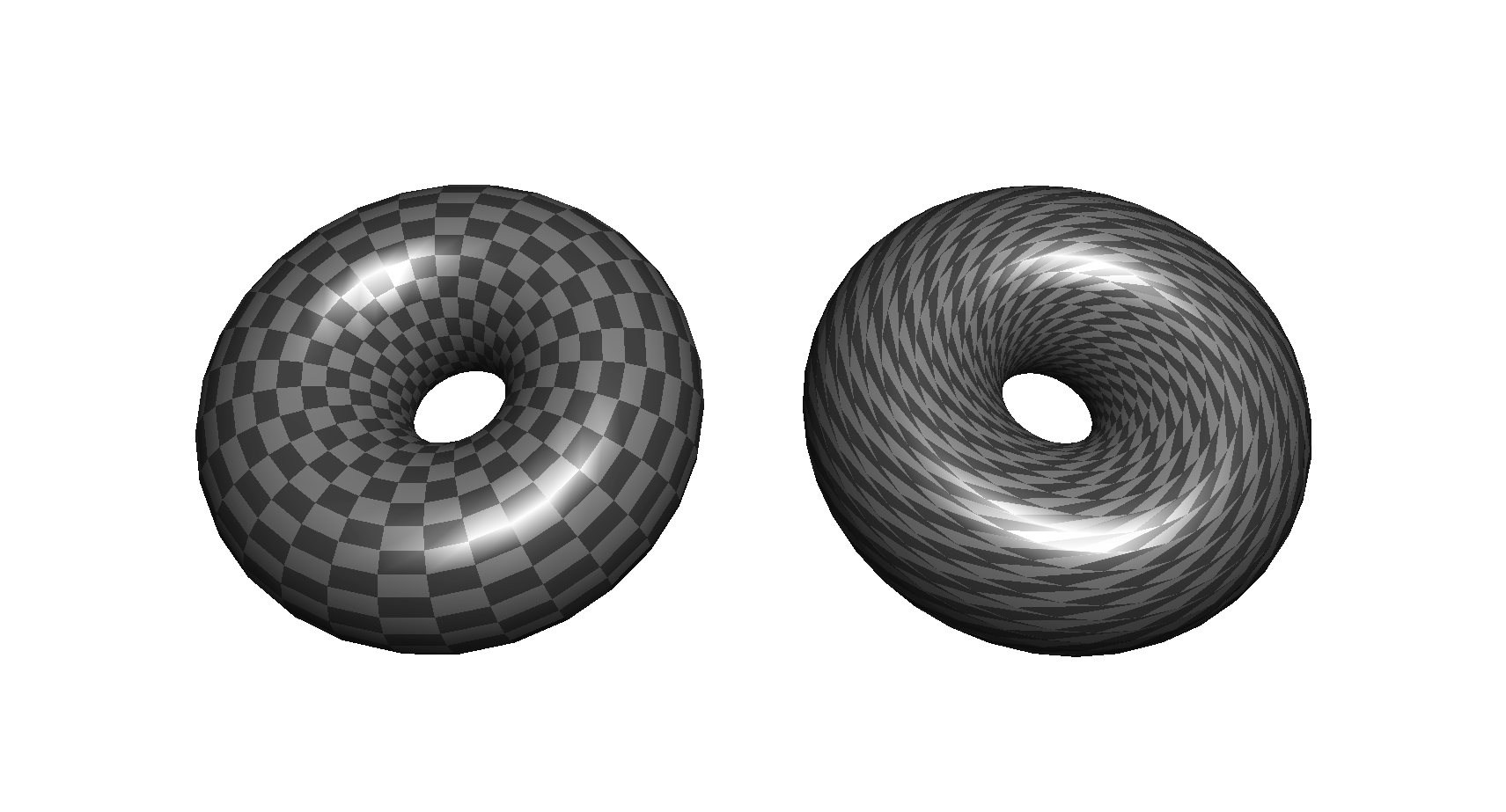}
       \end{center}
   \caption{Heegaard splitting of lens space is achieved by gluing together two solid tori. One of these tori is generally twisted by the action of the modular group $\mathrm{SL}(2;\mathbb{Z})$. By precisely aligning and joining the meshes drawn on the two tori in the figure, a lens space is obtained.}
   \label{fig:heegaard-1}
\end{figure}

The Heegaard splitting of genus 0, i.e., that into two (3D) balls $H_0$, yields only $S^3$. 
The genus-1 case represents the first nontrivial example known as \emph{the lens space} $L(q;p)$, where $p$ and $q$ are coprime integers. In this case, the gluing map $\varphi$ is classified by the orientation-preserving mapping class group ${\rm MCG}_+$ of the torus $\mathbb{T}^2$ isomorphic to $\mathrm{SL}(2;\mathbb{Z})$. See Fig.~\ref{fig:heegaard-1} for pictorial explanations. Precisely speaking, we have the following commutative diagram
\begin{equation}
\begin{tikzcd}
M=L(q;p) & X=H_\gamma \arrow[l, hookrightarrow]  \\
Y=H_\gamma  \arrow[u, hookrightarrow] & \partial = \Sigma_\gamma \arrow[u, hookrightarrow,"\varphi"] \arrow[l, hookrightarrow] 
\end{tikzcd}
\end{equation}
where the homotopy class of gluing map $\varphi$ is determined by a $2\times 2$ matrix in $\mathrm{SL}(2;\mathbb{Z})$:
\begin{equation}
\begin{pmatrix}
  p&r\\q&s
\end{pmatrix},\quad ps-qr=1,\quad p,q,r,s \in \mathbb{Z}.
\end{equation}
It turns out that $r_{X,Y}$ in the diagram of ${\rm Hom}$ sets is injective, so given any $\phi$ in ${\rm Im}$ the block size is $|r_X^{-1}(\phi)|=|r_Y^{-1}(\phi)|=1$. We thus only have to determine the number of blocks $|{\rm Im}|$, which is given by $|{\rm Hom}(\pi_1(L(q;p)),G)|=|{g\in G:g^q=1}|$ since $\pi_1(L(q;p)=\mathbb{Z}_q$. In terms of irreps, this quantity can be evaluated as $|\mathrm{Im}|=\sum_{\alpha \in{\rm Rep}(G)} d_\alpha\, \nu^\alpha_q$, where we have used the higher Frobenius–Schur indicator $\nu_q^\alpha$ defined as follows:
\begin{equation}
  \nu_q^\alpha=\frac{1}{|G|}\sum_{g\in G} \chi^\alpha (g^q).
\end{equation}
In particular, taking $q=1$ such that $L(1;p)=S^3$, we get a trivial example (in the sense of Eq.~(\ref{tri})) where the fundamental group of subsystem is nontrivial.

It appears strange that the outcome solely hinges upon the value of $q$, yet the Reidemeister-Singer theorem \cite{reidemeisterZurDreidimensionalenTopologie1933,Singer1933ThreedimensionalMA,fomenko2013algorithmic,saveliev2011lectures} provides an interpretation of this.
For the Heegaard splitting of a fixed genus, that theorem claims the double coset $\mathfrak{H}_\gamma \backslash \mathrm{MCG^+} /\mathfrak{H}_\gamma$ is the 1-to-1 mapping with the equivalence classes of resulting manifold by the Heegaard splitting. 
Here, $\mathfrak{H}_\gamma$ is the subgroup of the mapping class group $\mathrm{MCG}^+$ that can be extended to the interior of genus $\gamma$ handlebody.
In the case of genus 1, we can take the generator of $\mathfrak{H}_1$ as $T$ transformation of the full modular group $\mathrm{SL}(2;\mathbb{Z})$. Computing the double coset reads that the pair $(q,p\,\mathrm{mod}\,q)$ determines the Reidemeister-Singer equivalence class. The fact that $p\,\mathrm{mod}\, q$ does not appear in the entanglement structure reflects a significant feature of the lens spaces: they cannot be solely determined by the fundamental group. 

In fact, the discussion above can be easily extended to the Heegaard splitting into $\gamma\geq 2$ handlebodies. It turns out that the block size is $|r_{X/Y}^{-1}(\phi)|=1$, and thus the number of blocks is given by $|{\rm Hom}(\pi_1(M),G)|$. While $\pi_1(M)$ is generally much more complicated in 3D compared to 2D, it enjoys many nice properties and cannot be an arbitrary finitely presented group \cite{MA13}. Unlike in higher dimensions, we expect that finding a general representation-theory-based formula like $R_{\gamma,0}$ and $K_{k,0}$ for $|{\rm Hom}(\pi_1(M),G)|$ in 3D may not be so hopeless.

\subsection{$n$-dimensional Torus $\mathbb{T}^n$}

As a final example, we consider $M = \mathbb{T}^n=(S^1)^{\times n}$ ($n\ge2$) to show that our framework is applicable to arbitrary dimensions. Two subsystems are chosen as
\begin{equation}
X=\left[0,\pi \right]^{\times k} \times (S^1)^{\times (n-k)}
\end{equation}
(for $1\le k\le n$; here $S^1=[0,2\pi]/\{0\sim1\}=2\pi\mathbb{R}/\mathbb{Z}$) and $Y=\overline{\mathbb{T}^n \backslash X}$ with common boundary $\partial=\partial_k\times (S^1)^{\times (n-k)}$ ($\partial_k$: boundary of $[0,\pi]^{\times k}$). If $k=1$, $\partial$ has two connected components, i.e., $|A|=2$. Otherwise, there is a single base point as the boundary is 
connected.
For $k \geq 2$, we find that $W$ matrix takes the form
\begin{equation}
  \bigoplus_{
    \phi \in \mathrm{Comm}_{n-k}(G)
    }
  \bigoplus_{j=1}^{|G|^{|V_\partial|-1}}
  \mathbb{C}^{
    |G|^{|V_X|} \times |\mathrm{Comm}_{k}(\mathrm{C}_\phi)||G|^{|V_Y|}
  },
  \label{hdt2}
\end{equation}
where ${\rm Comm}_m(G)=\{(g_1,...,g_m)\in G^{\times m}: g_ig_j=g_jg_i, \forall 1\le i<j\le m\}$ consists of $m$ mutually commuting group elements, and ${\rm C}_\phi=\{g\in G: g g_i = g_i g, \forall i=1,...,m\}$ is the centralizer with respect to $\phi=(g_1,...,g_m)$. Only when $k=1$ does $W$ exhibit an exceptional behavior:
\begin{equation}
  \bigoplus_{
    \substack{\phi \in \mathrm{Comm}_{n-1}(G)}
  }
  \bigoplus_{j=1}^{|[\phi]| |G|^{|V_\partial|-2}}
  \mathbb{C}^{
    |\mathrm{C}_{\phi}
    ||G|^{|V_X|} \times |\mathrm{C}_{\phi}
    ||G|^{|V_Y|}
  },
  \label{hdt1}
\end{equation}
where $[\phi]$ is the orbit set upon component-wise conjugate $G$-action, thus $|G|=|[\phi]||{\rm C}_\phi|$ according to the orbit-stabilizer theorem. Setting $n=2$, we can check that Eq.~(\ref{hdt1}) (Eq.~(\ref{hdt2})) reproduce the previous result for torus under non-trivial (trivial) bipartition.

\section{Gauge invariance and topological entanglement entropy}\label{sec-tee}
While gauge transformation was used in the derivation of the main result (\ref{bs}), our general setup only requires loop symmetries and is \emph{not} a gauge theory. In this sense, gauge transformation is more like a technique rather than physical essence. On the other hand, we can definitely add further constraint on top of loop symmetries, and a natural choice is to impose gauge invariance. This further constraint turns out to be strong enough to kill almost all the macroscopic degrees of freedom, leaving only a few allowed many-body states. In fact, our setup becomes a topological gauge theory, which reduces to the Kitaev quantum double model in 2D \cite{kitaevFaulttolerantQuantumComputation1997}, a prototypical example of topological order and topological quantum code. We will see that not only the well-known results about the topological entanglement entropy \cite{kitaevTopologicalEntanglementEntropy2006,levinDetectingTopologicalOrder2006}
can be reproduced, but also more refined information about the full entanglement spectrum can be extracted from our framework.

\subsection{Modified Framework}
We start from Eq.~(\ref{bs}) or (\ref{tg}) and consider the effect of gauge invariance on the entanglement structure. The gauge invariance within $X$ or $Y$ enforces each geometric block $\mathbb{C}^{|G|^{|V_X|} \times  |G|^{|V_Y|}}$ to be $1_{|G|^{|V_X|}\times |G|^{|V_Y|}}$ ($1_{m\times n}$: $m\times n$ matrix with all the entries being $1$), which is of rank $1$. This is because the entries are enforced to be equal by gauge invariance whenever their corresponding basis are related by a gauge transformation. Since the basis are obtained from gauge transforming a fixed representative, all of them are related by gauge transformations and thus the entries should be identical. Following the same reasoning, we know that the gauge invariance at the boundaries except for the base points enforce $|G|^{|V_\partial|-|A|}$ blocks to be identical. So far, the block structure has been simplified into 
\begin{equation}
\left(\bigoplus_{\phi\in {\rm Im}} \mathbb{C}^{|r_X^{-1}(\phi)| \times |r_Y^{-1}(\phi)| }\right)\otimes\left( \bigoplus^{|G|^{|V_\partial|-|A|}}_{j=1} 1_{|G|^{|V_X|}\times |G|^{|V_Y|}}\right),
\end{equation}
where the geometric part may be rewriten as $\mathbb{1}_{|G|^{|V_\partial|-|A|}}\otimes 1_{|G|^{|V_X|}\times |G|^{|V_Y|}}$ ($\mathbb{1}_n$: $n\times n$ identity), highlighting a macroscopically large degeneracy proportional to the boundary Hilbert-space dimension. We can clearly see that all the geometric degrees of freedom are killed by gauge invariance. The only remaining task is to analyze the effect of gauge invariance on the based points, which may modify the topological degrees of freedom. 

We recall that the $\Hom$-applied Seifert-van Kampen diagram (\ref{HomvK}) for determining the topological degrees of freedom is a pullback (fiber product) in the category of sets. 
To incorporate gauge invariance, we promote this diagram to a pullback in the \emph{category of $G$-sets} consisting of sets equipped with group actions and morphisms being $G$-equivariant maps. Precisely speaking, when there are $|A|$ base points, we consider the group action of $G^{\times |A|}$ corresponding to the gauge transformations on the based points. The action of $\boldsymbol{g}=(g_1,g_2,...,g_{|A|})\in G^{\times |A|}$ on a Hom element $\varphi$ is given by 
\begin{equation}
\boldsymbol{g}\cdot\varphi(f_{ij})= g_j \varphi(f_{ij}) g_i^{-1},
\label{eq-gt}
\end{equation}
where $f_{ij}$ is a morphism in the fundamental groupoid from the $i$th base point to the $j$th base point ($i,j=1,2,...,|A|$). If $i=j$, the action reduces to a conjugation (inner automorphism). 
The problem of determining the gauge-invariant topological degrees of freedom has so far been formalized as a problem of finding the $G$-orbits in a fiber product, which we analyze in detail as follows.

Similar to the geometric part, gauge invariance on the topological part may identify different blocks in the direct sum and introduce further constraints within each block. However, this is usually not so simple as the geometric case, where all the entries are identified as there is only a single orbit generated by group actions. To separate the inter- and intra-block effects, we first determine the stabilizer subgroup $G_\phi=\{\boldsymbol{g}\in G^{\times |A|}:\boldsymbol{g}\cdot\phi=\phi\}$ for a given block label $\phi\in{\rm Im}$. Thanks to the $G$-equivariant nature of the morphisms, the group elements outside the subgroup will identify the blocks labeled by other elements in the orbit $[\phi]$. The subgroup itself 
may still have nontrivial actions on each block $\mathbb{C}^{|r_X^{-1}(\phi)| \times |r_Y^{-1}(\phi)| }$, which can be formally worked out using representation theory. 
Let $D^{X,Y}_\phi$ be the representation for this intra-block action that permutates rows/columns following the group action on the underlying set. 
The action of $G_\phi$ on the vector space $\mathbb{C}^{|r_X^{-1}(\phi)| \times |r_Y^{-1}(\phi)| }$ is then given as 
\begin{equation}
\begin{split}
&\boldsymbol{g}_\phi\cdot W_\phi
=D^X_\phi (\boldsymbol{g}_\phi)W_\phi D^Y_\phi (\boldsymbol{g}_\phi)^\dagger,\\
\end{split}
\label{rep}
\end{equation} 
where $\boldsymbol{g}_\phi\in G_\phi$ and $W_\phi$ is an element in $\mathbb{C}^{|r_X^{-1}(\phi)| \times |r_Y^{-1}(\phi)| }$ (i.e., $|r_X^{-1}(\phi)| \times |r_Y^{-1}(\phi)|$ matrix under some basis). Recall that $D^{X,Y}_\phi$ is a group representation, we can always decompose it into
\begin{equation}
\begin{split}
D^X_\phi (\boldsymbol{g}_\phi)&=\bigoplus_{\boldsymbol{\alpha}_\phi} \mathbb{1}_{x_{\boldsymbol{\alpha}_\phi}}\otimes D_{\boldsymbol{\alpha}_\phi} (\boldsymbol{g}_\phi),\\
D^Y_\phi (\boldsymbol{g}_\phi)&=\bigoplus_{\boldsymbol{\alpha}_\phi} \mathbb{1}_{y_{\boldsymbol{\alpha}_\phi}}\otimes D_{\boldsymbol{\alpha}_\phi} (\boldsymbol{g}_\phi),
\end{split}
\end{equation}
where $\boldsymbol{\alpha}_\phi$ is an irrep of $G_\phi$ with dimension $d_{\boldsymbol{\alpha}_\phi}$, and $x_{\boldsymbol{\alpha}_\phi}/y_{\boldsymbol{\alpha}_\phi}$ denotes its multiplicity in the representation on $\mathbb{C}^{|r_{X/Y}^{-1}(\phi)|}$. According to character theory and the realness of $D^{X,Y}_\phi$, these non-negative integers can be calculated by the following formula:
\begin{equation}
\begin{split}
x_{\boldsymbol{\alpha}_{\phi}}&=\frac{1}{|G_\phi|}\sum_{\boldsymbol{g}_\phi \in G_\phi}\chi^{\boldsymbol{\alpha}_\phi}(\boldsymbol{g}_\phi)\Tr D_\phi^X(\boldsymbol{g}_\phi),\\
y_{\boldsymbol{\alpha}_{\phi}}&=
\frac{1}{|G_\phi|}\sum_{\boldsymbol{g}_\phi \in G_\phi}\chi^{\boldsymbol{\alpha}_\phi}(\boldsymbol{g}_\phi)\Tr D_\phi^Y(\boldsymbol{g}_\phi).
\end{split}
\label{xyap}
\end{equation}
Using Schur's lemma, we know that each $G_\phi$-invariant element in $\mathbb{C}^{|r_X^{-1}(\phi)| \times |r_Y^{-1}(\phi)| }$ is further constrained to be $\bigoplus_{\boldsymbol{\alpha}_\phi} \mathbb{C}^{x_{\boldsymbol{\alpha}_\phi}\times y_{\boldsymbol{\alpha}_\phi}}\otimes \mathbb{1}_{d_{\boldsymbol{\alpha}_\phi}}$. Recalling that the blocks labeled by those $\phi$ in the same orbit $[\phi]$ will be enforced to be identical, we finally conclude that the gauge invariant topological part reads
\begin{equation}
  \bigoplus_{[\phi]\in {\rm Im}/G^{\times |A|}}\mathbb{1}_{|[\phi]|} \otimes\left(\bigoplus_{\boldsymbol{\alpha}_{\phi}\in {\rm Rep}(G_\phi)} \mathbb{C}^{x_{\boldsymbol{\alpha}_{\phi}}\times y_{\boldsymbol{\alpha}_{\phi}}}\otimes \mathbb{1}_{d_{\boldsymbol{\alpha}_{\phi}}}\right). 
\label{W}
\end{equation}  
Note that $\phi$ can be any representative in $[\phi]$ because $G_\phi$'s, as well as the corresponding $G$-sets $r_{X/Y}^{-1}(\phi)$ equipped with the action of $G_\phi$, are isomorphic for $\phi$'s in the same orbit. We may further constrain the direct sum range of $\boldsymbol{\alpha}_{[\phi]}$ to be those with $x_{\boldsymbol{\alpha}_{[\phi]}}>0$ and $y_{\boldsymbol{\alpha}_{[\phi]}}>0$, as some irrep may not appear in $D^{X,Y}_\phi$ and thus do not contribute a block.

Let us examine Eq.~(\ref{W}) in some simple cases. Suppose $G$ is Abelian such that the group actions on the boundary ${\rm Hom}$ set are all trivial, we have $G_\phi =G^{\times |A|}$, $[\phi]=\{\phi\}$. On the other hand, due to the inter-base-point paths, we have $\Tr D^{X/Y}_\phi(\boldsymbol{g})=|r^{-1}_{X/Y}(\phi)|\delta_{\boldsymbol{g}}$, where $\delta_{\boldsymbol{g}}$ is $1$ if all the components in $\boldsymbol{g}$ are the same but otherwise vanishes. Using the fact that irreps of direct product groups are tensor products of individual's irreps and the Pontryagin duality, we can explicitly evaluate $x_{\boldsymbol{\alpha}_{[\phi]}}$ and $y_{\boldsymbol{\alpha}_{[\phi]}}$ to be $|r_{X/Y}^{-1}(\phi)|/|G|^{|A|-1}$, provided the product of components in $\boldsymbol{\alpha}_{[\phi]}$ gives the trivial irrep. Overall, the topological part is obtained to be 
\begin{equation}
 \bigoplus_{\phi\in{\rm Im}}\bigoplus^{|G|^{|A|-1}}_{j=1} \mathbb{C}^{|G|^{1-|A|}|r^{-1}_X(\phi)|\times |G|^{1-|A|}|r^{-1}_Y(\phi)|},
 \label{abtop}
\end{equation}
whose total number of degrees of freedom is reduced by a factor $|G|^{|A|-1}$ compared to Eq.~(\ref{tg}) and thus coincides with $|{\rm Hom}(\pi_1(M),G)|$. We then consider the case of general $G$ but a trivial subsystem $X$ with connected boundary. Now ${\rm Im}$ only consists of the trivial group element and only the trivial irrep contributes to the direct sum. Therefore, only the geometric part survives in $\rho_X=WW^\dagger$, i.e., $\rho_X\propto \mathbb{1}_{|G|^{|V_\partial|-1}}$, implying that the entanglement entropy is $(|V_\partial|-1)\ln |G|$ for any normalized state. This result reproduces the negative correction $\ln |G|$, well known as the topological entanglement entropy (TEE) in 2D, to the entanglement area law, and is applicable to arbitrary dimensions. The same result actually applies to any $M$ with trivial fundamental group and connected $\partial$, even if $X$ and $Y$ are nontrivial.

\subsection{Topological entanglement entropy}

As already noted in many previous studies \cite{dongTopologicalEntanglementEntropy2008,zhangQuasiparticleStatisticsBraiding2012}, the topological entanglement entropy for a topologically nontrivial bipartition may change considerably compared to the trivial case as discussed above. It may even depend on the explicit choice of many-body state, as local indistinguishability only works for the trivial bipartition. These works are usually restricted to specific groups or/and manifolds. 
In fact, the analysis of topological entanglement entropy for non-Abelian topological gauge theory in general dimensions and general bipartitions has been a challenging problem. Refining our algorithm provides a non-heuristic alternative for rigorous calculation of topological entanglement entropy for (generally non-Abelian) topological gauge theory in general dimension and general bipartition.
Below, we briefly describe how to compute the topological entanglement entropy within our modified framework. 

Let us first focus on the so-called minimally entangled states \cite{zhangQuasiparticleStatisticsBraiding2012} whose $W$ matrix is determined by a rank-1 element in some topological block $\mathbb{C}^{x_{\boldsymbol{\alpha}_\phi}\times y_{\boldsymbol{\alpha}_\phi}}$ in Eq.~(\ref{W}). Here minimal should be understood in a local sense, in contrast to the global minimally entangled state within a block with minimal $|[\phi]|d_{\boldsymbol{\alpha}_\phi}$. It is clear that a complete set of mutually orthogonal 
minimally entangled states, whose superpositions can generate any gauge-invariant loop-symmetric states, are labeled by $([\phi], \boldsymbol{\alpha}_{[\phi]},x,y)$ with $x=1,2,...,x_{\boldsymbol{\alpha}_{[\phi]}}$ and $y=1,2,...,y_{\boldsymbol{\alpha}_{[\phi]}}$. The corresponding reduced density matrix $\rho_X\propto \mathbb{1}_{|[\phi]|}\otimes\mathbb{1}_{d_{\boldsymbol{\alpha}_\phi}}\otimes \mathbb{1}_{|G|^{|V_\partial|-|A|}}$, so the corresponding TEE reads 
\begin{equation}
\Gamma= |A| \ln |G| - \ln (|[\phi]|d_{\boldsymbol{\alpha}_{[\phi]}}).
\label{TEE}
\end{equation}
If $G$ is Abelian, we have $|[\phi]|=d_{\boldsymbol{\alpha}_{[\phi]}}=1$ and $\gamma=|A|\ln |G|$, consistent with Eq.~(\ref{abtop}) that does not introduce further degeneracy. The same result has been obtained in Ref.~\cite{groverEntanglementEntropyGapped2011}, where $|A|$ is interpreted as the $0$th Betti number. In general, the second term in Eq.~(\ref{TEE}) reduces the TEE in a way that depends on the topology of (sub)system.
 
We move on to consider a general gauge-invariant loop-symmetric state specified by the topological part 
\begin{equation}
\bigoplus_{[\phi]\in {\rm Im}/G^{\times |A|}}\mathbb{1}_{|[\phi]|} \otimes\left(\bigoplus_{\boldsymbol{\alpha}_{\phi}\in {\rm Rep}(G_\phi)} \Psi_{\boldsymbol{\alpha}_\phi}\otimes \mathbb{1}_{d_{\boldsymbol{\alpha}_{\phi}}}\right),
\end{equation}
where $\Psi_{\boldsymbol{\alpha}_\phi}$ is an element in $\mathbb{C}^{x_{\boldsymbol{\alpha}_{\phi}}\times y_{\boldsymbol{\alpha}_{\phi}}}$. With the normalization constant 
\begin{equation}
\mathcal{N}:=\sum_{[\phi]\in \mathrm{Im}/G^{\times |A|}} |[\phi]|\left[\sum_{\boldsymbol{\alpha}_{[\phi]}\in {\rm Rep}(G_{[\phi]})} d_{\boldsymbol{\alpha}_{[\phi]}} \Tr (\Psi_{\boldsymbol{\alpha}_{[\phi]}} \Psi_{\boldsymbol{\alpha}_{[\phi]}}^\dagger)\right]    
\end{equation}
and small-block specta
${}^\exists U_{\boldsymbol{\alpha}_{[\phi]}},\, U_{\boldsymbol{\alpha}_{[\phi]}} \Psi_{\boldsymbol{\alpha}_{[\phi]}} \Psi_{\boldsymbol{\alpha}_{[\phi]}}^\dagger U_{\boldsymbol{\alpha}_{[\phi]}}^\dagger = \mathrm{diag} (|\psi_{\boldsymbol{\alpha}_{[\phi]},i}|^2)$, the correction term to the na\"ive TEE $|A|\ln |G|$ turns out to be
\begin{equation}
\begin{split}
\sum_{[\phi]\in \mathrm{Im}/G^{\times |A|}} |[\phi]|
\sum_{\boldsymbol{\alpha}_{[\phi]}\in {\rm Rep}(G_{[\phi]})} d_{\boldsymbol{\alpha}_{[\phi]}} 
\sum_{i=1}^{x_{\boldsymbol{\alpha}_{[\phi]}}} \frac{|\psi_{\boldsymbol{\alpha}_{[\phi]},i}|^2}{\mathcal{N}}\ln \frac{|\psi_{\boldsymbol{\alpha}_{[\phi]},i}|^2}{\mathcal{N}}.
\end{split}
\end{equation}
This result returns to Eq.~(\ref{TEE}) if all $|\psi_{\boldsymbol{\alpha}_{[\phi]},i}|$'s but one vanish.

As an illustration of the general results above, we consider the case of torus and its nontrivial bipartition as shown in Fig.~\ref{fig:example-torus-2torus-d6}(a). 
Given $\phi\in{\rm Im}=\{(c_1,c_2)\in G^{\times 2}: [c_1]=[c_2]\}$, the stabilizer group is simply the product of the centralizer group of each holonomy, i.e.,
$G_\phi = \mathrm{C}_{c_1}\times \mathrm{C}_{c_2}$. Its action on $r_{X/Y}^{-1}(\phi)=\{f\in G: fc_1f^{-1}=c_2\}$ is given by $\boldsymbol{g}\cdot f= g_2fg_1^{-1}$. Following the Peter-Weyl theorem \cite{baezSpinNetworksGauge1996},
one finds $|[\phi]|=|[c]|^2$ and $d_{\bm{\alpha}_\phi}=d_{\alpha_{c}}^2$ with $[c]=[c_1]=[c_2]$ and $\alpha_c$ an irrep of ${\rm C_c}$. The TEE of a minimally entangled state is given by $\gamma=2\ln |G|-2\ln (|[c]|d_{\alpha_{c}})$. Note that $|[c]|d_{\alpha_{c}}$ is nothing but the anyon quantum dimension in the Kitaev quantum double model \cite{kitaevFaulttolerantQuantumComputation1997}. More generally, given an arbitrary ground sate of the Kitaev quantum double model, the TEE reads
\begin{equation}
\Gamma=2\ln |G|-\sum_{[c],\alpha\in{\rm C}_c} |[c]|^2 d_{\alpha}^2 
\frac{|\psi_{c,\alpha}|^2}{\mathcal{N}}\ln \frac{|\psi_{c,\alpha}|^2}{\mathcal{N}},
\label{T2TEE}
\end{equation}
which is consistent with the previous result based on the standard replica trick \cite{dongTopologicalEntanglementEntropy2008} \footnote{Their paper does not mention the explicit block structure of the density matrix, and since the multiplicity of the amplitudes does not appear, the amplitudes differ by a quantum dimensional factor from our convention, but they are consistent.}. 

Let us make a comparison with the non-orientable counterpart, i.e., 
bipartition of the Klein bottle to two tubes as shown in Fig.~\ref{fig:klein-bottle-bipartition}(a). 
One boundary holonomy is reversed during the gluing process. Therefore, $\phi=(c_1,c_2)\in{\rm Im}$ is enforced to satisfy $[c_1]=[c_2]=[c]$ for some $|[c^{-1}]|=|[c]|$. While formally the same result $\Gamma=2\ln |G|-2\ln (|[c]|d_{\alpha_{c}})$ as well as Eq.~(\ref{T2TEE}) can be derived, the sum range of $[c]$ is now restricted to those conjugacy classes satisfying $[c]=[c^{-1}]$. Finally, we recall that a Klein bottle can also be decomposed into two M\"obius bands which are themselves non-orientable (see Fig.~\ref{fig:klein-bottle-bipartition}(b)). It is not hard to find $\Gamma=\ln |G|-\ln (|[c]|d_{\alpha_c})$ in this case, though the allowed $[c]$ and $\alpha_c$ are not obvious. One necessary constraint is that $c$ should be $g^2$ for some $g\in G$, but this is not sufficient in general. In the following subsection, we will see that it is the $S$ matrix (together with the Frobenius-Schur indicator if non-orientable) that determines whether a block labeled by $[\phi]$ and $\boldsymbol{\alpha}_\phi$ appears.

\subsection{Li-Haldane correspondence}

As highlighted in the seminal work by Li and Haldane \cite{liEntanglementSpectrumGeneralization2008}, the full entanglement spectrum of a quantum many-body state contains much richer information than any specific entanglement entropy measure. In particular, they found that the entanglement spectrum of the Moore-Read state shares the same structure as the associated RCFT \cite{mooreClassicalQuantumConformal1989,mooreLecturesRCFT1990}.
This so-called Li-Haldane correspondence is expected for general 2D topologically ordered states, just like the well established correspondence between the underlying topological field theory and RCFT \cite{wittenQuantumFieldTheory1989,fuchsTFTConstructionRCFT2002,fuchsTFTConstructionRCFT2004}. However, the Li-Haldane correspondence remains largely a conjecture (except for some specific cases like free fermions \cite{LF10}), probably because the standard replica-trick-based analysis of entanglement in topological field theories focuses more on the entropy rather than the full spectrum. In contrast, our framework deals direct with the full spectrum and allows us to prove the Li-Haldane conjecture, at least for the Kitaev quantum double model.

All we have to do is to determine $[\phi]$, $\boldsymbol{\alpha}_\phi$, $x_{\boldsymbol{\alpha}_\phi}$ and $y_{\boldsymbol{\alpha}_\phi}$ for general bipartition in 2D. Suppose the boundary consists of $|A|=n$ disjoint cycles, we have $\phi=(c_1,c_2,...,c_n)\in G^{\times n}$. Accordingly, $[\phi]$ is determined by the conjugacy classes $[c_i]$ of individual components, and $G_\phi={\rm C}_{c_1}\times {\rm C}_{c_2}\times\cdots\times{\rm C}_{c_n}$. Since the irreps of direct product groups are tensor products of those of individual groups, we know that $\chi^{\boldsymbol{\alpha}_\phi}(\boldsymbol{g}_\phi)=\prod^n_{j=1}\chi^{\alpha_j}(g_j)$, where $\boldsymbol{g}_\phi=(g_1,g_2,...,g_n)$ with $g_j\in{\rm C}_{c_j}$ and $\boldsymbol{\alpha}_\phi=(\alpha_1,\alpha_2,...,\alpha_n)$ with $\alpha_j\in{\rm Rep}({\rm C}_{c_j})$. To determine $x_{\boldsymbol{\alpha}_\phi}$ and $y_{\boldsymbol{\alpha}_\phi}$, we only have to calculate $\Tr D_\phi^{X/Y}(\boldsymbol{g}_\phi)$ in Eq.~(\ref{xyap}), which is the number of gauge-invariant elements in $r_{X,Y}^{-1}(\phi)$. Here comes the explicit dependence on the topology of $X,Y$. We leave the detailed derivations in Appendix~\ref{app:xy2D} and present the final results for $x_{\boldsymbol{\alpha}_\phi}$ and $y_{\boldsymbol{\alpha}_\phi}$. Suppose $X$ is an orientable surface $X$ with genus $\gamma_X$ (and $n$ boundaries), we find that
\begin{equation}
x_{\boldsymbol{\alpha}_{\phi}}=\sum_{[g]}\sum_{\beta\in {\rm Rep}({\rm C}_g)}  S_{(g,\beta),(1,1)}^{2-2\gamma_X-n}
\prod^n_{j=1}S_{(g,\beta),(c_j,\alpha_j)},
\label{xapR}
\end{equation}
where $S_{(g,\beta),(c,\alpha)}$ is given by
\begin{equation}
S_{(g,\beta),(c,\alpha)}=\frac{1}{|{\rm C}_g||{\rm C}_c|}\sum_{\substack{h\in G,\\h^{-1}gh\in {\rm C}_c}}\chi^\beta (hch^{-1})\chi^\alpha (h^{-1}gh).
\end{equation}
One can easily check that $S_{(g,\beta),(1,1)}=|{\rm C}_g|^{-1}|G|^{-1}\sum_{h\in G}\chi^\beta(1)=|{\rm C}_g|^{-1}d_\beta$. In fact, $S$ is nothing but the modular $S$ matrix for finite group \cite{costeFiniteGroupModular2000}, or more precisely, the Drinfel'd double of the finite group $G$ \cite{dijkgraafQUASIHOPFALGEBRAS,costeFiniteGroupModular2000}. On the TQFT side, Eq.~(\ref{xapR}) gives the Hilbert space dimensions for $n$ anyons labeled by $(c_j,\alpha_j)$ on a genus $\gamma_X$ surface. Here we demonstrate that this is exactly the number of remaining degrees of freedom in the entanglement spectrum. On the RCFT side, which is classical  according to Moore and Seiberg \cite{mooreClassicalQuantumConformal1989}, this result (\ref{xapR}) is interpreted as a generalized Verlinde formula for the fusion of primary fields. We have thus verified the Li-Haldane correspondence.

Our framework also allows us to deal with non-orientable manifolds. Following similar calculations, we obtain the following result for a non-orientable surface $X$ with $k_X$ crosscaps (and $n$ boundaries):
\begin{equation}
x_{\boldsymbol{\alpha}_{[\phi]}}=\sum_{[g]}\sum_{\beta\in {\rm Rep}({\rm C}_g)} \iota_\beta^{k_X}S^{2-k_X-n}_{(g,\beta),(1,1)}\prod^n_{j=1}S_{(g,\beta),(c_j,\alpha_j)}.
  \label{xapK}
\end{equation}
Note that the same result has been derived in the context of TQFT \cite{barkeshliReflectionTimeReversal2020}. Here we confirmed that this information is perfectly inherited by the entanglement spectrum on a non-orientable subsystem, suggesting that the Li-Haldane correspondence does not require orientability. Now we are able to answer the question posed in the end of the previous subsection. Taking $k_X=n=1$ in Eq.~(\ref{xapK}), we know that a block labeled by $[c]$ and $\alpha$ appears if and only if
\begin{equation}
\sum_{[g]}\sum_{\beta\in {\rm Rep}({\rm C}_g)} \iota_\beta S_{(g,\beta),(c,\alpha)} >0.
\end{equation}
Finally, we remark that by taking $k_X=k$ and $n=0$ in Eq.~(\ref{xapK}), we will obtain the ground-state degeneracy of the Kitaev quantum double model living on $N_k$:
\begin{equation}
\mathfrak{K}_k= \sum_{[g]}\sum_{\beta\in {\rm Rep}({\rm C}_g)} \iota_\beta^{k}\left(\frac{|{\rm C}_g|}{d_\beta}\right)^{k-2},
\label{degK}
\end{equation}
where we have replaced $S_{(g,\beta),(1,1)}$ by $|{\rm C}_g|^{-1}d_\beta$. This should be compared to the well-known result for genus-$\gamma$ surfaces:
\begin{equation}
\mathfrak{R}_\gamma= \sum_{[g]}\sum_{\beta\in {\rm Rep}({\rm C}_g)} \left(\frac{|{\rm C}_g|}{d_\beta}\right)^{2\gamma-2},
\label{degR}
\end{equation}
which follows from Eq.~(\ref{xapR}) with $\gamma_X=\gamma$ and $n=0$. Similar to Eqs.~(\ref{RRR}), (\ref{KKK}) and (\ref{KRK}), one can check the consistency of degrees of freedom between the above results (\ref{degK}), (\ref{degR}) and Eqs.~(\ref{xapR}), (\ref{xapK}).

\section{Summary and Outlook}\label{sec-closing}

We have introduced a systematic framework, grounded in the Seifert–van Kampen theorem, that enables a rigorous analysis of the entanglement structure for arbitrary bipartitions of ${\rm Rep}(G)$ loop-symmetric many-body states living on general manifolds.
We find that the entanglement block structure contains a topological part and a geometric part (\ref{tg}), while degneracy is not enforced. We provide a general algorithm for determining this block structure and shows how it can be executed for low-dimensional manifolds and simple high-dimensional manifolds. Our framework can be refined to incorporate further constraints, such as gauge invariance which promotes the setup to be topological gauge theories. 
Within the modified framework, we have reproduced known results including the topological entanglement entropy and the Li–Haldane correspondence in 2D. As the framework itself is applicable to arbitrary manifolds, we are able to identify which result is not dimension or topology specific and explore entanglement structure unique to higher dimensions.

Our work paves the way towards the ambitious goal of extending the Wigner-Dyson's classical results on group symmetries \cite{wigner1931gruppentheorie,dyson1962threefold} to arbitrary generalized symmetries. This includes at least two directions: one is general higher-form symmetries beyond loop symmetries, the other is general categorical symmetries beyond ${\rm Rep}(G)$ symmetries (c.f. \cite{asaedaExoticSubfactorsFinite1999}). We note that the entanglement entropy of Abelian higher-form gauge theories has been systematically investigated in previous works (see, e.g., \cite{ibieta-jimenezTopologicalEntanglementEntropy2020}), but the full entanglement block structure remains unclear, let alone the impact of generally non-invertible categorical symmetries. 
We expect that, as the Seifert–van Kampen theorem concerns the fundamental group, its extension to higher homotopy groups \cite{lurieKerodon,brown2004nonabelian} would allow one to compute the entanglement structure for general higher-form gauge theories through the same algorithmic procedure developed here. 
We may also need to work in the Tannaka-Krein dual picture \cite{deligne1982tannakian} of the current framework, where the groupoid morphisms are replaced by functors between the category of (higher-homotopy) local systems and that of symmetries. 
This is especially necessary for proving the Li-Haldane correspondence for general 2D topological order. 
Such an extension would necessarily involve higher-categorical formulations \cite{lurieKerodon}, demanding more sophisticated mathematical input, yet it is expected to expose fine-grained topological features beyond fundamental group. 

Another important and complementary direction is the generalization to multipartite entanglement.
While bipartite entanglement is completely characterized by the $W$ matrix, a natural generalization suggests that $N$-partite entanglement should be encoded in an $N$th-order $W$ tensor.
While tensors of rank three or higher generally lack a spectral decomposition, there do exist local-unitary invariant scalar measures analogous to the entanglement entropy, which are referred to as the multi-entropy in the literature \cite{liuMultiWavefunctionOverlap2024}.
The present framework makes use of the Seifert–van Kampen theorem as it exactly concerns the bipartition of a topological space. 
We expect that a multipartite generalization of the Seifert–van Kampen theorem would enable a systematic analysis on the impact of generalized symmetries on the multipartite $W$-tensor. 
This is particularly important for studying various quantum anomalies in higher dimensions, whose profound connections to multipartite entanglement have become clear in recent studies \cite{lessaMixedStateQuantumAnomaly2025}.

\acknowledgements{
  We are grateful to Rintaro Masaoka, Sogen Ikegami, Yuki Furukawa, Takamasa Ando, Ching-Yu Yao, Ken Shiozaki, and Yuxuan Guo for valuable discussions.
  H.Y. thanks Ippo Orii for a fruitful discussion on the modular tensor categories and the connection between non-orientable surfaces and TQFTs.
  The authors thank the Yukawa Institute for Theoretical Physics at Kyoto University. Discussions during the YITP workshop YITP-I-25-02 on "Recent Developments and Challenges in Tensor Networks: Algorithms, Applications to Science, and Rigorous Theories" were useful to complete this work.
  H.Y. acknowledges support from FoPM, WINGS Program, the University of Tokyo. Z.G. acknowledges support from the University of Tokyo Excellent Young Researcher Program and from JST ERATO Grant No. JPMJER2302, Japan.
}

\emph{Note added.---}
While finalizing this paper, we become aware of the appearance of relevant research \cite{lamasHigherformEntanglementAsymmetry2025,liu2025entanglementsumrulehigherform}. 
The former explores entanglement asymmetry with respect to 1-form $\mathbb{Z}_2$ symmetry and non-trivial partition of a torus. 
The latter investigates the bipartite entanglement entropy of symmetric eigenstates in a quantum lattice model with finite Abelian higher-form symmetry.

\bibliographystyle{unsrtnat}
\bibliography{cite}
\clearpage

\appendix
\onecolumngrid
\begin{center}
\textbf{\large Supplemental Materials}
\end{center}
\setcounter{figure}{0}
\renewcommand{\thefigure}{S\arabic{figure}}

\section{Loop Symmetry and Flat Configuration}
\label{app:lsfc}
In this section, we elucidate the equivalence between $\mathrm{Rep}(G)$ loop-symmetric configurations and flat configurations.

We recall that $\mathrm{Rep}(G)$ loop symmetry is defined by a MPO whose building block is four-leg tensor $D^\alpha_{ij} (g)\delta_{g,g'}$. The physical indices are labeled by $g,g'$ and the virtual indices are labeled by $i,j$, for a fixed irrep $\alpha$. By contracting a sequence of these tensors into a loop, one obtains the MPO. Explicit form of the MPO is
\begin{equation}
    \sum_{g_1,g_2,\cdots, g_\ell} \chi^\alpha (g_\ell \cdots g_2 g_1) \ket{g_1,g_2,\cdots, g_\ell}\bra{g_1,g_2,\cdots, g_\ell},
\end{equation}
where $\chi^\alpha$ is the character of the irrep $\alpha$.
This MPO acts on the state $\ket{\{h\}}$ as
\begin{equation}
\begin{split}
    \left[\sum_{g_1,g_2,\cdots, g_\ell} \chi^\alpha (g_\ell \cdots g_2 g_1) \ket{g_1,g_2,\cdots, g_\ell}\bra{g_1,g_2,\cdots, g_\ell}\right]\ket{\{h\}}&\\
    =\chi^\alpha (h_\ell \cdots h_2 h_1) \ket{\{h\}}&.
\end{split}
\end{equation}
Note that this MPO detects the \emph{conjugacy class} of the entire loop holonomy since a character takes the same value on all elements of a conjugacy class. However, for any finite group, the conjugacy class of the identity element $1$ is $\{1\}$.

\begin{figure}[htb]
\begin{center}
       \includegraphics[width=7cm, clip]{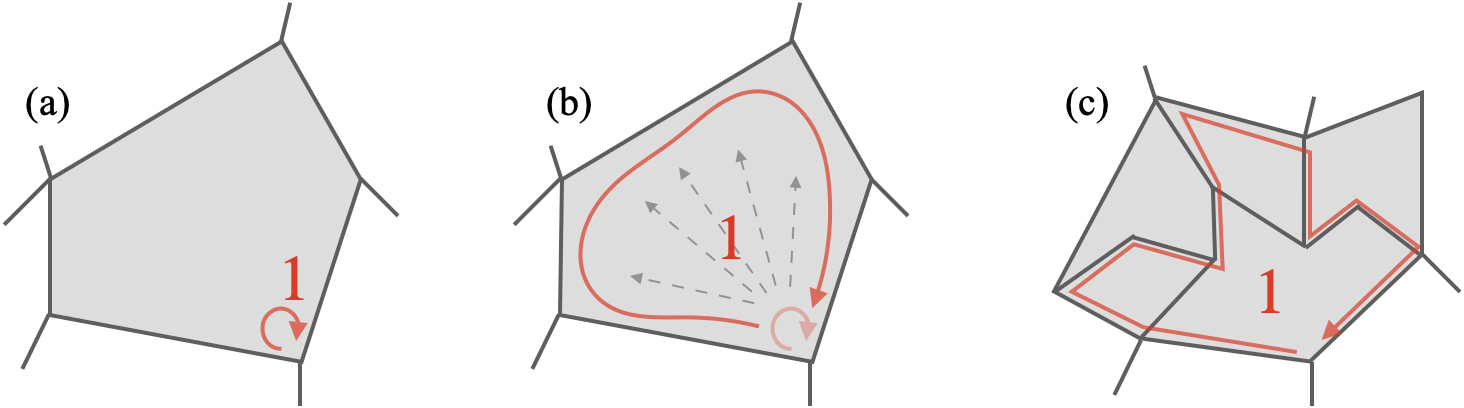}
       \end{center}
   \caption{
   Equivalence between $\mathrm{Rep}(G)$ loop-symmetric configurations and flat configurations.
   (a) Set the initial path to be a point. The holonomy along a point is defined as $1$.
   (b) The holonomy, along the path deformed by a plaquette move. Since the conjugacy class of $1$, the holonomy along the deformed path is also $1$.
   (c) Requiring the Wilson loop to be topological, all the holonomies along arbitrary plaquettes is set to $1$. Conversely, such configuration makes the Wilson loop topological.
   }
   \label{fig:flat-config}
\end{figure}

Next, we analyze the consequence of topological MPO.
Let us restate the claim more precisely: for any irrep and any closed loop, the MPO along that loop is topological if and only if the holonomy along every contractible loop is $1$.
See Fig~.\ref{fig:flat-config} for pictorial explanations.
Allowing topological move implies $\chi^\alpha(g_\ell)=\chi^\alpha(g_{\ell'})$ for ${}^\forall \alpha$, where $g_\ell \, (g_{\ell'})$ denotes the holonomy along the loop $\ell \, (\ell')$.
From the Schur orthogonality for the characters, we find
\begin{equation}
\begin{split}
    \sum_{\alpha \in \mathrm{Rep}(G)} \chi^\alpha (g_\ell)\overline{\chi^\alpha (g_{\ell'})} &= |C_{g_\ell}|\delta_{[g_\ell],[g_{\ell'}]}\\
    =\sum_{\alpha \in \mathrm{Rep}(G)} |\chi^\alpha (g_\ell)|^2 &= |C_{g_\ell}|,
\end{split}
\end{equation}
implying $\delta_{[g_\ell],[g_{\ell'}]}=1\Leftrightarrow [g_\ell]=[g_{\ell'}]$. Here, $[g]$ denotes the conjugacy class of element $g$.
Since we have assumed this relation holds for any closed loop, we particularly choose a point (as a length-zero loop) and a plaquette including that point.
Since the holonomy along point must be $1$, we can rewrite this as $g_\ell=1$, implying flat configuration.

Conversely, showing that plaquette move of the Wilson loop are always permitted in flat configuration is straightforward; plaquette move does not change $\chi^\alpha (h_\ell \cdots h_2 h_1) \ket{\{h\}}$. This completes the proof of equivalence between flat configuration and loop-symmetric configuration.

\section{Mathematical Details Omitted in the Proof of the Main Theorem}

In this section, we provide mathematical background and details of the main theorem.
First, we begin by stating in mathematical terms a setup where manifolds and their discretizations naturally share the same properties. We then supplement the details omitted in the main proof. Finally, the alternative ``proof'' of the main theorem, described in the language of non-Abelian cohomology, is discussed.

\subsection{Good Discretization of $M$}

Let $M$ be a smooth manifold. Choose an open cover $\mathcal U=\{U_i\}_{i\in I}$ of $M$ such that every finite intersection $U_{i_0}\cap\cdots\cap U_{i_k}$ is either empty or contractible (i.e. a good cover). Such a cover exists for smooth manifolds (see theorem 5.1 in \cite{bott2013differential}).

Define the \emph{nerve} $N(\mathcal U)$ to be the abstract simplicial complex with vertex set $I$ and with a $k$--simplex $\{i_0,\dots,i_k\}$ whenever $U_{i_0}\cap\cdots\cap U_{i_k}\neq\varnothing$. Let $|N(\mathcal U)|$ denote its geometric realization.

By the Nerve Theorem, the canonical map $|N(\mathcal U)|\to M$ is a homotopy equivalence (since all finite intersections in the cover are contractible). See proposition 4.G.2 and proposition 4.G.3 in \cite{hatcherAlgebraicTopology2002} for the reference. Consequently, $|N(\mathcal U)|$ and $M$ have the same homotopy type.

Fundamental groupoids are homotopy invariant: if $X\to Y$ is a homotopy equivalence, then for any set $A\subset X$ that meets every path component, the induced functor $\Pi_1(X,A)\to \Pi_1(Y,f(A))$ is an equivalence of groupoids (see 6.5.10 corollary 1 in \cite{brownTopologyGroupoids2006}.). Applying this to the homotopy equivalence $|N(\mathcal U)|\to M$, we obtain an equivalence of groupoids
\begin{equation}
\pi_1\bigl(|N(\mathcal U)|,A\bigr)\ \simeq\ \pi_1\bigl(M,A'\bigr),    
\end{equation}
for appropriate choices of $A\subset |N(\mathcal U)|$ and $A'\subset M$ meeting each path component.

Therefore, for purposes such as computing non-Abelian first cohomology with finite coefficients (via flat $H$-bundles or representations of the fundamental groupoid), one may replace $M$ by the simplicial model $|N(\mathcal U)|$ without loss of homotopical information.

\subsection{Equivalence between the set of configurations for fixed topological block and the set of gauge transformation on $V\backslash A$}

In this section, we  provide a sketch to prove the equivalence between the set of configurations for fixed topological block and the set of gauge transformation on $V\backslash A$. This is an omitted mathematical detail in the proof described in the main text. This equivalence directly guarantees that the block structure of the $W$ matrix is determined by our general algorithm.

We start the proof of the equivalence with the following lemma:
If and only if two flat configurations $\ket{\{g\}}$ and $\ket{\{g'\}}$ have same holonomy along the path the pair of base points (the loop with one base point is also considered), they can be transformed into each other by gauge transformations other than those at the base points.

\begin{figure}
\begin{center}
       \includegraphics[width=7cm, clip]{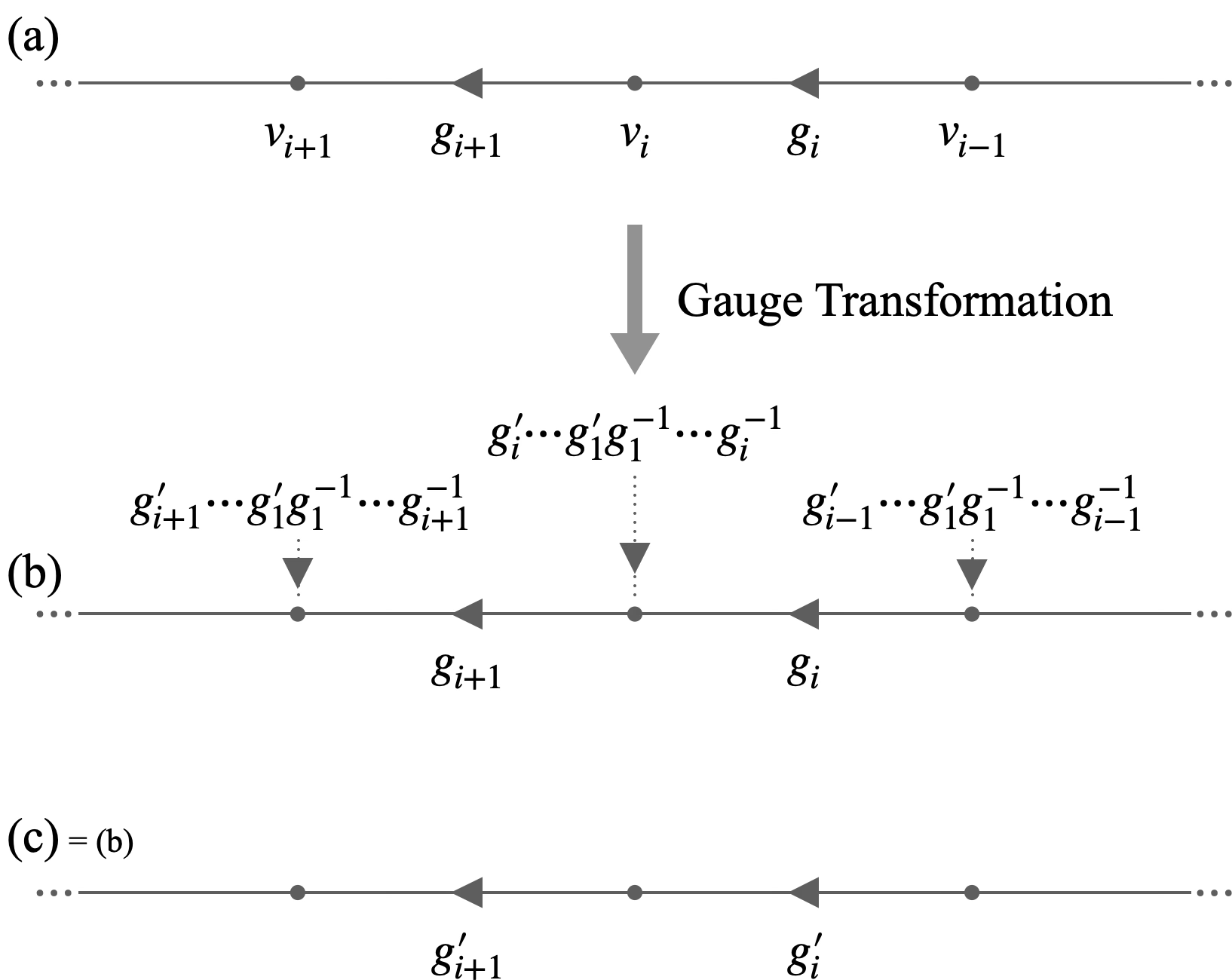}
       \end{center}
   \caption{
   Procedure for gauge transformations of vertices on $V_p \backslash A$. 
   (a) Determine the initial configuration as shown. 
   (b) Perform gauge transformations on the vertices as illustrated. 
   (c) Performing gauge transformations as shown allows transitions between two configurations to proceed smoothly in most cases. However, consistency around the vertices is not yet guaranteed at this stage.
   }
   \label{fig:gauge-transformation}
\end{figure}

\begin{figure}
\begin{center}
       \includegraphics[width=7cm, clip]{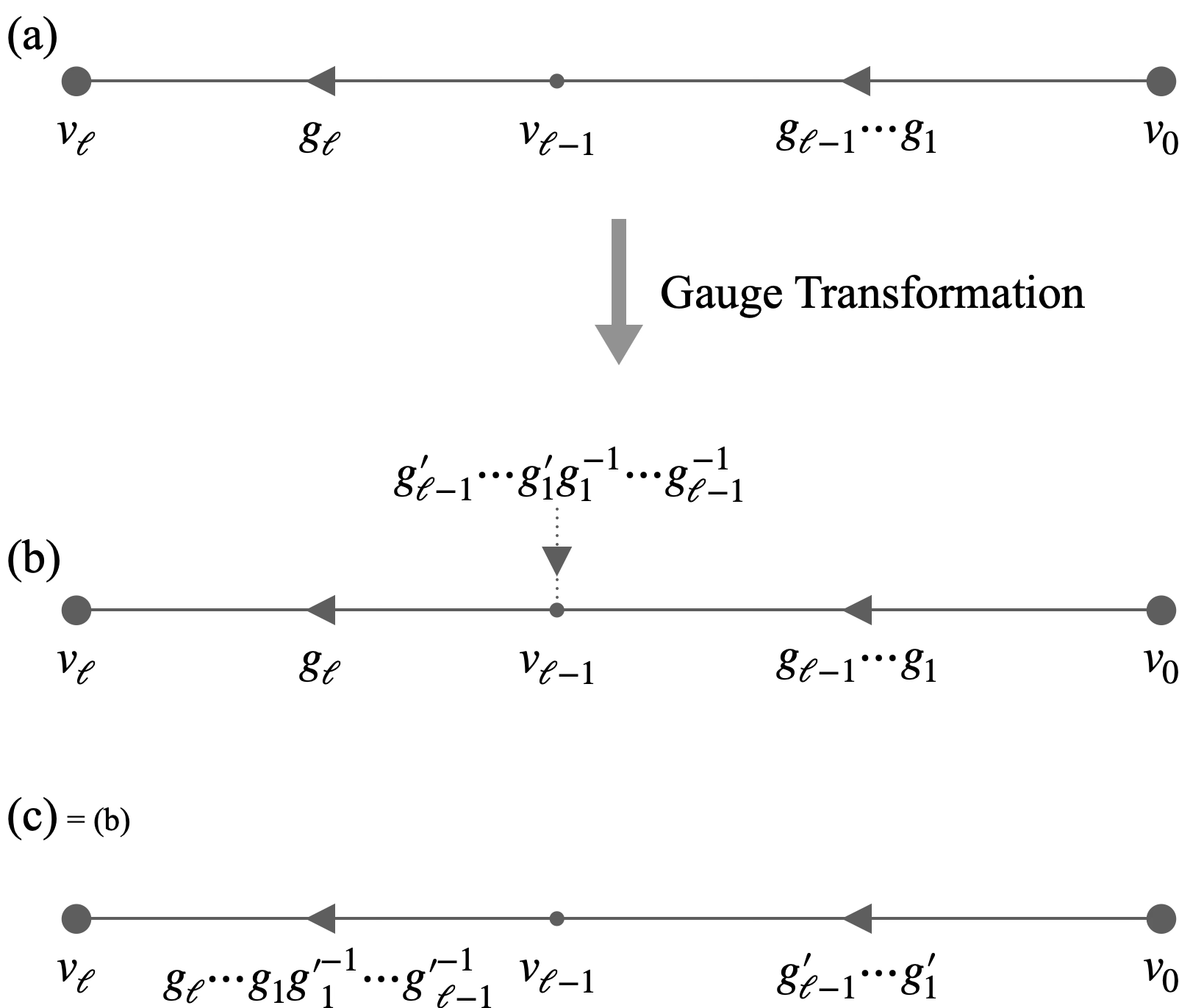}
       \end{center}
   \caption{Verifying consistency when transitioning between two configurations via gauge transformations on non-base-point vertices. 
   (a) Focus only on the neighborhood of final base point in the initial configuration. 
   (b) Execute the gauge transformation on $V_p\backslash A$.
   (c) Attempting transitions by arbitrarily selecting two configurations and applying gauge transformations generally reveals inconsistencies near the final base point. It becomes clear that transitions to arbitrary configurations are possible only when the initial configuration and global holonomy coincide.}
   \label{fig:gauge-transformation-global}
\end{figure}

To prove this, first we fix ${}^\forall \textrm{(base-point pair)}\in A$.
We focus on ${}^\forall$path between the two base points.
We number the vertices and edges as $v_i,e_i$ where $v_0$ is the first base point and $e_1$ is the edge outgoing from $v_0$. Let $l$ be the length of the path.
Align the orientation of the edges to be outgoing from $v_i$ and incoming into $v_{i+1}$.
We are considering two configurations $\ket{\{g\}}$ and $\ket{\{g'\}}$ on edges.
One can check that gauge transformation $g'_k\cdots g'_1 (g_k\cdots g_1)^{-1}$ on the $k$-th vertex can transfer $\ket{\{g\}}$ to $\ket{\{g'\}}$, ignoring $e_l$.
The local holonomy on $e_l$ is 
\begin{equation}
  g_{l}\cdots g_1 (g'_{l-1}\cdots g'_1)^{-1}
\end{equation}
after the gauge transformation.
Thus, if and only if $\ket{\{g\}}$ and $\ket{\{g'\}}$ have same global holonomy along this path ($g_l\cdots g_1=g'_l\cdots g'_1$), we have
\begin{equation}
  g_{l}\cdots g_1 (g'_{l-1}\cdots g'_1)^{-1}=g'_l.
\end{equation}
See Fig.~\ref{fig:gauge-transformation} and Fig.~\ref{fig:gauge-transformation-global} for pictorial explanations.
For a loop outgoing from/coming in the same base point, we can prove the statement in the same manner.

Next, we verify that this also works for spanning tree. Choose one of the base points as the root and arbitrarily select a spanning tree $T$ such that the other base points become leaves.
Applying the same gauge transformation procedure as for paths or loops, starting from the root and proceeding sequentially, we observe that two configurations sharing the same holonomy among base points will again be connected by gauge transformations. 
Conversely, as with paths or loops, two configurations possessing different holonomies will not be connected. 
Note that vertices that are not base points may also become leaves in general.
In such a case, by selecting gauge transformations on leaves that are not base points, we can always ensure that holonomy remains consistent, so this corner case does not break the proof.

\begin{figure}
\begin{center}
       \includegraphics[width=7cm, clip]{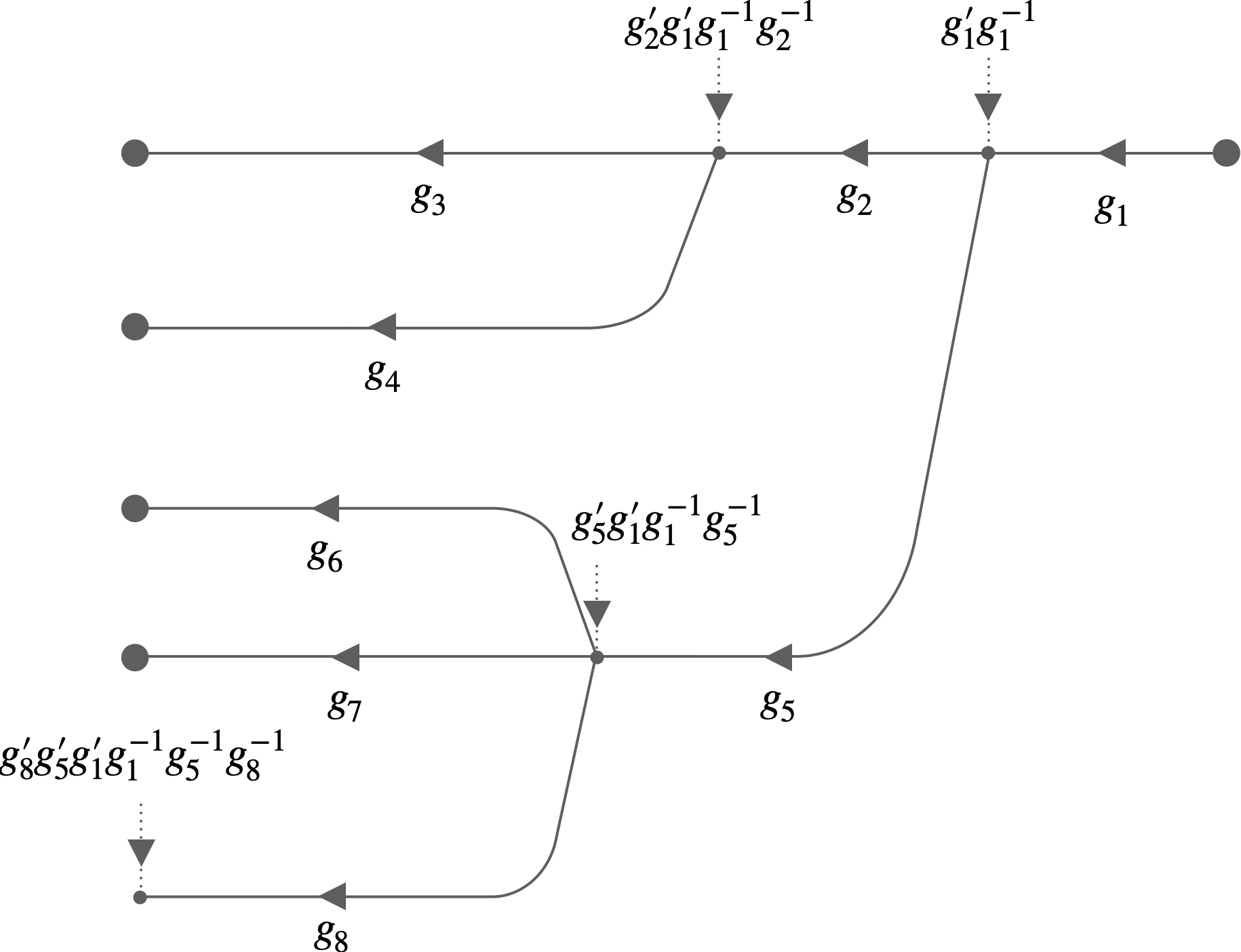}
       \end{center}
   \caption{Consistency check for gauge transformations with respect to spanning trees.
   By selecting one base point as the root and initiating gauge transformations from there, we understand that—just as in the case of paths or loops—two configurations can be transformed if the holonomy between base points matches, and cannot be transformed if it does not match.}
   \label{fig:gauge-transformation-spanningtree}
\end{figure}

The question is whether the holonomies of edges not included in the spanning tree can be connected by gauge transformations.
Fig.~\ref{fig:gauge-transformation-spanningtree} illustrates this situation.

Finally, we confirm that the same fact holds for edges not included in the spanning tree. Taking any edge $e \notin T$, by the definition of a spanning tree, the two vertices connected by edge $e$ are included in the spanning tree.

\begin{figure}
\begin{center}
       \includegraphics[width=7cm, clip]{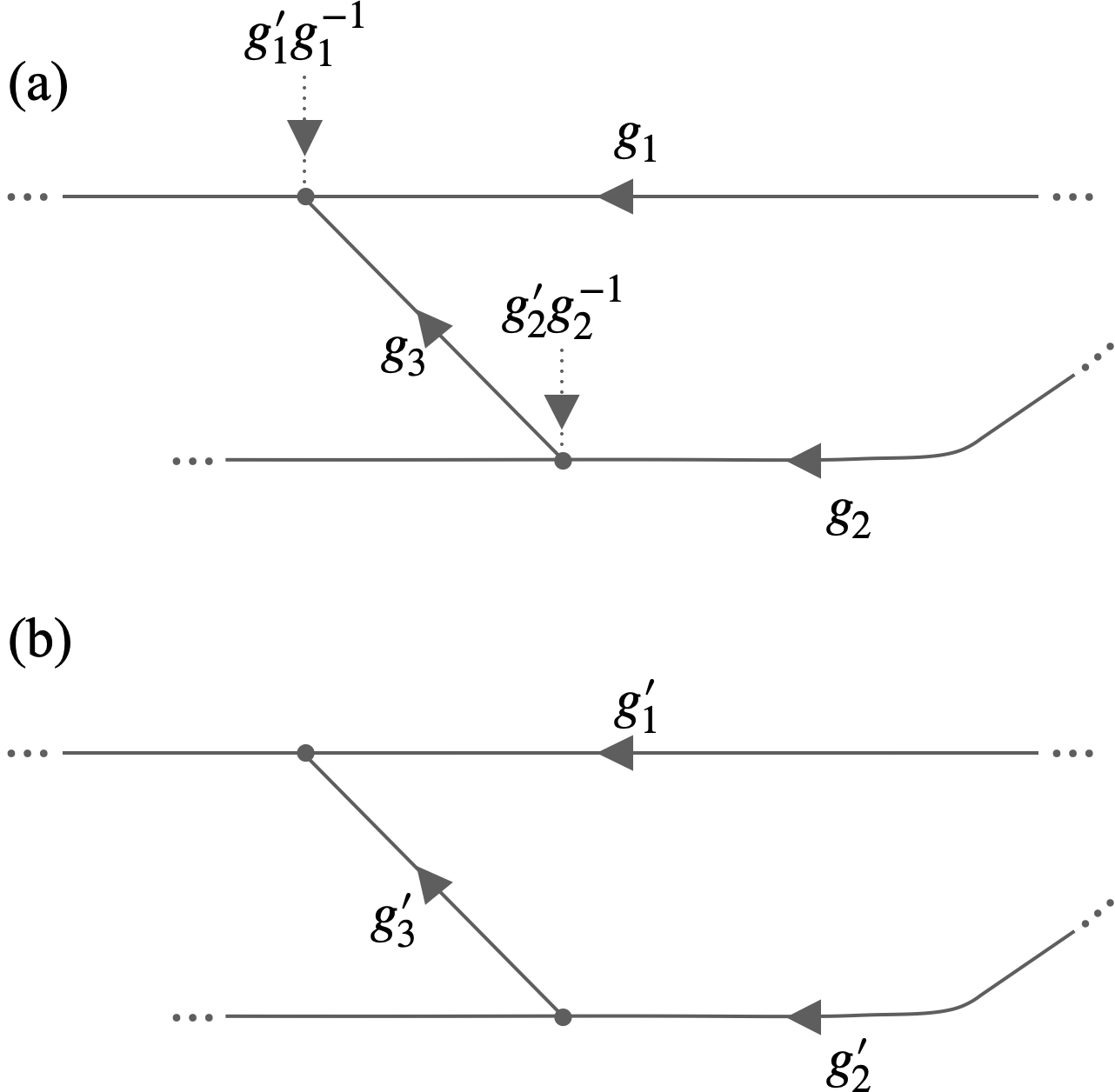}
       \end{center}
   \caption{Any two configurations with the same holonomy between base points can be transformed into each other via gauge transformations other than those at the base points, including edges not contained in the spanning tree. (a) is the original configuration, (b) is the target configuration. Performing the gauge transformation depicted in (a) yields compatibility $g_1^{-1}g_3 g_2={g'}_1^{-1}g'_3 g'_2$ equivalent to the gauge transformation $g'_3=g'_1g_1^{-1}g_3g_2 {g'}_2^{-1}$ derived from the shared holonomy between base points.}
   \label{fig:gauge-transformation-all}
\end{figure}

Now, consider two configurations $\ket{\{g\}}, \ket{\{g'\}}$ defined not only on $T$ but also on other regions, such that the holonomy between any two base points is identical.
Any edge $e$ not contained in $T$ can connect two vertices in $T$. Let us denote these vertices as $v_1$ and $v_2$.
We denote the holonomy from the root to these two vertices as $g_i$ and $g_i'$ for each configuration, respectively.
Furthermore, we denote the holonomy of $e$ from $v_2$ to $v_1$ as $g_{1,2}$ and $g'_{1,2}$ for each configuration.
Since the holonomies between the base points are identical for the two configurations, the relation $g_1^{-1}g_3 g_2={g'}_1^{-1}g'_3 g'_2$ holds.
On the other hand, performing the same operation as the gauge transformation applied to the spanning tree replaces $g_1$ with $g'_1$, $g_2$ with $g'_2$, and $g_3$ with $g'_1g_1^{-1}g_3g_2 {g'}_2^{-1}$, thus requiring the compatibility condition $g_3=g'_1g_1^{-1}g_3g_2 {g'}_2^{-1}$.
This is equivalent to the relation derived earlier from the identical holonomy between base points. Therefore, two configurations with identical holonomy between base points can be transformed into each other via gauge transformation.
Conversely, if the holonomy between base points is not identical, we can see that at some stage—path, loop, or spanning tree—the gauge transformation cannot transition it. Alternatively, there exists edge(s) ${}^\exists e$ in the set of edges not included in the spanning tree that does not satisfy the required compatibility condition.
Fig.~\ref{fig:gauge-transformation-all} provides pictorial explanations.
Now the lemma is proved.

Furthermore, we state the following lemma: For the same configuration $\ket{\{g\}}$, applying different gauge transformations $\{h\},\{h'\}$ on $V\backslash A$ will always yield a different path configuration. Here $V_p$ denotes the set of vertices on the path.

\begin{figure}
\begin{center}
       \includegraphics[width=15cm, clip]{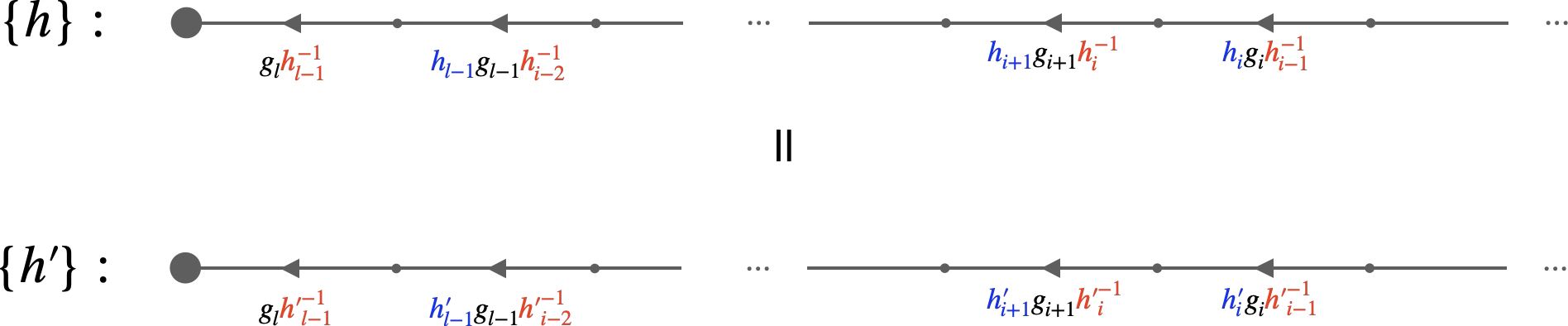}
       \end{center}
   \caption{Graphical proof that if two ways of gauge transformations acting on the same state along a path, when applied on $V_p \backslash A$, return the same state, the gauge transformations at each vertex coincide.
   This diagram can be understood by reading from left to right. Since edges connected to the left base point only receive action from the right side, if the edge holonomy coincides, the gauge transformations also coincide (red indicates that gauge transformation coincidence has been established). Using this, since the left action among the holonomies of the adjacent edges is already proven to be the same (blue indicates gauge transformations already proven to be identical), the right action—that is, the gauge transformation on the adjacent vertex—must also be equal. Repeating this process shows that the gauge transformations on all vertices are identical.} 
   \label{fig:gauge-transform-isomorphism}
\end{figure}

Taking the dual of this statement is equivalent to showing that “if the actions of two gauge transformations on $V\backslash A$ on the same state yield the same state, then the two gauge transformations are identical at each vertex.” 
We again start from focusing on the case of a path or a loop. Fig.~\ref{fig:gauge-transform-isomorphism} illustrates the proof.
First, an edge $e^*$ connected to the base point receives only either a right or left action from the gauge transformation on vertex $v^*$ on the path one step away from the base point. 
On such an edge, if two gauge transformations on $v^*$ assign the same holonomy to that edge, then the gauge transformations on $v^*$ must be identical. 
Next, consider the adjacent edge $e'$ of the edge just examined, which is included in the path. 
This edge receives transformations on both sides, but we know one side's transformation is identical. 
Therefore, if the holonomy of $e'$ remains unchanged under two gauge transformations for the same state, the gauge transformation on the vertex other than $v^*$ must also be identical. 
Repeating this process shows that the gauge transformations must be identical on all vertices.

We now move on to the proof for the general case. Actually, it can be easily understood that same fact holds for the spanning tree. (Even when there are leaves that are not the base points, one can easily see the same procedure works. Such cases do not break the proof.)
Next, the conditions required by the edges outside the spanning tree will be considered.
The condition that holonomy coincides for both gauge transformations applied to ${}^\forall $edges outside the spanning tree may appear to be an excessive constraint.
However, since all gauge transformations for vertices inside the spanning tree are already determined, we see that this condition is always satisfied.
This completes the proof for the second lemma.

The two lemmas above shows that there exists a bijection between the set of states with fixed holonomy between base points and the set of gauge transformations on $V\backslash A$ under a certain choice of a representative configuration, and that their sizes coincide.
This identifies the geometric block as $\bigoplus_{i=1}^{|G|^{|V_\partial|-|A|}} \mathbb{C}^{|G|^{|V_X|}\times |G|^{|V_Y|}}$.

Note that gauge transformations over base points violate what were stated above. 
First, considering a path between two base points, a gauge transformation over the base points can alter the holonomy between them (taking a loop with a single base point, we comprehend that gauge transformation on that base point can change the holonomy within the range of the conjugacy class).
Performing gauge transformation $g_{i/f}\in G$ on the initial/final vertex (namely the base points) maps the path holonomy $g_\ell$ as $g_\ell\mapsto g_f g_\ell g_i^{-1}$. 
Furthermore, gauge transformations along a loop involving one base point can sometimes yield the same state despite performing different gauge transformations. Take a loop with one base point and assume that all the edges have holonomy $1\in G$. 
Performing $g\in G$ on all the vertices uniformly, we obtain the same configuration.

For these reasons, performing gauge transformations at the base point causes deviations from the topological blocks of the $W$ matrix and breaks the one-to-one correspondence between gauge transformations and states, necessitating separate treatment.

Note that not all topological blocks are permitted here.
For example, in the case of a torus, the holonomy of $a$-loop and $b$-loop must commute.
This circumstance not only makes the structure of the W matrix even more sparse (beyond the constraints imposed by loop symmetry), but also gives rise to the mystery of the size of the topological blocks.
To find possible topological block, we enumerate the possible global holonomies over $X$ and $Y$ that is compatible in $M$.
The image $\mathrm{Im} \subseteq \Hom (\pi_1(\partial,A),G)$ where the $\Hom$-applied diagram becomes commutative satisfies this condition, and the way it is realized on $X/Y$ can be seen from the preimage to $X/Y$.
This completely determines the block structure; $\oplus_{\phi \in \mathrm{Im}}$ and $|r_X^{-1}(\phi)|\times |r_Y^{-1}(\phi)|$.

Let us translate the above explanation into a mathematically elegant statement.
A set $X$ is called a $G$-set if a group action $\mu:G\times X\to X$ is defined on it and satisfies the following conditions:
\begin{equation}
\begin{split}
    {}^\forall x\in X, \quad &\mu(1,x)=x,\\
    {}^\forall x\in X, {}^\forall g,h\in G, \quad &\mu(g,\mu(h,x))=\mu(gh,x).
\end{split}
\end{equation}
If for some $x \in X$, $gx=x \Rightarrow g=1$ holds, then the group action is called free. Furthermore, if for all $x,y\in X$, there exists some $g$ such that $gx=y$, the group action is called transitive. A free and transitive group action is called a $G$-torsor. Fixing $x\in X$, the $G$-torsor on it is known to be isomorphic to $X$.

This ``proof'' essentially demonstrates that the set of basis in a geometric block is a $G$-torsor under the gauge transformation as $G$-action. This explanation can be understood more elegantly from the perspective of non-Abelian cohomology, which will be discussed in the next section.

\subsection{Non-Abelian Cohomology Perspective}

Here we explain how to understand, in cohomological terms, the homotopy-theoretic proof of the main theorem of this paper. For readers less familiar with cohomology, we proceed by exploiting an analogy with classical electromagnetism.

Our setup, namely flat configurations, is the analogue (in the language of electromagnetism) of a theory with no static magnetic field on any local (contractible) region. Recall that the magnetic field is given as the curl of a vector potential. The absence of static magnetic field means that the curl of the vector potential vanishes, i.e.\ $\bm{B}=\mathrm{rot}\,\bm{A}=0$. This means that the magnetic flux through any infinitesimal contractible loop $\ell$ vanishes. Passing to an integral form allows us to enclose finite regions, and we may write
\begin{equation}
\int_{S}\mathrm{rot}\,\bm{A}\cdot d\bm{S}\;=\;\oint_{\ell}\bm{A}\cdot d\bm{l}\;=\;0.
\end{equation}
Note that the integral need not vanish for noncontractible loops.

Recall also that, in quantum mechanics, the vector potential produces a phase associated with a path. Concretely, we consider the quantity $\exp\bigl(i\int \bm{A}\cdot d\bm{l}\bigr)$. In our setting, the phase for any contractible loop equals $1$, whereas for noncontractible loops it is in general not $1$. The quantity $\exp\bigl(i\int \bm{A}\cdot d\bm{l}\bigr)$ composes under path concatenation and takes values in the group $U(1)$, called the gauge group in this context, whence electromagnetism is a $U(1)$ gauge theory. Moreover, the condition that every contractible loop encloses zero flux implies that, as long as the endpoints are fixed, the phase is invariant under homotopy of the path. A gauge theory with this property is called a topological gauge theory. Our object of analysis replaces this $U(1)$ by a finite (in general non-Abelian) group $G$.

Writing the identity $\int_{S}\mathrm{rot}\,\bm{A}\cdot d\bm{S}=\oint_{\ell}\bm{A}\cdot d\bm{l}$ in the language of differential forms gives
\begin{equation}
(S,dA)\;=\;\int_{S}dA\;=\;\int_{\partial S}A\;=\;(\partial S,A).
\end{equation}
The statement that “the integral of $d$ (acting on forms) over a region equals the integral of the form over the boundary $\partial$ of the region” is Stokes' theorem. Stokes' theorem is not specific to electromagnetism: it always holds for the pairing $(\cdot,\cdot)$ between singular homology (a suitable discretization of a manifold together with the boundary operator $\partial$) and de~Rham cohomology (real differential forms together with the exterior derivative $d$). 

Next, recall gauge transformations. A gauge transformation adds the exterior derivative of a $0$-form field to the vector potential $A$, namely $A\mapsto A+d \lambda$ where $\lambda$ is 0-form. Since the phase along a path is given by the integral of $A$, the contribution of the gauge term is the difference of the $0$-form gauge field evaluated at the endpoints. Therefore, phases associated with loops are invariant under gauge transformations.

Our setup simply replaces the gauge group $U(1)$ by a finite group $G$ (possibly non-Abelian) and reproduces the above structure in a discrete model. Let us restate the setup. 
We fix a manifold and a good discretization of it.
$C^p(M,G)$ denotes the set of $p$-cochains on $M$ with $G$-valued coefficients. In our setup, it suffices to take $p$ up to $2$:
A $0$-form field ($0$-cochain $c^0 \in C^0(M,G)$) assigns a group element to each vertex, $1$-form field ($1$-cochain $c^1 \in C^1(M,G)$) assigns a group element to each oriented edge, and a $2$-form field ($2$-cochain $c^2 \in C^2(M,G)$) assigns a group element to each oriented face. 
We impose the “no magnetic field” constraint on every plaquette $p$. An analogue of the exterior derivative for finding the magnetic field has not yet been defined, but recalling Stokes' theorem, it would likely be as follows: 
\begin{equation}
(p,\ \delta^1 c^{1})\;=\;(\partial p,\ c^{1})\;=\;\prod_{e\subset\partial p}\bigl(c^{1}_{e}\bigr)^{\varepsilon(e)}\;=\;1,
\end{equation}
where $\varepsilon(e)=\pm1$ according as the edge $e$ appears with the same or opposite orientation as the boundary $\partial p$. 
Now, $\delta^1$ for $1$-cochain takes the value $\prod_{e\subset\partial p}\bigl(c^{1}_{e}\bigr)^{\varepsilon(e)}$ on a plaquette $p$ and is understood as the magnetic flux on $p$.
In the non-Abelian case one must fix the base point of the loop and the order of the product following the orientation of $\partial p$. 
Elements in $\mathrm{Ker} \delta$ ($\mathrm{Im} \delta$) are called cocycle (coboundary). One may view this cocycle condition as the discrete analogue of $\int_{S}dA=\int_{\partial S}A$ (Stokes' theorem) in the continuum. Next, we see the role of coboundary in our setup.

A $G$-valued $0$-cochain $c^{0}\in C^{0}(K;G)$ defines a gauge transformation. For an oriented edge $e:s(e)\to t(e)$ it acts by
\begin{equation}
((\delta^0 c^{0})\circ z^{1})_{e}\;=\;c^{0}_{t(e)}\,z^{1}_{e}\,\bigl(c^{0}_{s(e)}\bigr)^{-1}.
\end{equation}
In the case $G$ Abelian, this operation is nothing but the modulation by coboundary.
This action commutes with $\delta^{1}$ and therefore preserves flatness (the “product around each plaquette equals $1$” constraint). For a path $\gamma=e_{1}^{\varepsilon_{1}}\cdots e_{\ell}^{\varepsilon_{\ell}}$ we define the holonomy
\begin{equation}
\mathrm{Hol}_{z^{1}}(\gamma)\;:=\;\prod_{j=1}^{\ell}\left(z^{1}_{e_{j}}\right)^{\varepsilon_{j}}.
\end{equation}
If $z^{1}\in Z^{1}$, then $\mathrm{Hol}_{z^{1}}(\gamma)$ is invariant under homotopy of $\gamma$ with fixed endpoints (deformations are compensated by products over boundaries of $2$-simplices). The effect of a gauge transformation is
\begin{equation}
\mathrm{Hol}_{\,(\delta^0 c^{0})\circ z^{1}}(\gamma)\;=\;c^{0}_{t(\gamma)}\ \mathrm{Hol}_{z^{1}}(\gamma)\ \bigl(c^{0}_{s(\gamma)}\bigr)^{-1},
\end{equation}
i.e.\ conjugation by the endpoint values; consequently, phases on loops ($s(\gamma)=t(\gamma)$) are gauge invariant.

Now, let us outline the proof of the main theorem using this setup.
From now on we choose a set of base points $A\subset K^{(0)}$ and we fix the “groupoid-level” holonomy on it. Concretely, we fix a map from the $A$-subgroupoid of the fundamental groupoid $\pi_{1}(K)$ to $G$:
\begin{equation}
\phi:\ \pi_{1}(K,A)\ \longrightarrow\ G.
\end{equation}
We consider the set of flat configurations whose holonomy restricts to $\phi$:
\begin{equation}
\mathcal{C}_{\phi}\;:=\;\Bigl\{\,z^{1}\in Z^{1}(K;G)\ \Bigm|\mathrm{Hol}_{z^{1}}\big|_{\pi_{1}(K,A)}=\phi\,\Bigr\}.
\end{equation}
Let $\Gamma_{A}\;:=\;\{\,c^{0}\in C^{0}(K;G)\mid c^{0}\big|_{A}=1\,\}$ be the subgroup of “gauge transformations trivial on the base points”.

Although it is essentially the same as the explanation described above, we now state the key claim and ``prove'' it in non-Abelian cohomology (cochain) terms: the action $(c^{0},z^{1})\mapsto (\delta^0c^{0})\circ z^{1}$ of $\Gamma_{A}$ on $\mathcal{C}_{\phi}$ is free and transitive. Preservation is immediate from the above formulas. For transitivity, take $z^{1},z^{1\prime}\in\mathcal{C}_{\phi}$. For each vertex $v$, choose a base point $a(v)\in A$ in the same connected component and a path $p_{v}:a(v)\to v$, and set
\begin{equation}
c^{0}_{v}\;:=\;\mathrm{Hol}_{z^{1\prime}}(p_{v})\ \mathrm{Hol}_{z^{1}}(p_{v})^{-1}.
\end{equation}
Changing $p_{v}$ alters $c^{0}_{v}$ by $\mathrm{Hol}_{z^{1\prime}}(\ell)\,\mathrm{Hol}_{z^{1}}(\ell)^{-1}$ for a loop $\ell$ based at $a(v)$; since $z^{1}$ and $z^{1\prime}$ have the same restriction $\phi$ on $\pi_{1}(K,A)$, this factor equals $1$, so $c^{0}_{v}$ is well-defined. If $v\in A$, take $p_{v}$ to be the constant path and get $c^{0}_{v}=1$, whence $c^{0}\in\Gamma_{A}$. For an edge $e:s\to t$ we may choose $p_{t}=p_{s}\cdot e$ and compute
\begin{equation}
c^{0}_{t}
=\mathrm{Hol}_{z^{1\prime}}(p_{s})\,z^{1\prime}_{e}\ \bigl(\mathrm{Hol}_{z^{1}}(p_{s})\,z^{1}_{e}\bigr)^{-1}
= z^{1\prime}_{e}\ c^{0}_{s}\ (z^{1}_{e})^{-1},
\end{equation}
equivalently $z^{1\prime}_{e}=c^{0}_{t}\,z^{1}_{e}\,(c^{0}_{s})^{-1}$, i.e.\ $z^{1\prime}=(\delta^0 c^{0}) \circ z^{1}$. Thus the action is transitive. For freeness, suppose $c^{0}\in\Gamma_{A}$ satisfies $(\delta^0 c^{0}) \circ z^{1}=z^{1}$. For any path $p:a\to v$ with $a\in A$ we get
\begin{equation}
\mathrm{Hol}_{z^{1}}(p)\;=\;\mathrm{Hol}_{\,c^{0}\cdot z^{1}}(p)\;=\;c^{0}_{v}\,\mathrm{Hol}_{z^{1}}(p)\,(c^{0}_{a})^{-1}\;=\;c^{0}_{v}\,\mathrm{Hol}_{z^{1}}(p),
\end{equation}
hence $c^{0}_{v}=1$ for all $v$, i.e.\ $c^{0}=\id$. Therefore
\begin{equation}
\mathcal{C}_{\phi}\;=\;\{\,(\delta^0 c^{0}) \circ z^{1}\mid c^{0}\in\Gamma_{A}\,\},
\end{equation}
and $\mathcal{C}_{\phi}$ is a $\Gamma_{A}$-torsor.
Moreover, as long as we fix $\phi$ at the topological level (i.e.\ on all paths in $\pi_{1}(K,A)$), this torsor collapses to a single point (equivalently, any two representatives are related by a gauge transformation trivial on $A$).

As explained in the previous section, this discussion is independent of the chosen good discretization. Refining a good cover (and hence the nerve) does not change the homotopy type of $\pi_{1}$, and the flatness constraint on the $2$-skeleton is preserved; the formulas for holonomy and gauge action carry over verbatim. Thus the “homotopy statement” (conjugacy classes of representations of $\pi_{1}$) and the “cohomology statement” (1-cocycles modulo the action of 0-cochains) are rigorously equivalent within this discrete setting.
For a more detailed mathematical explanation, please refer to \cite{olumNonAbelianCohomologyVan1958,ivanov2023nonabeliancohomologyseifertvankampen}.

A $G$-local system can be understood equivalently as a principal $G$-bundle equipped with a flat connection, so that parallel transport along paths is well defined and homotopy invariant.
The classical Riemann--Hilbert correspondence identifies isomorphism classes of such local systems with conjugacy classes of homomorphisms from the fundamental group to $G$; put differently, the global content of a flat connection is its monodromy~\cite{deligne2006equations,kobayashi1996foundations,husemollerFibreBundles1994}.
The resulting set admits two compatible readings: a representation-theoretic one (as monodromy data along loops) and a geometric/analytic one (as gauge-equivalence classes of flat $G$-connections).
There also exists a broader framework, the Non-Abelian Hodge correspondence, which organizes these readings within a unified perspective~\cite{simpson1992higgs}.

\section{The Fundamental Groupoid and Entanglement in $S^1$}
\label{app:S1}

We argued in the main text that the entanglement structure of theories with loop symmetry is fully revealed by the fundamental groupoid and the Seifert-van Kampen theorem. In this section, we understand why the fundamental groupoid, rather than the fundamental group, is necessary through a partition of $S^1$. Note, however, that the discretization of $S^1$ does not contain plaquettes, so no symmetry acts. Indeed, it can be verified that the resulting entanglement structure matches the outcome when no symmetry is imposed. The biparition is described as the inclusion maps:
\begin{equation}
\begin{tikzcd}
M=S^1 & X=S^1\backslash \{-\ii\} \arrow[l, hookrightarrow]  \\
Y=S^1\backslash \{\ii\}\arrow[u, hookrightarrow] & \partial = S^1\backslash \{\pm \ii\} \arrow[l, hookrightarrow]  \arrow[u, hookrightarrow]
\end{tikzcd}
\end{equation}
Two connected components of the boundary appear under bipartition of $S^1$.
We take $A=\{\pm 1\}$. Then,
\begin{equation}
  \Ob(\pi_1(\partial,A))=\{\pm 1\}, \quad \Hom(\pm 1,\pm 1)=\{\id_{\pm 1}\}, \quad \Hom(\pm 1,\mp 1)=\emptyset.
\end{equation}
Pictorially, this fundamental groupoid is trivial with respect to two base points, as follows:
\begin{center}
\begin{tikzpicture}
\fill (-1,0) circle (2pt) node[below] {$-1$};
\fill (1,0) circle (2pt) node[below] {$1$};
\draw[->, shorten >=3pt, shorten <=3pt] (-1,0) .. controls (-2,-0.5) and (-2,0.5) .. (-1,0) node[midway,above] {$\id_{-1}$};
\draw[->, shorten >=3pt, shorten <=3pt] (1,0) .. controls (2,-0.5) and (2,0.5) .. (1,0) node[midway,above] {$\id_{1}$};
\end{tikzpicture}
\end{center}
In contrast, the fundamental groupoids $\pi_1(X,A) \simeq \pi_1(Y,A)$ have two nontrivial holonomy $q$ and its inverse $q^{-1}$ that bridge two base points, as follows:
\begin{center}
\begin{tikzpicture}
\fill (-1,0) circle (2pt) node[below,yshift=-2mm] {$-1$};
\fill (1,0) circle (2pt) node[below,yshift=-2mm] {$1$};
\draw[->, shorten >=3pt, shorten <=3pt] (-1,0) .. controls (-2,-0.5) and (-2,0.5) .. (-1,0) node[midway,above] {$\id_{-1}$};
\draw[->, shorten >=3pt, shorten <=3pt] (1,0) .. controls (2,-0.5) and (2,0.5) .. (1,0) node[midway,above] {$\id_{1}$};
\draw[->, shorten >=3pt, shorten <=3pt] (-1,0) .. controls (-1.0,0.5) and (1.0,0.5) .. (1,0) node[midway,above] {$q^{-1}$};
\draw[->, shorten >=3pt, shorten <=3pt] (1,0) .. controls (1.0,-0.5) and (-1.0,-0.5) .. (-1,0) node[midway,above] {$q$};
\end{tikzpicture}
\end{center}
To put it more precisely, it can be understood as a small category possessing the following information:
\begin{equation}
\begin{split}
  \Ob(\pi_1(X,A))=\{\pm \ii\}, &\quad \Hom(\pm 1,\pm 1)=\{\id_{\pm 1}\},\\
  \Hom(1,-1)=\{q\}, &\quad \Hom(-1,1)=\{q^{-1}\},\\
  q^{-1}\circ q=\id_{1}, &\quad q\circ q^{-1}=\id_{-1}.
\end{split}
\end{equation}
Finally, the fundamental groupoid of the entire manifold $M$ is generated by the holonomy $q_{X/Y}$ as follows:
\begin{equation}
\begin{split}
  \Ob(\pi_1(M,A))=\{\pm 1\}, &\quad \Hom(\pm 1,\pm 1)=\mathbb{Z}\\
  \Hom(1,-1)=\{q_X \circ \mathbb{Z},q_Y \circ \mathbb{Z}\}, &\quad \Hom(-1,1)=\{q_X^{-1} \circ \mathbb{Z},q_Y^{-1} \circ \mathbb{Z}\}\\
  q_{X/Y}^{-1}\circ q_{X/Y}=\id_{1}, &\quad q_{X/Y}\circ q_{X/Y}^{-1}=\id_{-1}.
\end{split}
\end{equation}
This fundamental groupoid can be depicted as follows:
\begin{center}
\begin{tikzpicture}
\fill (-1,0) circle (2pt) node[right] {$-1$};
\fill (1,0) circle (2pt) node[left] {$1$};
\draw[->, shorten >=3pt, shorten <=3pt] (-1,0) .. controls (-2,-0.5) and (-2,0.5) .. (-1,0) node[midway,left] {$\id_{-1}$};
\draw[->, shorten >=3pt, shorten <=3pt] (-1,0) .. controls (-4,-1.5) and (-4,1.5) .. (-1,0) node[midway,left] {$1_{-1}$};
\draw[->, shorten >=3pt, shorten <=3pt] (-1,0) .. controls (-6,-2.5) and (-6,2.5) .. (-1,0) node[midway,left] {$\cdots \quad 2_{-1}$};
\draw[->, shorten >=3pt, shorten <=3pt] (1,0) .. controls (2,-0.5) and (2,0.5) .. (1,0) node[midway,right] {$\id_{1}$};
\draw[->, shorten >=3pt, shorten <=3pt] (1,0) .. controls (4,-1.5) and (4,1.5) .. (1,0) node[midway,right] {$1_1$};
\draw[->, shorten >=3pt, shorten <=3pt] (1,0) .. controls (6,-2.5) and (6,2.5) .. (1,0) node[midway,right] {$2_{1}\quad \cdots$};
\draw[->, shorten >=5pt, shorten <=5pt] (-1,0) .. controls (-1.0,0.5) and (1.0,0.5) .. (1,0) node[midway,above] {$q_{X/Y}^{-1}$};
\draw[->, shorten >=5pt, shorten <=5pt] (-1,0) .. controls (-2.0,1.5) and (2.0,1.5) .. (1,0) node[midway,above] {$q_{X/Y}^{-1}\circ 1_{-1}$};
\draw[->, shorten >=5pt, shorten <=5pt] (-1,0) .. controls (-3.0,2.5) and (3.0,2.5) .. (1,0) node[midway,above] {$q_{X/Y}^{-1}\circ 2_{-1}$};
\draw[->, shorten >=5pt, shorten <=5pt] (1,0) .. controls (1.0,-0.5) and (-1.0,-0.5) .. (-1,0) node[midway,below] {$q_{X/Y}$};
\draw[->, shorten >=5pt, shorten <=5pt] (1,0) .. controls (2.0,-1.5) and (-2.0,-1.5) .. (-1,0) node[midway,below] {$q_{X/Y}\circ 1_1$};
\draw[->, shorten >=5pt, shorten <=5pt] (1,0) .. controls (3.0,-2.5) and (-3.0,-2.5) .. (-1,0) node[midway,below] {$q_{X/Y}\circ 2_1$};
\draw (0,3) node{$\vdots$};
\draw (0,-3) node{$\vdots$};
\end{tikzpicture}
\end{center}
From now on, we count up the block size. Given a finite group $G$, one finds that the $\Hom$-applied fundamental groupoids are just coloring by $G$ and they are as follows:
\begin{equation}
  \Hom (\pi_1(\partial,A),G)=\{\id\}, \quad \Hom (\pi_1(X,A),G)=G, \quad \Hom (\pi_1(Y,A),G)=G, \quad \textrm{and} \quad \Hom (\pi_1(M,A),G)=G^2.
\end{equation}
Now, the commutative diagram of the $\Hom$-applied fundamental groupoid can be constructed:
\begin{equation}
\begin{tikzcd}
G^2 \arrow[r, "\Pi_X"]\arrow[d, "\Pi_Y"'] & G \arrow[d, "r_X"] \\
G \arrow[r, "r_Y"'] & \{\id\}  \\
\end{tikzcd}
\end{equation}
Since only one element, $\id$, can be the image, the preimage is easily estimated as $|r_X^{-1}(\id)|=|r_Y^{-1}(\id)|=|G|$. Note that we have $|V_\partial|=|A|=2$.
Thus, the block structure breaks down to $\mathbb{C}^{|G|\cdot |G|^{|V_X|} \times |G|\cdot |G|^{|V_Y|}}.$
Suppose the entire system forms the chain of length $|E|$.
One finds that $|E|=|E_X|+|E_Y|$, $|E_X|=|V_X|+1$ and $|E_Y|=|V_Y|+1$, implying that the result can be interpreted as $\mathbb{C}^{|G|^{|E_X|} \times |G|^{|E_Y|}}.$
This result is consistent with the case of unconstrained entanglement.

\section{Computation of the Examples}

This section provides a detailed explanation, including the calculations for the specific examples introduced in the main text.

\subsection{Regular Rep Trick}

In the calculation of specific examples, we will explain a trick, which we call ``regular rep trick'' and will be repeatedly used as a tool for describing block sizes using representation theory data.

In many cases, fundamental groups are obtained by imposing constraint relations on free groups.
These constraint relations are preserved even when coloring is applied via the $\Hom(*,G)$ functor.
Consider the case where these constraint relations can be rewritten in the form of a $\textrm{group element product}=1$, and where the variable appears exactly twice within the relation.
Here, a variable refers to a holonomic degree of freedom that is not fixed.

First, consider cases where variables appear in relation to inverse elements within the relation. Specifically, let $g$ be a variable and $g_1,g_2$ be arbitrary constants taking values in $G$. 
The relation then takes the form of $g g_1 g^{-1} g_2 =1$. To evaluate the number of elements in $G$, represented by $g$, that satisfy this relation, we sum up $g$ with weight $\delta(g g_1 g^{-1} g_2)$. 
Here $\delta$ denotes the delta function of group, which takes value 1 when the argument is identity, or takes value 0 otherwise.

The group delta function is nothing other than the normalized version of the character of the regular representation.
That is, the delta function can be rewritten using the following decomposition:
\begin{equation}
    \delta(g g_1 g^{-1} g_2)=\frac{1}{|G|}\Tr D^\mathrm{reg} (g g_1 g^{-1} g_2)=\frac{1}{|G|}\sum_\alpha d_\alpha \Tr D^\alpha (g g_1 g^{-1} g_2)
\end{equation}
On the other hand, in the representation theory of finite groups, there exists grand orthogonality:
\begin{equation}
    \frac{1}{|G|}\sum_{g\in G} D^\alpha_{ij}(g)\overline{D^\alpha_{kl}(g)}=\frac{1}{d_\alpha}\delta_{\alpha \beta}\delta_{ik}\delta_{jl}.
\end{equation}
Summing up with $g$, we can count up the group elements that satisfy the constraint relation as follows:
\begin{equation}
\begin{split}
    \sum_{g\in G}\delta(g g_1 g^{-1} g_2)
    &=\frac{1}{|G|}\sum_\alpha d_\alpha \sum_{ijkl} \sum_{g\in G} D^\alpha_{ij} (g) D^\alpha_{jk} (g_1) \overline{D^\alpha_{l k} (g)} D^\alpha_{li}(g_2)\\
    &=\sum_\alpha d_\alpha \sum_{ijkl} \frac{\delta_{il}\delta_{jk}}{d_\alpha} D^\alpha_{jk} (g_1) D^\alpha_{li}(g_2)\\
    &=\sum_\alpha \chi^\alpha(g_1) \chi^\alpha(g_2).
\end{split}
\end{equation}

Next, consider cases where variables appear twice within the relation.
Specifically, let $g$ be a variable and $g_1,g_2$ be arbitrary constants taking values in $G$. 
The relation then takes the form of $g g_1 g g_2 =1$. 
Following the same procedure as above, we can proceed until we have
\begin{equation}
\begin{split}
    \sum_{g\in G}\delta(g g_1 g g_2)
    &=\frac{1}{|G|}\sum_\alpha d_\alpha \sum_{ijkl} \sum_{g\in G} D^\alpha_{ij} (g) D^\alpha_{jk} (g_1) D^\alpha_{k l} (g) D^\alpha_{li}(g_2).
\end{split}
\end{equation}
To proceed, we use the Frobenius-Schur indicator, defined as follows:
\begin{equation}
    \iota^\alpha=\frac{1}{|G|}\sum_g \chi^\alpha(g^2)=\begin{cases}
        1 \quad (\textrm{when we can take as } \overline{D^\alpha}=D^\alpha \textrm{ with specific basis}),\\
        0 \quad (\overline{D^\alpha}\ncong D^\alpha),\\
        -1 \quad (\textrm{otherwise}).
    \end{cases}
\end{equation}
The final case occurs when the complex conjugate representation and the original representation are unitarily equivalent but not identical without a change of basis. In such cases, the irrep is of even dimension, and it is known that the unitary transformation connecting the two can be taken as tensoring the Pauli $Y$ matrix with the identity matrix for some basis, namely
\begin{equation}
    \iota^\alpha=-1,\quad \textrm{ when } D^\alpha=\overline{Y D^\alpha Y^\dagger} \textrm{ where } Y=\mathbb{1}_\frac{d_\alpha}{2}\otimes \begin{pmatrix}
        0&-i\\
        i&0
    \end{pmatrix}.
\end{equation}
We can rewrite the sum above by flipping $\alpha\mapsto \overline{\alpha}$ that appeared in the second place, as follows:
\begin{equation}
\begin{split}
    \sum_{g\in G}\delta(g g_1 g g_2)
    &=\frac{1}{|G|}\sum_{\alpha: \iota^\alpha=1} d_\alpha \sum_{ijkl} \sum_{g\in G} D^\alpha_{ij} (g) D^\alpha_{jk} (g_1) \overline{D^\alpha_{k l} (g)} D^\alpha_{li}(g_2)\\
    &+\frac{1}{|G|}\sum_{\alpha: \iota^\alpha=0} d_\alpha \sum_{ijkl} \sum_{g\in G} D^\alpha_{ij} (g) D^\alpha_{jk} (g_1) \overline{D^{\overline{\alpha}}_{k l} (g)} D^\alpha_{li}(g_2)\\
    &+\frac{1}{|G|}\sum_{\alpha: \iota^\alpha=-1} d_\alpha \sum_{ijkl k'l'} \sum_{g\in G} D^\alpha_{ij} (g) D^\alpha_{jk} (g_1) \overline{Y_{k k'} D^{\alpha}_{k' l'} (g)Y_{l' l}} D^\alpha_{li}(g_2)\\
    &=\sum_{\alpha: \iota^\alpha=1} \chi^\alpha(g_1 g_2)
    +\sum_{\alpha: \iota^\alpha=-1} \sum_{ijkl} D^\alpha_{jk} (g_1) \overline{Y_{k i} Y_{j l}} D^\alpha_{li}(g_2)\\
    &=\sum_{\alpha} \iota^\alpha \chi^\alpha(g_1 g_2).
\end{split}
\end{equation}
Here we used $\overline{Y_{k i} Y_{j l}}=-\delta_{\lfloor k/2\rfloor \lfloor i/2\rfloor}\delta_{\lfloor j/2\rfloor \lfloor l/2\rfloor}\delta_{[i] [l]}\delta_{[k] [j]}\delta_{[i][k+1]}\delta_{[j][l+1]}$, where $[i]$ is equivalence class of the remainder of $i \,\mathrm{mod} \, 2$.

Examples that can be calculated in the same manner are listed below:
\begin{equation}
\begin{split}
  \sum_{g_1} \delta(g_1 g \bar{g_1} g')&=\sum_{\alpha}\chi^\alpha(g)\chi^\alpha(g')\\
  \sum_{g_1}\chi^\alpha(g_1)\chi^\alpha(\bar{g_1} g)&=\frac{|G|}{d_\alpha}\chi^\alpha (g)\\
  \sum_{g_1}\chi^\alpha(g_1 g \bar{g_1} g')&=\frac{|G|}{d_\alpha} \chi^\alpha (g)\chi^\alpha (g')\\
  \sum_{g_1,g_2}\chi^\alpha(g_1 g_2 \bar{g_1} \bar{g_2} g)&=\left(\frac{|G|}{d_\alpha}\right)^2 \chi^\alpha(g)\\
  \sum_g \chi^\alpha(g^2 g')&=\frac{|G|}{d_\alpha} \iota^\alpha \chi^\alpha(g').
\end{split}
\end{equation}
Now, we move on to the examples. The derived formula will be repeatedly used to calculate the block size for a specific example.

\subsection{Torus $\mathbb{T}^2$}
\label{app:torus}
We first consider the case of torus $\mathbb{T}^2=S^1\times S^1$.
Let $\Sigma_{\gamma,n}$ be the orientable surface with genus $\gamma$ and $n$ punctures.
There are several ways of decomposition:
\begin{equation}
  (\gamma_X,\gamma_Y,n)=(0,1,1),(1,0,1),(0,0,2).
\end{equation}
The entanglement in the second case is obtained by swapping $X$ and $Y$ in the first case; thus, there are essentially two ways of decomposition. In general, closed orientable surface has the following decomposition:
\begin{equation}
  \Sigma_{\gamma_X,n}\sqcup_{\sqcup^n S^1} \Sigma_{\gamma_Y,n} \mapsto \Sigma_{\gamma_X+\gamma_Y+n-1,0}.
\end{equation}
We repeatedly use this in the following discussions.

In the case of $(\gamma_X,\gamma_Y,n)=(0,1,1)$, we decompose the torus to the disk $\Sigma_{0,1}=D^2$ and the other.
The diagram of manifold partition is as follows:
\begin{equation}
\begin{tikzcd}
M=\Sigma_{1,0} & X=\Sigma_{0,1} \arrow[l, hookrightarrow]  \\
Y=\Sigma_{1,1}  \arrow[u, hookrightarrow] & \partial = S^1 \arrow[u, hookrightarrow] \arrow[l, hookrightarrow] 
\end{tikzcd}
\end{equation}
The morphisms are inclusion maps.

Applying the Seifert-van Kampen theorem, we obtain the diagram of fundamental groupoids. 
Because the boundary is a connected manifold, a single base point on the boundary is sufficient.
The diagram for the fundamental groupoids is as follows:
\begin{equation}
\begin{tikzcd}
\pi_1(\Sigma_{1,0},A)=\mathbb{Z} \oplus \mathbb{Z} & \pi_1(\Sigma_{0,1},A)=\{\id\}  \arrow[l, "p_X"] \\
\pi_1(\Sigma_{1,1},A)=F_2 \arrow[u, "p_Y"'] &\pi_1(S^1,A)=\mathbb{Z}\arrow[u, "i_X"] \arrow[l, "i_Y"']
\end{tikzcd}
\end{equation}
where $F_n$ is the free group with $n$ generators.
Concretely, the inclusion $i_X:S^1\hookrightarrow D^2$ induces the trivial map $i_{X*}:\mathbb{Z}\to 1$, while the inclusion $i_Y:S^1\hookrightarrow \Sigma_{1,1}$ sends the generator of $\mathbb{Z}$ to the boundary loop of $\Sigma_{1,1}$, which in $\pi_1(\Sigma_{1,1})\cong F_2=\langle a,b\rangle$ represents the commutator $[a,b]$.

Application of $\Hom(*,G)$ functor sends the fundamental groupoids to the category of sets as follows:
\begin{equation}
\begin{split}
\Hom (\mathbb{Z},G)&=G\\
\Hom(\{\id\},G)&=\{\id\}\\
\Hom(F_2,G)&=G^2\\
\Hom(\mathbb{Z}\oplus \mathbb{Z},G)&=\mathrm{Comm}_2(G).
\end{split}
\end{equation}
Here we defined the set of commuting elements $\mathrm{Comm}_n (G):=\{(g_1,\cdots,g_n)\in G^n|{}^\forall i,j=1,\cdots, n, \space g_i g_j = g_j g_i\}$. 
Note that $\Hom(\mathbb{Z}^{\oplus n},G)=\mathrm{Comm}_n(G)$ holds for arbitrary $n\geq 0$. Since $\Hom(*,G)$ is contravariant functor the orientations of morphisms are inverted, and we obtain the following diagram:
\begin{equation}
\begin{tikzcd}
\mathrm{Comm}_2(G) \arrow[d, "\Pi_Y"'] \arrow[r, "\Pi_X"] & \{\id\}  \arrow[d, "r_X"] \\
G^2  \arrow[r, "r_Y"'] & G \\
\end{tikzcd}
\end{equation}

A significant property of this diagram is $\Hom(\pi_1(X,A),G)=\{\id\}$.
This requires the image to be trivial; $\mathrm{Im}=\{\id \}.$
Obviously we have $r_X^{-1}(\id)=\{\id\}$. 
On the other hand, only the elements in $\mathrm{Comm}_2(G)$ can be compatible in $\Sigma_{1,0}$.
This requires the preimage about $Y$ to be $r_Y^{-1}(\id)=\mathrm{Comm}_2(G)$.
From the regular rep trick, we find
\begin{equation}
|\mathrm{Comm}_2(G)|=\sum_{g_1,g_2}\delta([g_1,g_2])=\sum_{\alpha,g_2}\chi^\alpha(g_2)\chi^\alpha(g_2^{-1})=|G||\mathrm{Irr}(G)|.
\end{equation}
Finally, we find that $W$ takes the form
\begin{equation}
  \bigoplus_{j=1}^{|G|^{|V_\partial|-1}}
  \mathbb{C}^{
    |G|^{|V_X|} \times |\mathrm{Irr}(G)||G|^{|V_Y|+1}
  }.
\end{equation}

\subsubsection{$(\gamma_X,\gamma_Y,n)=(0,0,2)$}

We examine the structure of the $W$ matrix using a partitioning method that differs topologically from the previous example.
The diagram of manifold partition is as follows:
\begin{equation}
\begin{tikzcd}
M=\Sigma_{1,0}   & X=\Sigma_{0,2} \arrow[l, hookrightarrow]  \\
Y=\Sigma_{0,2} \arrow[u, hookrightarrow] & \partial = S^1 \sqcup S^1 \arrow[l, hookrightarrow] \arrow[u, hookrightarrow]
\end{tikzcd}
\end{equation}
We place one base points on each connected components of $\partial$: $|A|=2$.
\begin{equation}
\begin{tikzcd}
\pi_1(\Sigma_{1,0},A) & \pi_1(\Sigma_{0,2},A) \arrow[l, "p_X"] \\
\pi_1(\Sigma_{0,2},A) \arrow[u, "p_Y"'] & \pi_1(S^1\sqcup S^1,A) \arrow[l, "i_Y"'] \arrow[u, "i_X"] 
\end{tikzcd}
\end{equation}
The fundamental groupoid of the union of two copies of the circle $S^1$ is generated by two independent generators. Each generator can be freely chosen from $G$. Consequently, the homomorphism group $\Hom (\pi_1(S^1\sqcup S^1,A),G)$ is isomorphic to $G^2$ as a set.

Next we focus on $\pi_1(X=\Sigma_{0,2},A)$. This fundamental groupoid requires an additional generator $b_X$ which connects two base points.
Since two base points are connected, two boundary holonomies are conjugate to each other. This is expressed as the constraint $a_1=b_X a_2 b_X^{-1}$.
Then, we only have the freedom to choose $a_1$ and $b_X$ independently.
Consequently, $\Hom (\pi_1(\Sigma_{0,2},A),G)\simeq G^2$. Similarly, $\pi_1(Y=\Sigma_{0,2},A)$ is generated by two generators. The holonomies on boundaries are related by a constraint $a_1=b_Y^{-1} a_2 b_Y$.

Finally, the groupoid $\pi_1(\Sigma_{1,0},A)$ is determined to be generated by four paths depicted as follows
\begin{center}
\begin{tikzpicture}
\fill (-1,0) circle (2pt) node[below,yshift=-2mm] {$1$};
\fill (1,0) circle (2pt) node[below,yshift=-2mm] {$2$};
\draw[->, shorten >=3pt, shorten <=3pt] (-1,0) .. controls (-2,-0.5) and (-2,0.5) .. (-1,0) node[midway,above] {$a_1$};
\draw[->, shorten >=3pt, shorten <=3pt] (1,0) .. controls (2,-0.5) and (2,0.5) .. (1,0) node[midway,above] {$a_2$};
\draw[->, shorten >=3pt, shorten <=3pt] (-1,0) .. controls (0,0.5) .. (1,0) node[midway,above] {$b_X$};
\draw[->, shorten >=3pt, shorten <=3pt] (1,0) .. controls (0,-0.5) .. (-1,0) node[midway,above] {$b_Y$};
\end{tikzpicture}
\end{center}
with the constraints $a_1=b_X a_2 b_X^{-1}$ and $a_1=b_Y^{-1} a_2 b_Y$.
This simplifies to $[a_1,b_X b_Y]=[a_2,b_X^{-1} b_Y^{-1}]=1$, implying that two holonomies on the torus are commutative and $\Hom (\pi_1(\Sigma_{1,0},A),G)\simeq \mathrm{Comm}_2(G)\times G$. The diagram of the $\Hom$-applied fundamental groupoids is as follows:
\begin{equation}
\begin{tikzcd}
\mathrm{Comm}_2(G)\times G = \{(a_1,b_X,b_Y)\} \arrow[d, "\Pi_Y"'] \arrow[r, "\Pi_X"] & G^2  = \{(a_1,b_X)\} \arrow[d, "r_X"] \\
G^2 = \{(a_1,b_Y)\} \arrow[r, "r_Y"'] & G^2  = \{(a_1,a_2)\} \\
\end{tikzcd}
\end{equation}
Here, $r_{X}:(a_1,b_{X})\mapsto (a_1,b_X^{-1}a_1 b_X)$ and $r_{Y}:(a_1,b_Y)\mapsto (a_1,b_Ya_1 b_Y^{-1})$, thus $\mathrm{Im}$ is labeled by $(a_1,a_2)$ which satisfies ${}^\exists g\in G, g a_1 g^{-1} = a_2$.
For each $a_1$, the size of the conjugacy class is given by the orbit-stabilizer theorem $|[g]|=|G|/|\mathrm{C}_g|$ where $\mathrm{Conj}(g)$ is the conjugacy class of $g$ and $\mathrm{C}_g$ is the centralizer of $g$.
Thus, the decomposition is given as follows:
\begin{equation}
  \bigoplus_{c\in G}
  \bigoplus_{j=1}^{|[c]| |G|^{|V_\partial|-2}} \mathbb{C}^{|{\rm C}_c||G|^{|V_X|}\times |{\rm C}_c||G|^{|V_Y|}},
\end{equation}
Here, the size of $\mathrm{C}_g$ can be calculated from the regular rep trick as follows:
\begin{equation}
  |\mathrm{C}_g|=\sum_{g'}\delta(gg'\bar{g}\bar{g'})=\frac{1}{|G|}\sum_{g'}\Tr D^\mathrm{reg}(gg'\bar{g}\bar{g'})=\frac{1}{|G|}\sum_{\alpha,g'}d_\alpha \,\Tr D^\alpha(gg'\bar{g}\bar{g'})=\sum_{\alpha}|\chi^\alpha(g)|^2
\end{equation}

\subsection{Genus $\gamma$ Surface}
\label{app:ggs}
\begin{figure}
\begin{center}
       \includegraphics[width=10cm, clip]{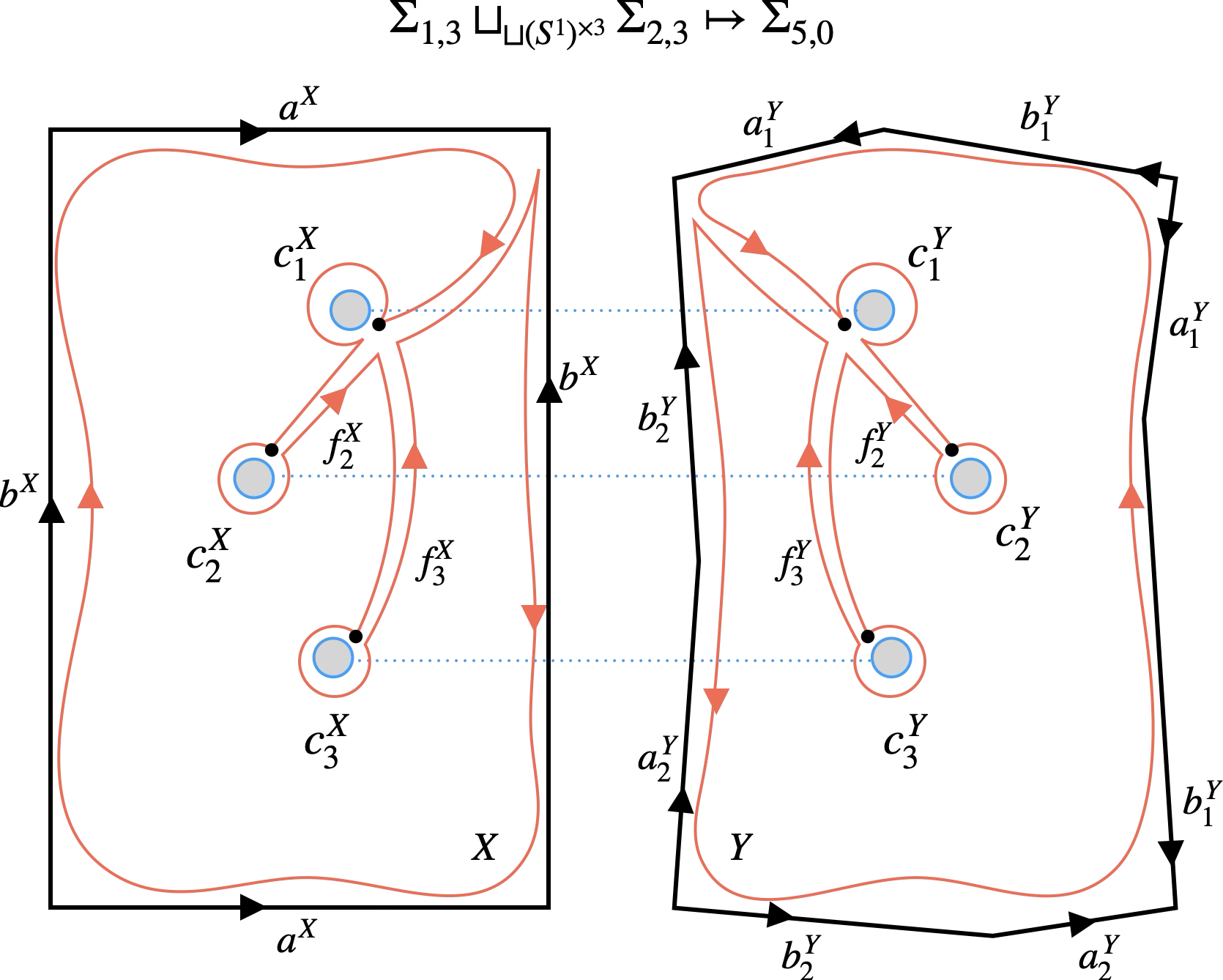}
       \end{center}
   \caption{Two surfaces with a common boundary are glued together. On each surface, the red loop is contractible, so the holonomic information captured by that loop yields consistency conditions within each manifold. If the consistency conditions for both surfaces are satisfied, the overall consistency condition is also satisfied.} 
   \label{fig:orientable-surface-gluing}
\end{figure}

The torus can be regarded as an example of a connected, oriented, closed, and compact two-dimensional manifold — a genus-$\gamma$ surface \cite{gallierGuideClassificationTheorem2013,francisConwaysZIPProof1999}.
So we now consider general bipartition of genus $\gamma$ surfaces.
As we explained in the previous example, such surfaces admit the decomposition $\Sigma_{\gamma_X,n}\sqcup_{\sqcup^n S^1} \Sigma_{\gamma_Y,n} \mapsto \Sigma_{\gamma_X+\gamma_Y+n-1,0}$.
From now on, we consider this general decomposition.
We set $\gamma=\gamma_X+\gamma_Y+n-1$ and place $n$ base points on each connected components of the boundary.
The fundamental groupoids forms the commutative diagram as follows:
\begin{equation}
\begin{tikzcd}
\pi_1(\Sigma_{\gamma,0},A)=\frac{F_{2\gamma}}{\prod_{i=1}^{\gamma}[a_i,b_i]} \,\&\, 2n-2\textrm{ paths }  & \pi_1(\Sigma_{\gamma_X,n},A)=\frac{F_{2\gamma_X+2n-2}}{\prod_{i=1}^{\gamma_X}[a^X_i,b^X_i]\prod_{i=1}^n f^X_i c^X_i (f^X_i)^{-1}}\arrow[l, "p_X"] \\
\pi_1(\Sigma_{\gamma_Y,n},A)=\frac{F_{2\gamma_Y+2n-2}}{\prod_{i=1}^{\gamma_Y}[a^Y_i,b^Y_i]\prod_{i=1}^n f^Y_i c^Y_i (f^Y_i)^{-1}}\arrow[u, "p_Y"'] & \pi_1(\sqcup^n S^1,A) \arrow[l, "i_Y"'] \arrow[u, "i_X"] 
\end{tikzcd}
\end{equation}
where $a_i,b_i$ are the holonomies along $i$th a-cycle and b-cycle in each manifolds.
$c_i$ are the holonomy along $i$th boundary and $f_i$ denotes the one-way path holonomy from the base point $i$ to the base point $1$
Here, holonomies must satisfy the conditions in each manifolds as $\mathrm{rel}_X(a^X,b^X,c,f^X)=\mathrm{rel}_Y(a^Y,b^Y,c,f^Y)=\mathrm{rel}_{M}(a,b,c,f)=1$ for ${}^\exists \{a,b,f\}$, where
\begin{equation}
\begin{split}
  \mathrm{rel}_X&(a^X,b^X,c,f^X):=[a^X_1,b^X_1]\cdots [a^X_{\gamma_X}, b^X_{\gamma_X}]
  f^X_n c_n (f^X_n)^{-1}\cdots f^X_1 c_1 (f^X_1)^{-1}\\
  \mathrm{rel}_Y&(a^Y,b^Y,c,f^Y):=[a^Y_1,b^Y_1]\cdots [a^Y_{\gamma_Y}, b^Y_{\gamma_Y}]
  f^Y_n c_n (f^Y_n)^{-1}\cdots f^Y_1 c_1 (f^Y_1)^{-1} \\
  \mathrm{rel}_{M}&(a,b,c,f):=\left(\mathrm{rel}_X(a^X,b^X,c,f^X)\right)\left(\mathrm{rel}_Y(a^Y,b^Y,c,f^Y)\right)^{-1}.
\end{split}
\end{equation}
This comes from the fact that any contractible loop on the surface has trivial holonomy, as depicted in Fig.~\ref{fig:orientable-surface-gluing}.
The denominator of the group fraction represents the constraint (to be 1) imposed on the group of the numerator.
$\pi_1(\sqcup^n S^1,A)$ consists of $n$ independent loops that does not share base points.
We fix $f^{X/Y}_1=1$, since the first base point is set as the starting point of paths $f^{X/Y}_i$.
The condition $\mathrm{rel}_{M}(a,b,c,f)=1$ is automatically satisfied if $\mathrm{rel}_X(a^X,b^X,c,f^X)=\mathrm{rel}_Y(a^Y,b^Y,c,f^Y)=1$.

The commutative diagram of $\Hom$-applied fundamental groupoid is as follows:
\begin{equation}
\begin{tikzcd}
\frac{G^{2\gamma_X+2\gamma_Y+3n-3}}{\prod_{i=1}^{\gamma}[a_i,b_i]} \arrow[d, "\Pi_Y"'] \arrow[r, "\Pi_X"] & G^{2\gamma_X+2n-2}  \arrow[d, "r_X"] \\
G^{2\gamma_Y+2n-2}  \arrow[r, "r_Y"'] & G^n \\
\end{tikzcd}
\end{equation}
For an element $(c_1,\cdots,c_n)\in G^n$, we evaluate the size of the blocks by using the regular rep trick. 
The block structure of the density matrix is identified as
\begin{equation}
  \bigoplus_{\{c\}} \bigoplus_{j=1}^{|V_\partial|-n} \mathbb{C}^{R_{\gamma_X,n}(c)|G|^{|V_X|}\times R_{\gamma_Y,n}(c)|G|^{|V_Y|}}
\end{equation}
where $r^{-1}_*(c)=\{\{a,b,d\}|\mathrm{rel}_{*}(a,b,c,d)=1\}$.
Then, the block size for $X$ is computed as follows:
\begin{equation}
\begin{split}
  \sum_{\{a,b,f\}}\delta (\mathrm{rel}_X (a,b,c,d))\\
  &=\sum_{\alpha\in{\rm Rep}(G),\{f\}}\left(\frac{|G|}{d_\alpha}\right)^{2\gamma_X -1}\chi^\alpha (f^X_n c_n (f^X_n)^{-1}\cdots f^X_1 c_1 (f^X_1)^{-1})\\
  &=\sum_{\alpha \in{\rm Rep}(G)} \left(\frac{|G|}{d_\alpha}\right)^{2\gamma_X+n-2}\prod_{i=1}^n \chi^\alpha (c_i)=:R_{\gamma_X,n}(c).
\end{split}
\end{equation}
Similarly, we obtain the block size for $Y$ as follows:
\begin{equation}
  R_{\gamma_Y,n}(c)=\sum_{\alpha \in{\rm Rep}(G)} \left(\frac{|G|}{d_\alpha}\right)^{2\gamma_Y+n-2}\prod_{i=1}^n \chi^\alpha (c_i).
\end{equation}
Thus, even for the general case of a genus $\gamma$ surface, the block structure can be computed algorithmically if one possesses information from representation theory, particularly the character table.

To appreciate the power of this result, let us fix $G = D_6$ and explicitly compute the block structure for two examples: $\Sigma_{0,2}\sqcup_{(s^1)^2} \Sigma_{0,2}\mapsto\Sigma_{1,0}$ and $\Sigma_{1,1}\sqcup_{s^1} \Sigma_{1,1}\mapsto\Sigma_{2,0}$.

$D_6$ has two irreps of dimension 1 and one irrep of dimension 2. 
The character table is shown in Table \ref{tab:d6charactertable}.
\begin{table}[b]
  \caption{Character table of $D_6$.}
  \label{tab:d6charactertable}
  \centering
  \begin{tabular}{c|cccccc}
      & $1$ & $r$ & $r^2$ & $s$ & $sr$ & $sr^2$ \\
    \hline
    trivial rep $\mathbb{1}$& $1$ & $1$ & $1$ & $1$ & $1$ & $1$ \\
    sign rep $\mathbb{1}$'& $1$ & $1$ & $1$ & $-1$ & $-1$ & $-1$ \\
    standard rep $\mathbb{2}$& $2$ & $-1$ & $-1$ & $0$ & $0$ & $0$ \\
  \end{tabular}
\end{table}
We make use of the formula
\begin{equation}
\begin{split}
    R_{\gamma_X,n}(c)&=R_{\gamma_Y,n}(c)
    =\sum_{\alpha\in \mathrm{Rep}(G)} \left(\frac{|G|}{d_\alpha}\right)^{2\gamma_{X/Y}+n-2}\prod_{i=1}^n \chi^\alpha (c_i)\\
    &=6^0\cdot 1+6^0\cdot 1+ 3^0\cdot \chi^\mathbb{2} (c_1)^2.
\end{split}
\end{equation}
Here we used the fact that the holonomies along two boundaries are conjugate to each other and the character is the class function with respect to conjugacy class. One can easily check that
$|r^{-1}(1)|=6,\,|r^{-1}(r)|=|r^{-1}(r^2)|=3,|r^{-1}(s)|=|r^{-1}(sr)|=|r^{-1}(sr^2)|=2$.
The choice of boundary holonomies are just the square of the order of conjugacy classes.
The block structure of $W$ turns out to be
\begin{equation}
    \left(\mathbb{C}^{6\times 6}\oplus 4 \mathbb{C}^{3\times 3}\oplus 9 \mathbb{C}^{2\times 2}\right)\otimes \bigoplus_{j=1}^{|G|^{|V_\partial|-|A|}}\mathbb{C}^{|G|^{|V_X|} \times |G|^{|V_Y|}}.
    \label{eq:example-d6-torus-nontrivial-bipartition}
\end{equation}

We can derive the same result from the analysis of centralizer groups, although it is essentially equivalent to the prior analysis.
We recall another form of the block structure:
\begin{equation}
  \bigoplus_{c\in G}
  \bigoplus_{j=1}^{|[c]| |G|^{|V_\partial|-2}} \mathbb{C}^{|{\rm C}_c||G|^{|V_X|}\times |{\rm C}_c||G|^{|V_Y|}},
\end{equation}
The centralizers for each group element in $D_6$ are enumerated in Table \ref{tab:d6centralizers}.
\begin{table}[t]
  \caption{Centralizers for elements in $D_6$.}
  \label{tab:d6centralizers}
  \centering
  \begin{tabular}{cccc}
    \hline
    Elements in $D_6$ & Centralizer group \\ 
    \hline
    $1$ & $D_6$ \\
    $r$ & $\{1,r,r^2\}\simeq \mathbb{Z}_3$ \\
    $r^2$ & $\{1,r,r^2\}\simeq \mathbb{Z}_3$ \\
    $s$ & $\{1,s\}\simeq \mathbb{Z}_2$ \\
    $sr$ & $\{1,sr\}\simeq \mathbb{Z}_2$ \\
    $sr^2$ & $\{1,sr^2\}\simeq \mathbb{Z}_2$ \\
    \hline
  \end{tabular}
\end{table}
From Table \ref{tab:d6centralizers}, one can immediately comprehend that $W$ takes the form of Eq.~(\ref{eq:example-d6-torus-nontrivial-bipartition}).

We move on to the case of $\Sigma_{1,1}\sqcup_{s^1} \Sigma_{1,1}\mapsto\Sigma_{2,0}$. Applying the formula, we obtain
\begin{equation}
    |r^{-1}_X (c)|=|r^{-1}_Y (c)|=6^{2+1-2}\cdot 1+6^{2+1-2}\cdot \chi^\mathbb{1'}(c)+ 3^{2+1-2}\cdot \chi^\mathbb{2} (c).
\end{equation}
One can easily check that
$|r^{-1}(1)|=18,\,|r^{-1}(r)|=|r^{-1}(r^2)|=9,|r^{-1}(s)|=|r^{-1}(sr)|=|r^{-1}(sr^2)|=0$.
The blocks are only labeled by the boundary holonomy, implying $W$ takes the form 
\begin{equation}
    \left(\mathbb{C}^{18\times 18}\oplus 2 \mathbb{C}^{9\times 9}\right)\otimes \bigoplus_{j=1}^{|G|^{|V_\partial|-|A|}}\mathbb{C}^{|G|^{|V_X|} \times |G|^{|V_Y|}}.
\end{equation}

The coefficients and block sizes revealed here are nontrivial, and calculating them without our algorithm is difficult.

\subsection{Real 2D Projected Space}

Now that we have explored the general case of orientable two-dimensional manifolds, we turn to the case of non-orientable manifolds. A connected, non-orientable, closed, and compact two-dimensional topological manifold is completely classified and known as the surface with crosscaps.
Henceforth, $N_{k,n}$ will denote the two-dimensional surface with $k$ crosscaps and $n$ boundary circles. As a warm up, we will focus on two primary cases real 2D projected space and the Klein bottle. 
After that, we will discuss general bipartition of general non-orientable surfaces.

Consider real 2D projected space $\mathbb{RP}^2$.
Assuming $X$ and $Y$ both connected, topologically nontrivial bipartition does not exist.
We consider the bipartition below:
\begin{equation}
\begin{tikzcd}
M=\mathbb{RP}^2 & X=N_{0,1} \arrow[l, hookrightarrow]  \\
Y=N_{1,1}  \arrow[u, hookrightarrow] & \partial = S^1 \arrow[u, hookrightarrow] \arrow[l, hookrightarrow] 
\end{tikzcd}
\end{equation}
Note that $N_{0,1}$ is nothing but the disk $D^2$, and $N_{1,1}$ is exactly the Möbius band.
The boundary is connected, thus we need only one base point.
The fundamental groupoids for each manifolds are given as follows:
\begin{equation}
\begin{split}
\pi_1(\mathbb{RP}^2,A)&=\mathbb{Z}_2\\
\pi_1(N_{0,1},A)&=\{\id\}\\
\pi_1(N_{1,1},A)&=\pi_1(S^1,A)=\mathbb{Z}.
\end{split}
\end{equation}
The diagram for the $\Hom$-applied fundamental groupoids is given as follows:
\begin{equation}
\begin{tikzcd}
\{g\in G|g^2=1\} \arrow[d, "\Pi_Y"'] \arrow[r, "\Pi_X"] & \{\id\}  \arrow[d, "r_X"] \\
G  \arrow[r, "r_Y"'] & G \\
\end{tikzcd}
\end{equation}
Every element in $\Hom(\pi_1(M,A),G)$ is mapped to the trivial element by $\Pi_X$ and $r_X$ maps $\id\in \{\id\}$ to $1 \in G$.
Thus, $\mathrm{Im}$ contains only $\id$. The preimage about $r_X$ is trivial, whereas it for $r_Y$ is the set of element that the square is identity.
Using the regular rep trick, we find that the size of the preimage about $r_Y$ is $|r_Y^{-1}(\id)|=\frac{1}{|G|}\sum_{a} \Tr D^\mathrm{reg}(a^2)=\frac{1}{|G|}\sum_{a,\alpha,i,j} d_\alpha \, D_{i,j}^\alpha(a)D_{j,i}^\alpha(a)=\sum_{\alpha} d_\alpha \iota^\alpha$.
The matrix $W$ takes the form
\begin{equation}
  \bigoplus_{j=1}^{|G|^{|V_\partial|-1}}
  \mathbb{C}^{
    |G|^{|V_X|} \times |r_Y^{-1}(\id)||G|^{|V_Y|}
  }.
\end{equation}

\subsection{The Klein Bottle}

The next simplest non-orientable example is the Klein bottle, which exhibits three distinct bipartitions.
First, we focus on the bipartition into the disk $N_{0,1}=D^2$ and the other $N_{2,1}$ as the trivial case.
The diagram of manifold partition is as follows:
\begin{equation}
\begin{tikzcd}
M=N_{2,0} & X=D^2 \arrow[l, hookrightarrow]  \\
Y=N_{2,1}  \arrow[u, hookrightarrow] & \partial = S^1 \arrow[u, hookrightarrow] \arrow[l, hookrightarrow] 
\end{tikzcd}
\end{equation}
The fundamental groupoids for each manifolds are given as follows:
\begin{equation}
\begin{split}
\pi_1(N_{2,0},A)&=\frac{F_2}{a b a b^{-1}}\\
\pi_1(N_{0,1},A)&=\{\id\}\\
\pi_1(N_{2,1},A)&=F_2\\
\pi_1(S^1,A)&=\mathbb{Z}.
\end{split}
\end{equation}
Note that the constraint for the torus is $aba^{-1}b=1$, whereas for the torus, it is $aba^{-1}b^{-1}=1$.
From the same discussion above, we find that $\mathrm{Im}=\{\id\}$.
The diagram for the $\Hom$-applied fundamental groupoids is as follows:
\begin{equation}
\begin{tikzcd}
\frac{G^2}{a b a^{-1} b} \arrow[d, "\Pi_Y"'] \arrow[r, "\Pi_X"] & \{\id\}  \arrow[d, "r_X"] \\
G^2  \arrow[r, "r_Y"'] & G \\
\end{tikzcd}
\end{equation}
The size of the preimage is given by the regular rep trick, as $|r_Y^{-1}(\id)|=|G|\sum_{\alpha}(\iota^\alpha)^2.$
The matrix $W$ takes the form
\begin{equation}
  \bigoplus_{j=1}^{|G|^{|V_\partial|-1}}
  \mathbb{C}^{
    |G|^{|V_X|} \times |r^{-1}_Y (\id)||G|^{|V_Y|}
  }.
\end{equation}

Let us compare this result with topologically different bipartition:
\begin{equation}
\begin{tikzcd}
M=N_2 & X=N_{0,2} \arrow[l, hookrightarrow]  \\
Y=N_{0,2}  \arrow[u, hookrightarrow] & \partial = \sqcup^2 {S^1} \arrow[u, hookrightarrow] \arrow[l, hookrightarrow] 
\end{tikzcd}
\end{equation}
We need two base points.
This bipartition appeared when we calculated the nontrivial bipartition of torus. However, one boundary circle is glued with antipodal flip in this case, while the other boundary is not flipped. This antipodal flip leads to the non-orientable nature of $M$.
For the fundamental groupoids, refer to the above discussion or that of the torus.
The diagram for the $\Hom$-applied fundamental groupoids is as follows:
\begin{equation}
\begin{tikzcd}
\frac{G^2}{a b a^{-1} b} \arrow[d, "\Pi_Y"'] \arrow[r, "\Pi_X"] & \frac{G^3}{f_2^X c_1 (f_2^X)^{-1}c_2}  \cong G^2 \arrow[d, "r_X"] \\
\frac{G^3}{f_2^Y (c_1)^{-1}\, (f_2^Y)^{-1}c_2} \cong G^2 \arrow[r, "r_Y"'] & \{c_1,c_2\in G\}\cong G^2 \\
\end{tikzcd}
\end{equation}
where $c_i$ is the holonomy of $i$th connected component of the boundary (shared between $X$ and $Y$) and $f_2$ is holonomy along the path from the base point $2$ to the base point $1$.
The sizes of the preimages are given by summing up $f^{X/Y}_2$ as follows:
\begin{equation}
\begin{split}
  |r_X^{-1}(c)|&=\sum_{f_2^X} \delta(f_2^X c_1 (f_2^X)^{-1}c_2)=\sum_\alpha \chi^\alpha(c_1)\chi^\alpha(c_2),\\
  |r_Y^{-1}(c)|&=\sum_\alpha \overline{\chi^\alpha(c_1)}\chi^\alpha(c_2).
\end{split}
\end{equation}
Then, the matrix $W$ takes the form
\begin{equation}
  \bigoplus_{\{c\}}\bigoplus_{j=1}^{|G|^{|V_\partial|-1}}
  \mathbb{C}^{
    |r^{-1}_X (c)||G|^{|V_X|} \times |r^{-1}_Y (c)||G|^{|V_Y|}
  }
\end{equation}
From the Schur orthogonality, which is famous in the context of group representations, it is clear that 
$|r_X^{-1}(c)|=0$ if $[c_1]\neq [c_2^{-1}]$, whereas $|r_Y^{-1}(c)|=0$ if $[c_1]\neq [c_2]$.
Note that this is consistent with the data regarding the orientation of the boundary holonomy during the bonding process.

In the case of the torus, the two axes were treated equally, resulting in two topologically distinct divisions.
However, the other topologically distinct partition of the Klein bottle exists; a third alternative to divide it into two Möbius bands $N_{1,1}$. 
Since $\partial N_{1,1}=S^1$, only one base point $|A|=1$ is sufficient.
The diagram of manifold partition is as follows:
\begin{equation}
\begin{tikzcd}
M=N_2 & X=N_{1,1} \arrow[l, hookrightarrow]  \\
Y=N_{1,1}  \arrow[u, hookrightarrow] & \partial = {S^1} \arrow[u, hookrightarrow] \arrow[l, hookrightarrow] 
\end{tikzcd}
\end{equation}
For the fundamental groupoids, refer to the above discussion.
The diagram for the $\Hom$-applied fundamental groupoids is as follows:
\begin{equation}
\begin{tikzcd}
\frac{G^2}{aba^{-1}b} \arrow[d, "\Pi_Y"'] \arrow[r, "\Pi_X"] & G  \arrow[d, "r_X"] \\
G  \arrow[r, "r_Y"'] & G \\
\end{tikzcd}
\end{equation}
Note that $r_X$ and $r_Y$ are identity maps. Restriction comes from $\Pi_X$ and $\Pi_Y$.
In $X$ and $Y$, any contractible loops can be deformed to loop goes along the crosscap and the boundary; this yields the constraints $c^{-1}=\left(a^X\right)^2=\left(a^Y\right)^2$.
Then, defining $b:=(a^Y)^{-1}a^X$, the condition of the Klein bottle $a^X b (a^X)^{-1} b=a^Y b (a^Y)^{-1} b=1$
automatically follows.
The size of the preimage is given by
\begin{equation}
\begin{split}
  |r_X^{-1}(c)|&=\sum_{a^X}\delta\left(\left(a^X\right)^2 c\right)=\sum_\alpha \iota^\alpha \chi^\alpha(c)=|r_Y^{-1}(c)|.
\end{split}
\end{equation}

From previous examples, we can expect that the Frobenius-Schur indicator appears in the entanglement structure of non-orientable surface subsystems. In the next section, we consider general partitions of general non-orientable surfaces and show that this expectation is indeed correct.

\subsection{General Non-orientable Surface}
From the classification theorem for 2-dimensional topological surfaces \cite{gallierGuideClassificationTheorem2013,francisConwaysZIPProof1999}, it suffices to consider biparition of surface with $k$ crosscaps.
The possible bipartitions are classified into three cases: both orientable, orientable plus non-orientable, or both non-orientable.

First, we focus on the case of both orientable: we construct non-orientable surface from two orientable surfaces with $n$ punctures.
The data of antipodal flip is specified with the sequence of signs $\{s_i=\pm 1,\,i=1\cdots n\}$.
If all the signs indicate the same value, the resulting manifold $M$ is orientable and otherwise $M$ is non-orientable.
\begin{equation}
  \Sigma_{\gamma_X,n} \sqcup_{\sqcup^n S^1} \Sigma_{\gamma_Y,n} \mapsto N_{2\gamma_X+2\gamma_Y+2n-2}
\end{equation}

In this case the crosscap number of the entire manifold is defined as $k:=2\gamma_X+2\gamma_Y+2n-2$.
The diagram of manifold partition is as follows:
\begin{equation}
\begin{tikzcd}
M=N_{2\gamma_X+2\gamma_Y+2n-2} & X=\Sigma_{\gamma_X,n} \arrow[l, hookrightarrow]  \\
Y=\Sigma_{\gamma_Y,n}  \arrow[u, hookrightarrow] & \partial = \sqcup^n S^1 \arrow[u, hookrightarrow] \arrow[l, hookrightarrow] 
\end{tikzcd}
\end{equation}
The fundamental groupoids for each manifolds are given as follows:
\begin{equation}
\begin{split}
\pi_1(M,A)&=\frac{F_{2k}}{\prod_{i=1}^k \textrm{holonomy}_i^2} \textrm{ with paths connecting base points, }\\
\pi_1(\Sigma_{\gamma_X,n},A)&=\frac{F_{2\gamma_X+2n-1}}{\prod_{i=1}^{\gamma_X} [a^X_i, b^X_i] \prod_{i=1}^n f^X_i c_i (f^X_i)^{-1}}\\
\pi_1(\Sigma_{\gamma_Y,n},A)&=\frac{F_{2\gamma_Y+2n-1}}{\prod_{i=1}^{\gamma_Y} [a^Y_i, b^Y_i] \prod_{i=1}^n f^Y_i c_i (f^Y_i)^{-1}}\\
\pi_1(\partial,A)&=\mathbb{Z}^{\oplus n}.
\end{split}
\end{equation}
As done for orientable surfaces, we set $f_1 = 1$. 
Due to the constraint, one of the boundary holonomies can be expressed in terms of the other holonomies.
The diagram for the $\Hom$-applied fundamental groupoids is as follows:
\begin{equation}
\begin{tikzcd}
\Hom (\pi_1(M,A),G) \arrow[d, "\Pi_Y"'] \arrow[r, "\Pi_X"] & G^{2\gamma_X+2n-2}  \arrow[d, "r_X"] \\
G^{2\gamma_Y+2n-2}  \arrow[r, "r_Y"'] & G^n \\
\end{tikzcd}
\end{equation}
Henceforth, we will write as $c_i^Y=\left(c_i^X\right)^{s_i}=\left(c_i\right)^{s_i}$.
The sizes of preimages are given by recursively using the regular rep trick, as follows:
\begin{equation}
\begin{split}
  |r_X^{-1}(c)|
  &=\sum_{\{a,b,d\}}\delta([a_1,b_1]\cdots [a_{\gamma_X},b_{\gamma_X}]\times c_1\cdots c_n)\\
  &=\sum_{\{d\},\alpha}\left(\frac{|G|}{d_\alpha}\right)^{2\gamma_X-1}\chi^\alpha(c_1\cdots c_n)\\
  &=\sum_{\alpha}\left(\frac{|G|}{d_\alpha}\right)^{2\gamma_X+n-2}\chi^\alpha(c_1)\cdots \chi^\alpha(c^X_n)\\
  |r_Y^{-1}(c)|&=\sum_{\alpha}\left(\frac{|G|}{d_\alpha}\right)^{2\gamma_Y+n-2}\chi^\alpha(\left(c_1\right)^{s_1})\cdots \chi^\alpha(\left(c_n\right)^{s_n}).
\end{split}
\end{equation}
The matrix $W$ takes the form
\begin{equation}
\begin{split}
  \bigoplus_{\{c\}}\bigoplus_{j=1}^{|G|^{|V_\partial|-1}}
  \mathbb{C}^{
    |r_X^{-1} (c)||G|^{|V_X|} \times |r_Y^{-1} (c)||G|^{|V_Y|}
  }.
\end{split}
\end{equation}

Next, we focus the case of orientable plus non-orientable.
In this case the crosscap number of the entire manifold is defined as $k:=2\gamma_X+k_Y+2n-2$.
The diagram of manifold partition is as follows:
\begin{equation}
\begin{tikzcd}
M=N_{2\gamma_X+k_Y+2n-2} & X=\Sigma_{\gamma_X,n} \arrow[l, hookrightarrow]  \\
Y=N_{k_Y,n}  \arrow[u, hookrightarrow] & \partial = \sqcup^n S^1 \arrow[u, hookrightarrow] \arrow[l, hookrightarrow] 
\end{tikzcd}
\end{equation}
The manifold $M$ obtained in this way is always non-orientable, thus we do not need to consider the boundary flip.
The difference from the former case is the fundamental groupoid and its Hom of $Y$;
\begin{equation}
\begin{split}
\pi_1(Y,A)&=\frac{F_{k_Y+2n-1}}{\prod_{i=1}^{k_Y} (a^Y_i)^2 \prod_{i=1}^n f^Y_i c_i (f^Y_i)^{-1}}\\
\Hom(\pi_1(Y,A),G)&=G^{k_Y+2n-2}.
\end{split}
\end{equation}
The diagram for the $\Hom$-applied fundamental groupoids is as follows:
\begin{equation}
\begin{tikzcd}
\Hom (\pi_1(M,A),G) \arrow[d, "\Pi_Y"'] \arrow[r, "\Pi_X"] & G^{2\gamma_X+2n-2}  \arrow[d, "r_X"] \\
G^{k_Y+2n-2}  \arrow[r, "r_Y"'] & G^n \\
\end{tikzcd}
\end{equation}
Again, it can be checked that the holonomies satisfying the conditions in $X$ and $Y$ also satisfy the condition of $M$: $\Im$ is the set of $(c_1,\dots,c_n)$ that satisfies the conditions in $X$ and $Y$.
We find that the sizes of blocks are given by
\begin{equation}
\begin{split}
  |r_X^{-1}(c)|&=\sum_{\alpha}\left(\frac{|G|}{d_\alpha}\right)^{2\gamma_X+n-2}\chi^\alpha(c^X_1)\cdots \chi^\alpha(c^X_n)\\
  |r_Y^{-1}(c)|
  &=\sum_{\alpha} (\iota^\alpha)^k \left(\frac{|G|}{d_\alpha}\right)^{k_Y-n-2} \chi^\alpha(c_1) \cdots \chi^\alpha(c_n).
\end{split}
\end{equation}
The $W$ matrix takes the same form as the case of both orientable. Note that the Frobenius-Schur indicator appeared.

The remaining case is when both $X$ and $Y$ are originally non-orientable. In this case, it follows that $M$ is also non-orientable when $k_X,k_Y \geq 1$.
In this case the crosscap number of the entire manifold is defined as $k:=k_X+k_Y+2n-2$. The diagram of manifold partition is as follows:
\begin{equation}
\begin{tikzcd}
M=N_{k_X+k_Y+2n-2} & X=N_{k_X,n} \arrow[l, hookrightarrow]  \\
Y=N_{k_Y,n}  \arrow[u, hookrightarrow] & \partial = \sqcup^n S^1 \arrow[u, hookrightarrow] \arrow[l, hookrightarrow] 
\end{tikzcd}
\end{equation}
The diagram for the $\Hom$-applied fundamental groupoids is as follows:
\begin{equation}
\begin{tikzcd}
\Hom (\pi_1(M,A),G) \arrow[d, "\Pi_Y"'] \arrow[r, "\Pi_X"] & G^{k_X+2n-2}  \arrow[d, "r_X"] \\
G^{k_Y+2n-2}  \arrow[r, "r_Y"'] & G^n \\
\end{tikzcd}
\end{equation}
The matrix $W$ has the same form as two cases above, differing only in block size. These are given below:
\begin{equation}
\begin{split}
  |r_X^{-1}(c)|&=\sum_{\alpha} (\iota^\alpha)^k \left(\frac{|G|}{d_\alpha}\right)^{k_X-n-2} \chi^\alpha(c_1) \cdots \chi^\alpha(c_n)\\
  |r_Y^{-1}(c)|&=\sum_{\alpha} (\iota^\alpha)^k \left(\frac{|G|}{d_\alpha}\right)^{k_Y-n-2} \chi^\alpha(c_1) \cdots \chi^\alpha(c_n)
\end{split}
\end{equation}
Comparing the three results, we see that the Frobenius-Schur indicator appears when the submanifold $X/Y$ is non-orientable.

\subsection{Heegaard Splitting of General 3D Manifolds}

We provided results for general divisions of two-dimensional manifolds in the previous section.
Nevertheless, the general division of three-dimensional manifolds presents a substantial computational obstacle.
Therefore, we present only the well-known case;
Every 3D closed orientable manifold admits the splitting into two genus $\gamma$ handlebodies, referred to as the \emph{Heegaard splitting}.

The Heegaard splitting of genus 0 yields only $S^3$.
We begin with the genus 1 case, which represents the first nontrivial example.

\subsubsection{Heegaard Splitting of Lens Spaces into $H_1$}

The 3-manifold that can be split into two genus 1 handlebodies is \emph{the lens space}.
The gluing map $\varphi$ is classified by the mapping class group of the torus $\mathbb{T}^2$ isomorphic to $\mathrm{SL}(2;\mathbb{Z})$. The diagram of manifold partition is as follows:
\begin{equation}
\begin{tikzcd}
M & X=H_1 \arrow[l, hookrightarrow]  \\
Y=H_1  \arrow[u, hookrightarrow] & \partial = \mathbb{T}^2 \arrow[u, hookrightarrow] \arrow[l, hookrightarrow] 
\end{tikzcd}
\end{equation}
The boundary is always connected, thus we need only one base point.
The fundamental group of the solid torus is $\mathbb{Z}$.
The diagram of the fundamental group is as follows:
\begin{equation}
\begin{tikzcd}
\pi_1(M) & \mathbb{Z} \arrow[l,"p_X"]  \\
\mathbb{Z}  \arrow[u,"p_Y"] & \mathbb{Z}^2 \arrow[u, "i_X"] \arrow[l, "i_Y"] 
\end{tikzcd}
\end{equation}
To understand this diagram precisely, we provide detailed explanation below.
Let $\sigma$ denote the trivialization of holonomies along meridians. The mapping $\phi$ is determined by the $2\times 2$ matrix
\begin{equation}
\begin{pmatrix}
  p&r\\q&s
\end{pmatrix},\quad ps-qr=1,\quad p,q,r,s \in \mathbb{Z}.
\end{equation}
The gluing map is defined as $i_X:= \sigma \circ \varphi,\quad i_Y:= \sigma$. Through injection $i_Y$, the meridian of $Y$ is transformed into a contractible loop in $H_1$, while the longitude remains unchanged.
In contrast, injection $i_X$ twists both the meridian and the longitude. 
Specifically, since the action of $\sigma$ is distributive, we obtain
\begin{equation}
  \begin{split}
    i_X&(\lambda)=\sigma(\lambda^p \mu^r)=\sigma(\lambda^p)\sigma(\mu^r)=\lambda^p\\
    i_X&(\mu)=\sigma(\lambda^q \mu^s)=\sigma(\lambda^q) \sigma(\mu^s)=\lambda^q\\
    i_Y&(\lambda)=\lambda\\
    i_Y&(\mu)=1
  \end{split}
\end{equation}
for longitude $\lambda$ and meridian $\mu$. Applying $\Hom (*,G)$ functor, we obtain the commutative diagram:
\begin{equation}
\begin{split}
\Hom (\pi_1(X,A),G)&=G\\
\Hom (\pi_1(Y,A),G)&=G\\
\Hom (\pi_1(\partial,A),G)&=\mathrm{Comm}_2(G)
\end{split}
\end{equation}

\begin{equation}
\begin{tikzcd}
\Hom(\pi_1(M,A),G) \arrow[d, "\Pi_Y"'] \arrow[r, "\Pi_X"] & G  \arrow[d, "r_X"] \\
G  \arrow[r, "r_Y"'] & \mathrm{Comm}_2(G) \\
\end{tikzcd}
\end{equation}
We exploit the well-known theorem of category theory; since $i_{X/Y}$ is surjective and $\Hom(*,G)$ induces its pre-composition, $r_{X/Y}$ is injective. Consequently, the block size turns out to be $|r_X^{-1}(\phi)|=|r_Y^{-1}(\phi)|=1$.

It remains to determine the size $|\mathrm{Im}|$ of the direct sum generated by $\phi \in \mathrm{Im}$.
Let $g_{X/Y}\in G$ is the coloring by the group $G$, then they must satisfy
\begin{equation}
\begin{split}
  (\sigma\circ \varphi)^* (g_X)&=(g_X^p,g_X^q)\in \mathrm{Comm}_2(G)\\
  \sigma^* (g_Y)&=(g_Y,1)\in \mathrm{Comm}_2(G).
\end{split}
\end{equation}
Thus, the fundamental group of $M_3$ is identified as follows:
\begin{equation}
\begin{split}
\pi_1(M_3)&=\langle g_X,g_Y | g_X^q=1 ,g_Y=g_X^p \rangle=\langle g_X | g_X^q=1 \rangle=\begin{cases}
  \mathbb{Z}_q & (|q| \geq 1)\\
  \mathbb{Z} & (q = 0 ).
\end{cases}
\end{split}
\end{equation}
The image can be derived from this, as follows:
\begin{equation}
  \begin{split}
    \mathrm{Im}&=\{\phi \in \Hom(\pi_1(\partial,A),G)|\phi(\ker \sigma)=\phi(\ker (\sigma\circ T))=1\}=\{g_X \in G|g_X^q=1\}.
  \end{split}
\end{equation}

We define the higher Frobenius–Schur indicator $\nu_n^\alpha$ as $\nu_n^\alpha=\frac{1}{|G|}\sum_{g\in G} \chi^\alpha (g^n)$.
The size $|\mathrm{Im}|$ can be determined as follows:
\begin{equation}
  \begin{split}
    |\mathrm{Im}|&=|\{g_X \in G|g_X^q=1\}|=\frac{1}{|G|}\sum_{\alpha,g \in G} d_\alpha \, \chi^\alpha(g^q)=\sum_{\alpha}d_\alpha\, \nu^\alpha_q.
  \end{split}
\end{equation}

It appears strange that the outcome solely hinges upon the value of $q$, yet the Reidemeister-Singer theorem \cite{reidemeisterZurDreidimensionalenTopologie1933,Singer1933ThreedimensionalMA,fomenko2013algorithmic,saveliev2011lectures} provides an interpretation of this.
For the Heegaard splitting of a fixed genus, that theorem claims the double coset $\mathfrak{H}_\gamma \backslash \mathrm{MCG^+} /\mathfrak{H}_\gamma$ is the 1-to-1 mapping with the equivalence classes of resulting manifold by the Heegaard splitting.
Here, $\mathfrak{H}_\gamma$ is the subgroup of the mapping class group $\mathrm{MCG}^+$ that can be extended to the interior of genus $\gamma$ handlebody.
In the case of genus 1, we can take the generator of $\mathfrak{H}_1$ as $T$ transformation of the full modular group $\mathrm{SL}(2;\mathbb{Z})$:
\begin{equation}
    T=
    \begin{pmatrix}
        1&1\\
        0&1
    \end{pmatrix}.
\end{equation}
Let us compute the double coset:
\begin{equation}
    T^k \begin{pmatrix}
        p&r\\
        q&s
    \end{pmatrix} T^l=
    \begin{pmatrix}
        1&k\\
        0&1
    \end{pmatrix}
    \begin{pmatrix}
        p&r\\
        q&s
    \end{pmatrix}
    \begin{pmatrix}
        1&l\\
        0&1
    \end{pmatrix}=
    \begin{pmatrix}
        p+kq&pl+klq+r+ks\\
        q&lq+s
    \end{pmatrix}
\end{equation}
One finds that $q$ is invariant under $T$ transformation, implying that $q$ is the label of the Reidemeister-Singer equivalence class $\mathfrak{H}_\gamma \backslash \mathrm{MCG^+} /\mathfrak{H}_\gamma$.

To be more precise, the invariant of $\mathfrak{H}_\gamma \backslash \mathrm{MCG^+} /\mathfrak{H}_\gamma$ is not only $q$, but also $p \,\mathrm{mod}\, q$. Indeed, one can check $p'=p+kq \equiv p \,\mathrm{mod}\, q$. Apperantly we see that $s'=s+lq \,\mathrm{mod}\, q$ is also an invariant of $\mathfrak{H}_\gamma \backslash \mathrm{MCG^+} /\mathfrak{H}_\gamma$, but this is not an independent degree of freedom; ${}^\forall \varphi \in \mathrm{SL}(2;\mathbb{Z})$ satisfies $\mathrm{det}\, \varphi=1$, we have $ps-qr=1\, \Rightarrow\,  ps \equiv 1 \,\mathrm{mod}\, q.$
This implies that $s$ is the inverse of $p$ when we interpret $\mathbb{Z}\, \mathrm{mod}\,q$ forms multiplicative group $\mathbb{Z}_q$. Thus, $s$ can be uniquely determined by $(q, p \, \mathrm{mod}\,q)$.
Moreover, $r$ is determined by the conditions $p s-qr=1$. Therefore, the independent label turns out to be $(q, p \, \mathrm{mod}\,q)$.

Conversely, we show that any two elements of $\mathrm{SL}(2;\mathbb{Z})$ characterized by the same $(q, p \bmod q)$ can be transformed into each other.
As proved above, $p\equiv q' \,\mathrm{mod}\, q$ induces $s'\equiv s \,\mathrm{mod}\, q$.
We can rewrite them as $p'-p= mq$ and $s'-s=nq$. 
$T$ transformations acting on both left and right can transform $(p,q,r,s)$ to $(p',q,r',s')$:
\begin{equation}
\begin{split}
    T^m 
    \begin{pmatrix}
        p&r\\
        q&s
    \end{pmatrix}
    T^n  =
    \begin{pmatrix}
        1&m\\
        0&1
    \end{pmatrix} 
    \begin{pmatrix}
        p&r\\
        q&s
    \end{pmatrix}
    \begin{pmatrix}
        1&n\\
        0&1
    \end{pmatrix}  =
    \begin{pmatrix}
        p+mq&pn+mnq+r+ms\\
        q&nq+s
    \end{pmatrix} =
    \begin{pmatrix}
        p'&pn+mnq+r+ms\\
        q&s'
    \end{pmatrix}.
\end{split}
\end{equation}
Finally, $r'=pn+mnq+r+ms$ can be uniquely determined by the condition $p's'-q r'=1$. This implies that $(q, p\, \mathrm{mod}\, q)$ is enough to characterize the class in $\mathfrak{H}_\gamma \backslash \mathrm{MCG^+} /\mathfrak{H}_\gamma$.

We can apply the discussion above for the Heegaard splitting into $\gamma\geq 2$: The block size will once again be $|r_{X/Y}^{-1}(\phi)|=1$, consistent with our previous discussion. However, it is hard to identify the mapping class for a given 3D manifold.

\subsection{$n$-dimensional Torus $\mathbb{T}^n$}

In this section, we consider $n$-dimensional torus $M = \mathbb{T}^n$.
In higher dimensions, it is extremely difficult to exhaustively examine all possible ways of partitioning a manifold. Therefore, we focus on the case where the manifold $X$ is chosen as
\begin{equation}
X=\left[0,\pi\right]^{\times k} \times (S^1)^{\times (n-k)}
\end{equation}
(for $1\le k\le n$; here $S^1=[0,2\pi]/\{0\sim 2\pi\}=2\pi\mathbb{R}/\mathbb{Z}$).
Then, $Y$ is chosen as $\mathbb{T}^n \backslash X$ with boundary compensated.

First, we compute the boundary of these manifolds. The boundary operator satisfies the Leibniz rule $\partial(M_1 \times M_2)=\partial M_1 \times M_2 \cup M_1\times \partial M_2$. Using this property, we find
\begin{equation}
\begin{split}
\partial &= \partial \left(\left[0,\pi\right]^{\times k} \times (S^1)^{\times (n-k)}\right)\\
&= \left(\bigcup_{j=0}^{k-1} \left[0,\pi\right]^{\times j} \times \{0,\pi\}\times  \left[0,\pi\right]^{\times k-j-1}\right) \times (S^1)^{\times (n-k)}.
\end{split}
\end{equation}
We find that the boundary $\partial$ is nothing but the product of $(k-1)$-dimensional sphere and $(n-k)$-dimensional torus:
\begin{equation}
\begin{split}
    k=1:\quad \partial&=\{0,\pi\} \times (S^1)^{\times (n-1)}=S^0\times(S^1)^{\times (n-1)},\\
    k=2:\quad \partial&=\left(\{0,\pi\}\times [0,\pi]\cup [0,\pi]\times \{0,\pi\} \right) \times (S^1)^{\times (n-2)}\cong(S^1)^{\times (n-1)},\\
    k=3:\quad \partial&=\left(\{0,\pi\}\times [0,\pi]^2\cup [0,\pi]\times \{0,\pi\}\times [0,\pi]\cup  \{0,\pi\}\times [0,\pi]^2\right) \times (S^1)^{\times (n-3)}\cong S^2\times (S^1)^{\times(n-3)},\\
    &\,\,\,\vdots\\
    k:\quad \partial&\cong S^{k-1}\times (S^1)^{\times(n-k)}.
\end{split}
\end{equation}
If $k= 1$, $\partial$ has two connected components; $|A|=2$.
Otherwise, the set of base points is order 1 as the boundary is connected.
Exceptional case is $k=2$, since the boundary is just $(n-1)$-dimensional torus, differing from other cases. Divide into three cases: $k=1, k=2,$ and $k\geq 3$, and compute each separately.

First we consider the case $k\geq 3$. The fundamental groupoids and their Hom are given as follows:
\begin{equation}
  \begin{split}
    \pi_1(M,A)&=\mathbb{Z}^n, \quad \Hom (\pi_1(M,A),G)=\mathrm{Comm}_n(G)\\
    \pi_1(X,A)&=\mathbb{Z}^{n-k}, \quad \Hom (\pi_1(X,A),G)=\mathrm{Comm}_{n-k}(G)\\
    \pi_1(Y,A)&=\mathbb{Z}^n, \quad \Hom (\pi_1(Y,A),G)=\mathrm{Comm}_n(G)\\
    \pi_1(\partial,A)&=\mathbb{Z}^{n-k} , \quad \Hom (\pi_1(\partial,A),G)=\mathrm{Comm}_{n-k}(G).
  \end{split}
\end{equation}
Recall the commutative diagram of the $\Hom$-applied fundamental groupoids:
\begin{equation}
\begin{tikzcd}
\Hom (\pi_1(M,A),G) \arrow[r, "\Pi_X"]\arrow[d, "\Pi_Y"']& \Hom(\pi_1(X,A),G) \arrow[d, "r_X"] \\
\Hom (\pi_1(Y,A),G) \arrow[r, "r_Y"'] & \Hom(\pi_1(\partial,A),G)   \\
\end{tikzcd}
\end{equation}
Since $r_X$ is trivial map, $|r_X^{-1}(\phi)|=1$. In contrast, $r_Y^{-1}(\phi)$ is a composite of $\phi$ and the elements $(g_1,\cdots,g_k)$ that ${}^\forall h \in \phi$ commutes with, implying that $W$ matrix takes the form
\begin{equation}
  \bigoplus_{
    \phi \in \mathrm{Comm}_{n-k}(G)
    }
  \bigoplus_{j=1}^{|G|^{|V_\partial|-1}}
  \mathbb{C}^{
    |G|^{|V_X|} \times |\mathrm{Comm}_{k}(\mathrm{C}_\phi)||G|^{|V_Y|}
  }.
\end{equation}

Next we explore the case of $k=2$.
The fundamental groupoids $\pi_1(M/X,A)$ and $\Hom(\pi_1(M/X,A),G)$ remain unchanged, while
\begin{equation}
  \begin{split}
    \pi_1(Y,A)&=
      F_2\times \mathrm{Z}^{n-2}, \quad \Hom (\pi_1(Y,A),G)=\{x,y,z_i\in G \,\, (i=3...n)|[x,z_i]=[y,z_i]=[z_i,z_j]=1\} \\
    \pi_1(\partial,A)&=\mathbb{Z}^{n-1}, \quad \Hom (\pi_1(\partial,A),G)=\mathrm{Comm}_{n-1}(G).
  \end{split}
\end{equation}
Note that $r_X$ is no longer a trivial map. However, holonomy along $S^1\simeq \partial ([0,\pi]^2)$ must be trivial to be in $\phi \in \mathrm{Im}$.
This constraints $\phi \in \mathrm{Comm}_{n-2}(G)$ and $|r^{-1}_X(\phi)|=1$.
Based on the same argument as in the previous example, $W$ matrix takes the form
\begin{equation}
  \bigoplus_{
    \phi \in \mathrm{Comm}_{n-2}(G)
    }
  \bigoplus_{j=1}^{|G|^{|V_\partial|-1}}
  \mathbb{C}^{
    |G|^{|V_X|} \times |\mathrm{Comm}_{2}(\mathrm{C}_\phi)||G|^{|V_Y|}
  }.
\end{equation}
In the end, this turns out to be the same as the previous case.

Finally we dive into the case $k=1$.
In this case the boundary is not connected we need two base points $|A|=2$.
Let $x,y$ be global holonomies between two base points in $X/Y$.
The $\Hom$-applied fundamental groupoid for $X,Y$ with two base points can be obtained as a product of the $\Hom$-applied fundamental group and $G$. We enumerate the fundamental groups and the $\Hom$-applied fundamental groupoids as follows:
\begin{equation}
  \begin{split}
    \pi_1(M,A)&=\frac{\mathbb{Z}^n\& \textrm{ path}^X_{1\to 2} \& \textrm{path}^Y_{2\to 1} }{\textrm{path}^Y_{2\to 1}\textrm{path}^X_{1\to 2}=\mathrm{hol}_n}, \quad \Hom (\pi_1(M,A),G)=\mathrm{Comm}_n(G)\times G\\
    \pi_1(X,A)&=\mathbb{Z}^{n-1} \& \textrm{ path}^X_{1\to 2}, \quad \Hom (\pi_1(X,A),G)=\mathrm{Comm}_{n-1}(G)\times G\\
    \pi_1(Y,A)&=\mathbb{Z}^{n-1} \& \textrm{ path}^Y_{2\to 1}, \quad \Hom (\pi_1(Y,A),G)=\mathrm{Comm}_{n-1}(G)\times G\\
    \pi_1(\partial,A)&=\mathbb{Z}^{n-1} \otimes \mathbb{Z}^{n-1}, \quad \Hom (\pi_1(\partial,A),G)=\mathrm{Comm}_{n-1}(G)\times \mathrm{Comm}_{n-1}(G)
  \end{split}
\end{equation}
where $\mathrm{hol}_n$ denotes the holonomy along the $n$th dimensional axis and the tensor product $\otimes$ means that there are two independent fundamental groups.

Here, two separated holonomies $\phi_1,\phi_2$ in $(\phi_1,\phi_2)\in \Hom (\pi_1(M,A),G)$ must satisfy that $x \phi_1 x^{-1}=y \phi_1 y^{-1}=\phi_2$ to fullfill $\phi \in \mathrm{Im}$, where $\phi_1,\phi_2$ are the $\mathrm{Comm}_{n-1}(G)$ part in $\Hom (\pi_1(X/Y,A),G)$. 
These observation concludes that $W$ matrix takes the form
\begin{equation}
  \bigoplus_{
    \substack{\phi_1,\phi_2 \in \mathrm{Comm}_{n-1}(G)\\
    \phi_1 \sim \phi_2
    }
  }
  \bigoplus_{j=1}^{|G|^{|V_\partial|-2}}
  \mathbb{C}^{
    |\mathrm{C}_{\phi_1}||G|^{|V_X|} \times |\mathrm{C}_{\phi_1}||G|^{|V_Y|}
  }.
\end{equation}

\section{Homotopy and non-Abelian cohomology}

\subsection{Non-Abelian $H^1(M,G)$ and the equivalence with $\Hom(\pi_1(M),G)/G$}

This section states that the first cohomology of a manifold with non-Abelian coefficients coincides with $\Hom(\pi_1(M,A),G)/G$. 
This has been known in pure mathematics for a long time, and the paper \cite{olumNonAbelianCohomologyVan1958}, which also mentions an elegant proof of the Seifert-van Kampen theorem, is informative. Specifically, \cite[Eq.~(5.3)]{olumNonAbelianCohomologyVan1958} is nothing but the $\Hom$-applied version of the Seifert-van Kampen theorem.

Fix throughout a finite group $G$ (hence discrete) and a smooth manifold $M$ (path connected unless explicitly stated otherwise). The goal is to explain, entirely in the finite-group case, why
\begin{equation}
H^1(M;G)\ \cong\ \Hom \bigl(\pi_1(M),G\bigr)\big/G,
\end{equation}
where the right-hand side is the set of group homomorphisms modulo conjugation in $G$. This identification is classical; see, for example, \cite{olumNonAbelianCohomologyVan1958}.

Pick a good cover $\{U_i\}$ of $M$ (all finite intersections are contractible; such covers exist for manifolds, e.g.\ by geodesically convex balls \cite[Thm.~5.1]{bott2013differential}). Because $G$ is discrete and each $U_{ij}$ is connected, any continuous $g_{ij}:U_{ij}\to G$ is constant; we therefore regard $g_{ij}\in G$. The non-Abelian Čech $1$-cocycles are families $(g_{ij})$ with
\begin{equation}
g_{ij}\,g_{jk}\,g_{ki}=1\qquad\text{on }U_{ijk},
\end{equation}
and $0$-cochains are families $(a_i)$ with $a_i\in G$, acting by gauge transformations
\begin{equation}
g'_{ij}=a_i\,g_{ij}\,a_j^{-1}.
\end{equation}
The (non-Abelian) first cohomology set is
\begin{equation}
H^1(M;G)\ :=\ Z^1(\{U_i\};G)\big/\{a_i\},
\end{equation}
and is independent of the chosen good cover. In the finite-group case this classifies principal $G$-bundles (equivalently, $G$-covering spaces) over $M$ \cite{olumNonAbelianCohomologyVan1958}.

Let $\mathbf{B}G$ denote the one-object category with $\Hom(*,*)\cong G$, and define the classifying space by
\begin{equation}
BG\ :=\ \bigl|N(\mathbf{B}G)\bigr|.
\end{equation}
For discrete $G$ one has $BG\simeq K(G,1)$, hence
\begin{equation}
[M,BG]\ \cong\ \Hom \bigl(\pi_1(M),G\bigr)\big/G,
\end{equation}
while the universal bundle $EG\to BG$ yields
\begin{equation}
H^1(M;G)\ \cong\ [M,BG]\qquad\text{(principal $G$-bundles as pullbacks of }EG\to BG\text{)}.
\end{equation}
Combining these equalities gives the desired identification. If $M$ is not path connected, replace $\pi_1(M)$ with the fundamental groupoid $\Pi_1(M)$ and obtain
\begin{equation}
H^1(M;G)\ \cong\ \mathrm{Fun} \bigl(\Pi_1(M),\mathbf{B}G\bigr)\big/\text{natural isomorphism},
\end{equation}
and, after choosing a base point set $A$ meeting each component, equivalently $H^1(M;G)\cong \Hom \bigl(\pi_1(M,A),G\bigr)/G$; see \cite[6.5.10 Corollary 1]{brownTopologyGroupoids2006}.

For later use it is convenient to reformulate everything on a CW $2$-skeleton $K$ of $M$. Let $V$ be the set of vertices and $\vec E$ the oriented edges (together with the opposite orientation $\bar e$ and the convention $h(\bar e)=h(e)^{-1}$). For a vertex field $a:V\to G$, the Kramers–Wannier (KW) map is the degree-$0$ non-Abelian coboundary
\begin{equation}
\delta^0:\ C^0(V,G)\longrightarrow C^1(\vec E,G),\qquad (\delta^0 a)(e:v\to w)=a(w)\,a(v)^{-1}.
\end{equation}
For a face $f$ with oriented boundary $\partial f=e_1\cdots e_k$, define the degree-$1$ coboundary by
\begin{equation}
(\delta^1 h)(f)=\prod_{j=1}^k h(e_j).
\end{equation}
Then $\delta^1(\delta^0 a)\equiv 1$ on every face, so KW-generated edge variables are flat. Vertex gauge acts by
\begin{equation}
h(e:u\to v)\ \longmapsto\ a(v)\,h(e)\,a(u)^{-1}.
\end{equation}
Imposing flatness $(\delta^1 h)\equiv 1$ on all faces and dividing by vertex gauge yields
\begin{equation}
Z^1(K;G)\big/G^{V}\ \cong\ \Hom \bigl(\pi_1(|K|),G\bigr)\big/G,
\end{equation}
and because $|K|\simeq M$ (e.g.\ by the Nerve Theorem for a good cover, \cite[Prop.~4G.2, Prop.~4G.3]{hatcherAlgebraicTopology2002}), this coincides with $H^1(M;G)$. Historical references for the KW transform and its non-Abelian generalizations include \cite{kramersStatisticsTwoDimensionalFerromagnet1941,kramersStatisticsTwoDimensionalFerromagnet1941a,wegnerDualityGeneralizedIsing1971,kogutIntroductionLatticeGauge1979,bellissardRemarkDualityNonAbelian1979,drouffeLatticeModelsSolvable1979,orlandDualityNonabelianLattice1980,tantivasadakarnHierarchyTopologicalOrder2023}.

\subsection{Abelian $H^n(M,A)$ and the equivalence with $\Hom(\pi_n(M),A)$?}

We have constructed a discrete setup for the topological gauge theory of 1-form gauge fields using the discretization of manifolds via nerves.
Nerves more generally contain information about higher-dimensional simplices, which sparks interest in the topological field theory of higher-form gauge fields.
Having witnessed the agreement between two theories, homotopy and cohomology, we naturally expect that higher-form topological gauge theory could be analyzed using higher homotopy or higher cohomology. We offer some comments on this. 

From a homotopy perspective, replacing the fundamental group part of the Seifert-van Kampen theorem with higher homotopy groups should make our algorithm applicable, allowing the entanglement block structure to be revealed mechanically.
Indeed, such a generalization of the Seifert-van Kampen theorem exists.
In the higher homotopy case, information about higher-dimensional gluing is required, and it is formulated using higher categories \cite{brown2004nonabelian,lurieKerodon}.

From a cohomology perspective, a straightforward extension of differentiation reveals that $\delta^{(n\geq 2)}$ is not well-defined for non-Abelian coefficients. This corresponds to the higher homotopy groups becoming Abelian, implying that the gauge group in the topological gauge theory of higher-form gauge fields must be Abelian. The entanglement entropy of such an Abelian higher-form topological gauge theory has been analyzed using the language of Abelian cohomology by \cite{ibieta-jimenezTopologicalEntanglementEntropy2020}, though its block structure remains unclear. Our result is an approach from homotopy restricted to 1-form gauge theory, applicable even in the non-Abelian case. While the connection to \cite{ibieta-jimenezTopologicalEntanglementEntropy2020} may seem unclear at first glance, since the first non-Abelian cohomology is isomorphic to $\Hom(\pi_1(M),G)/G$, our approach should seamlessly connect with that of \cite{ibieta-jimenezTopologicalEntanglementEntropy2020}. 

Incidentally, higher homotopy and cohomology do not necessarily coincide in general \cite{hatcherAlgebraicTopology2002}: for ${}^\exists n \geq 0, A \textrm{ finite Abelian, } \Hom(\pi_n (M),A) \ncong H^n(M,A).$
The entanglement block structure of higher-form topological gauge theories, specifically whether an algorithm based on the Seifert-van Kampen theorem for higher homotopy yields results consistent with \cite{ibieta-jimenezTopologicalEntanglementEntropy2020}, is likely to be one of the most interesting problems, not only physically but also mathematically.

\section{Number of orbits in a fiber product of $G$-sets}
In scenarios further requiring Gauss law constraint over loop symmetry, several holonomies are identified by the gauge transformation, leading commutative diagrams involving $\Hom(\pi_1(M),G)/G$, $\Hom(\pi_1(X),G)/G$, $\Hom(\pi_1(Y),G)/G$, and $\Hom(\pi_1(\partial),G)/G$. Using this, we can estimate the size of $\Hom(\pi_1(M),G)/G$, which can be generalized to the statement about pullbacks in general.
We provide an explicit formula for the number of orbits in the fiber product set, corresponding to the degeneracy of gauge invariant symmetric states. To simplify the notations we consider the following pullback (fiber product) diagram of $G$-sets:
\begin{equation}
\begin{tikzcd}
        X\times_Z Y \arrow[r, "\Pi_X"]  \arrow[d,"\Pi_Y"]       & X  \arrow[d,"r_X"]       \\
        Y \arrow[r, "r_Y"]       & Z  
\end{tikzcd}.
\end{equation}
We would like to know the number of orbits in $X\times_Z Y$. According to Burnside's lemma, we have ($X^g\equiv\{ x\in X| g\cdot x=x\}$)
\begin{equation}
\begin{split}
|X\times_Z Y/ G|&=\frac{1}{|G|}\sum_{g\in G}| X\times_Z Y^g|=\frac{1}{|G|}\sum_{g\in G}\sum_{z\in{\rm Im},g\cdot z=z}|r_X^{-1}(z)\times r_Y^{-1}(z)^g|=\frac{1}{|G|}\sum_{z\in{\rm Im}}\sum_{g\in G_z}|r_X^{-1}(z)\times r_Y^{-1}(z)^g| \\
&=\frac{1}{|G|}\sum_{z\in{\rm Im}}|G_z| |r_X^{-1}(z)\times r_Y^{-1}(z)/G_z|
=\sum_{z\in{\rm Im}}\frac{1}{|[z]|} |r_X^{-1}(z)\times r_Y^{-1}(z)/G_z|
=\sum_{[z]\in{\rm Im}/G}|r_X^{-1}(z)\times r_Y^{-1}(z)/G_z|.
\end{split}
\end{equation}
Comparing to Eq.~(\ref{W}), we point out that $[z]$ corresponds to $[\phi]$, while $|r_X^{-1}(z)\times r_Y^{-1}(z)/G_z|$ is estimated to be $\sum_{\boldsymbol{\alpha}_\phi\in {\rm Rep}(G_\phi)} x_{\boldsymbol{\alpha}_\phi}y_{\boldsymbol{\alpha}_\phi}$ using representation theory.

\section{Entanglement Entropy for General Ground State}

We recall the gauge invariant block structure:
\begin{equation}
  \mathbb{1}_{|G|^{V_|\partial|-|A|}}\otimes\left[\bigoplus_{[\phi]\in {\rm Im}/G^{\times |A|}}\mathbb{1}_{|[\phi]|} \otimes\left(\bigoplus_{\boldsymbol{\alpha}_{[\phi]}} \mathbb{C}^{x_{\boldsymbol{\alpha}_{[\phi]}}\times y_{\boldsymbol{\alpha}_{[\phi]}}}\otimes \mathbb{1}_{d_{\boldsymbol{\alpha}_{[\phi]}}}\right)\right]\otimes 1_{|G|^{|V_X|}\times |G|^{|V_Y|}}.
\end{equation}
The degree of freedom left $\sum_{[\phi]} \sum_{\boldsymbol{\alpha}_\phi\in{\rm Rep}(G_\phi)} x_{\boldsymbol{\alpha}_{[\phi]}}y_{\boldsymbol{\alpha}_{[\phi]}}$ is equivalent to the gauge-invariant ground state $|{\rm Hom}(\pi_1(M),G)/G|$.
If the spontaneous symmetry breaking does not takes place, the ground state can be the superposition of these states:
\begin{equation}
  W\propto \mathbb{1}_{|G|^{|V_\partial|-|A|}}
  \otimes
  \left[
    \bigoplus_{
      [\phi]\in {\rm Im}/G^{\times |A|}
    } \mathbb{1}_{|[\phi]|} \otimes 
    \left(
      \bigoplus_{\boldsymbol{\alpha}_{[\phi]}} 
        \Psi_{
          \boldsymbol{\alpha}_{[\phi]}
        }
      \otimes 
      \mathbb{1}_{
        d_{\boldsymbol{\alpha}_{[\phi]}}
      }
    \right)
  \right]
  \otimes
  1_{|G|^{|V_X|}\times |G|^{|V_Y|}},
\end{equation}
where $\Psi_{\boldsymbol{\alpha}_{[\phi]}}$ is $x_{\boldsymbol{\alpha}_{[\phi]}}\times y_{\boldsymbol{\alpha}_{[\phi]}}$-matrix. 
To normalize $W$, we have to multiply
\begin{equation}
  \left[|G|^{|V_\partial|-|A|}\times
  \sum_{[\phi]\in \mathrm{Im}/G^{\times |A|}} |[\phi]|\left(\sum_{\boldsymbol{\alpha}_{[\phi]}} d_{\boldsymbol{\alpha}_{[\phi]}} \Tr \Psi_{\boldsymbol{\alpha}_{[\phi]}} \Psi_{\boldsymbol{\alpha}_{[\phi]}}^\dagger\right)
  \times|G|^{|V_X|}\times |G|^{|V_Y|}\right]^{-1/2}.
\end{equation}
Let $\rho'$ be as follows:
\begin{equation}
  \rho'=\frac{\bigoplus_{
      [\phi]\in {\rm Im}/G^{\times |A|}
    } \mathbb{1}_{|[\phi]|} \otimes 
    \left(
      \bigoplus_{\boldsymbol{\alpha}_{[\phi]}} 
        \Psi_{
          \boldsymbol{\alpha}_{[\phi]}
        }
        \Psi_{
          \boldsymbol{\alpha}_{[\phi]}
        }^\dagger
      \otimes 
      \mathbb{1}_{
        d_{\boldsymbol{\alpha}_{[\phi]}}
      }
    \right)}{\sum_{[\phi]\in \mathrm{Im}/G^{\times |A|}} |[\phi]|\left(\sum_{\boldsymbol{\alpha}_{[\phi]}} d_{\boldsymbol{\alpha}_{[\phi]}} \Tr \Psi_{\boldsymbol{\alpha}_{[\phi]}} \Psi_{\boldsymbol{\alpha}_{[\phi]}}^\dagger\right)}.
\end{equation}
Then, the von Neumann entropy can be computed as
\begin{equation}
  S=|V_\partial| \ln |G| - |A| \ln |G| - \Tr \rho' \ln \rho'.
\end{equation}
Assuming ${}^\exists U_{\boldsymbol{\alpha}_{[\phi]}},\, U_{\boldsymbol{\alpha}_{[\phi]}} \Psi_{\boldsymbol{\alpha}_{[\phi]}} \Psi_{\boldsymbol{\alpha}_{[\phi]}}^\dagger U_{\boldsymbol{\alpha}_{[\phi]}}^\dagger = \mathrm{diag} (|\psi_{\boldsymbol{\alpha}_{[\phi]},i}|^2)$, the entanglement entropy for general setup turns out to be
\begin{equation}
\begin{split}
S=&|V_\partial| \ln |G| - |A| \ln |G|
-\sum_{[\phi]\in \mathrm{Im}/G^{\times |A|}} |[\phi]|
\sum_{\boldsymbol{\alpha}_{[\phi]}}d_{\boldsymbol{\alpha}_{[\phi]}} 
\sum_{i=1}^{x_{\boldsymbol{\alpha}_{[\phi]}}} \frac{|\psi_{\boldsymbol{\alpha}_{[\phi]},i}|^2}{\mathcal{N}}\ln \frac{|\psi_{\boldsymbol{\alpha}_{[\phi]},i}|^2}{\mathcal{N}}
\end{split}
\label{eq:ee-general}
\end{equation}
where $\mathcal{N}:=\sum_{[\phi]\in \mathrm{Im}/G^{\times |A|}} |[\phi]|\left(\sum_{\boldsymbol{\alpha}_{[\phi]}} d_{\boldsymbol{\alpha}_{[\phi]}} \Tr \Psi_{\boldsymbol{\alpha}_{[\phi]}} \Psi_{\boldsymbol{\alpha}_{[\phi]}}^\dagger\right)$ is normalization constant.
One can check the topological entanglement entropy for the minimally-entangled state is restored by setting $\psi_{\boldsymbol{\alpha}_{[\phi]},i}=\delta_{*,[\phi]}\delta_{*,\boldsymbol{\alpha}_{[\phi]}} \delta_{*,i}$, namely we have $\mathcal{N}=|[\phi]|d_{\boldsymbol{\alpha}_{[\phi]}}$, and the last correction term in Eq.~(\ref{eq:ee-general}) becomes $+\ln |[\phi]| d_{\boldsymbol{\alpha}_{[\phi]}}$.

\section{Properties of modular $S$ matrix}

Modular $S$ matrix appeared in the example of $(2+1)$-dimensional topological gauge theories. We briefly explain its properties used in our work.

Although modular $S$ matrix for finite group is defined in \cite{costeFiniteGroupModular2000}, the unitarity is not explicitly proven. Here, we prove the unitarity of $S$, following the method introduced in \cite{lusztigUNIPOTENTREPRESENTATIONSFINITE1979}.
We recall the definition of modular $S$ matrix.
\begin{equation}
S_{([a],\chi),([b],\chi')}=\frac{1}{|\mathrm{C}_a||\mathrm{C}_b|}\sum_{g\in G(a,b)} \overline{\chi(g b g^{-1})}\overline{\chi'(g^{-1} a g)}
\end{equation}
where $a,b\in G$ and $[g]$ is the conjugacy class of $[g]$ in $G$.
The range of summation is defined as $G(a,b)=\{g \in G: a g b g^{-1} = g b g^{-1} a\}$.
This definition follows the convention of \cite{costeFiniteGroupModular2000,etingofTensorCategories2015,simonTopologicalQuantum2023}.

This $S$ matrix is obviously symmetric and independent of the choice of the representative element $a,b$ of conjugacy classes $[a],[b]$. Hereafter, we often drop the bracket of conjugacy class from the index of the $S$ matrix.

Furthermore, this $S$ matrix is unitary. To check this, we conpute $S S^\dagger$.
\begin{equation}
  \begin{split}
  \left(S S^\dagger\right)_{(a,\chi),(b,\chi')}
  &=\frac{1}{|\mathrm{C}_a||\mathrm{C}_b|}\sum_{(c,\chi'')} \frac{1}{|\mathrm{C}_c|^2}
  \left[\sum_{g\in G(a,c)} \overline{\chi(g c g^{-1})}\overline{\chi''(g^{-1} a g)}\right]
  \left[\sum_{g'\in G(b,c)} \chi'(g' c g'^{-1})\chi''(g'^{-1} b g')\right]\\
  &=\sum_{g\in G(a,c),g'\in G(b,c),c}\frac{1}{|G||\mathrm{C}_a||\mathrm{C}_b||\mathrm{C}_c|}\overline{\chi(g c g^{-1})}
  \chi'(g' c g'^{-1})
  \sum_{\chi''} \overline{\chi''(g^{-1} a g)}\chi''(g'^{-1} b g')\\
  &=\sum_{g\in G(a,c),g'\in G(b,c),c}\frac{1}{|G||\mathrm{C}_a||\mathrm{C}_b||\mathrm{C}_c|}\overline{\chi(g c g^{-1})}
  \chi'(g' c g'^{-1})
  |\mathrm{C}_{c,g^{-1} a g}| \delta^{\mathrm{C}_c}_{[g^{-1} a g],[g'^{-1} b g']}.
  \end{split}
\end{equation}
Note that $\sum_{[c]}f(c)=\sum_c |\mathrm{C}_c|/|G| \times f(c)$ holds for the class function $f$.
We exploit the following identity that one can easily show by using the regular rep trick:
\begin{equation}
  |\mathrm{C}_{c,g^{-1} a g}|\delta^{\mathrm{C}_c}_{[g^{-1}a g],[g'^{-1} b g']}=\sum_{u\in \mathrm{C}_c, u g^{-1}a g u^{-1}= g'^{-1} b g'} 1.
\end{equation}
Then, $SS^\dagger$ becomes as follows:
\begin{equation}
  \begin{split}
  \left(S S^\dagger\right)_{(a,\chi),(b,\chi')}
  &=\sum_{
    \substack{g\in G(a,c),g'\in G(b,c),c\\u\in \mathrm{C}_c, u g^{-1}a g u^{-1}= g'^{-1} b g'}
  }\frac{1}{|G||\mathrm{C}_a||\mathrm{C}_b||\mathrm{C}_c|}\overline{\chi(g c g^{-1})}
  \chi'(g' c g'^{-1})
  \end{split}
\end{equation}
This is obviously zero unless $[a]=[b]$ in $G$. We can replace $b$ with $a$ since $S$ is class function. One finds that $g c g^{-1}, g' c g'^{-1} \in \mathrm{C}_a$ under the conjugation by $g' u g^{-1} \in \mathrm{C}_a$. We can replace $\chi(g c g^{-1}),\chi'(g' c g'^{-1})\mapsto \chi(c'),\chi'(c')$ since $\chi$ and $\chi'$ are irreps of $\mathrm{C}_a$.
\begin{equation}
  \begin{split}
  \left(S S^\dagger\right)_{(a,\chi),(b,\chi')}
  &=\sum_{
    \substack{g,g'\in G(a,c'),c'\\u\in \mathrm{C}_c', [a, g' u g^{-1}]=1}
  }\frac{\delta^{G}_{[a],[b]}}{|G||\mathrm{C}_a|^2|\mathrm{C}_{c'}|}\overline{\chi(c')}
  \chi'(c')
  \end{split}
\end{equation}
Retake the index of summation as $x':= g^{-1}x g, \tilde{g}=g^{-1} u^{-1} g'$, then the sum above reduces to
\begin{equation}
  \begin{split}
  \left(S S^\dagger\right)_{(a,\chi),(b,\chi')}
  &=\sum_{c'}\frac{\delta^{G}_{[a],[b]}}{|\mathrm{C}_a|}\overline{\chi(c')}
  \chi'(c')=\delta^{G}_{[a],[b]} \delta_{\chi,\chi'}.
  \end{split}
\end{equation}
This complets the proof of the unitarity.

Next, we consider the summation when one of $c$ is inverted.
In the same procedure we can evaluate this as follows:
\begin{equation}
\begin{split}
\sum_{(c,\chi'')}S_{(a,\chi),(c,\chi'')}\overline{S_{(b,\chi'),(c^{-1},\chi'')}}
&=\sum_{c'}\frac{\delta^{G}_{[a],[b]}}{|\mathrm{C}_a|}\overline{\chi(c')}
  \chi'(c'^{-1}).
\end{split}
\end{equation}
The character satisfies the property $\chi'(c'^{-1})=\overline{\chi'(c')}$, thus
the Schur orhotonality can be used if and only if the irrep $\overline{\chi'}$ is identical to $\chi'$. The Frobenius-Schur indicator detects this condition. Finally, we obtain
\begin{equation}
\begin{split}
\sum_{(c,\chi'')}S_{(a,\chi),(c,\chi'')}\overline{S_{(b,\chi'),(c^{-1},\chi'')}}
&=(\iota^\chi)^2 \delta^{G}_{[a],[b]}\delta_{\chi,\chi'}.
\end{split}
\label{eq-nonorientable-sip}
\end{equation}

\section{Gauge Invariance and Refined Block Structure in 2D manifolds}
\label{app:xy2D}
\subsubsection{Example: Genus $\gamma$ Surface}

For a general (connected) bipartition $M=X\cup Y$ with $M$ an orientable 2-manifold with genus $g$ and $|A|=n$, we would like to figure out the topological entanglement entropy as well as $x_{\boldsymbol{\alpha}_{[\phi]}}$ and $y_{\boldsymbol{\alpha}_{[\phi]}}$ (which determine the degeneracy) for a given $\phi=(c_1,c_2,...,c_n)\in{\rm Im}\subset G^{\times n}$. 

Note that $d_{\boldsymbol{\alpha}_{[\phi]}}$ is simply $\prod^n_{j=1}d_{\alpha_j}$ with $d_{\alpha_j}$ the dimension of an irrep of ${\rm C}_{c_j}$ 
(${\rm C}_g\equiv\{h\in G:[g,h]=1\}$ is the centralizer group of $g$), 
while $|[\phi]|=\prod^n_{j=1}|[c_j]|$ ($[g]$ denotes the set of conjugacy class of $g$) according to the orbit-stabilizer theorem. Therefore, the topological entanglement entropy of a minimally entangled state is simply given by $n\ln|G| - \sum^n_{j=1}\ln (d_{\alpha_j}|[c_j]|)$, provided $x_{\boldsymbol{\alpha}_{[\phi]}},y_{\boldsymbol{\alpha}_{[\phi]}}>0$ and $(c_1,c_2,...,c_n)\in{\rm Im}$. The criterion for the latter reads $\sum_{\alpha\in{\rm Rep}(G)} (|G|/d_\alpha)^{2g_S+n-2}\prod^n_{j=1}\chi_\alpha(c_j)>0$ for both $S=X,Y$. The only remaining task is to determine $x_{\boldsymbol{\alpha}_{[\phi]}}$ and $y_{\boldsymbol{\alpha}_{[\phi]}}$. 

We focus on $x_{\boldsymbol{\alpha}_{[\phi]}}$ since $y_{\boldsymbol{\alpha}_{[\phi]}}$ can be obtained similarly by replacing $\gamma_X$ with $\gamma_Y$. Formally, $x_{\boldsymbol{\alpha}_{[\phi]}}$ can be obtained from the orthogonality relation of characters:
\begin{equation}
x_{\boldsymbol{\alpha}_{[\phi]}}=\frac{\sum_{\{g_j\in {\rm C}_{c_j}\}^n_{j=1}} \chi_{\alpha_1}(g^{-1}_1)\cdots\chi_{\alpha_n}(g^{-1}_n)\Tr D^X_\phi(\boldsymbol{g})}{\prod^n_{j=1}|{\rm C}_{c_j}|},
\label{eq-f}
\end{equation}
where $[D^X_\phi(\boldsymbol{g})]_{\psi,\varphi}=\delta_{\psi,\boldsymbol{g}\cdot\varphi}$ with $\psi,\varphi\in r_X^{-1}(\phi)$. Given $\phi=(c_1,c_2,...,c_n)$, we can write down ($f_1\equiv 1$)
\begin{equation}
{\rm Hom}(\pi_1(X,A),G)=\left\{ a_1,b_1,a_2,b_2,...,a_{\gamma_X},b_{\gamma_X},c_1,c_2,...,c_n,f_2,...,f_n\in G: \prod^{\gamma_X}_{m=1}[a_m,b_m]\prod^n_{j=1}f_j^{-1} c_j f_j=1\right\},
\end{equation}
where $f_j$ denotes the assignment of a group element in $G$ to the trivial path from base point $1$ to $j$, $c_j$ is the holonomy of the $j$th boundary (with respect to base point $j$), while $a_m,b_m$'s are associated to base point $1$. Under such a presentation of Hom elements, $r_X$ is simply a projection onto $c_j$'s. Recall that the group action is given by Eq.~(\ref{eq-gt}), we have
\begin{equation}
\Tr D_\phi^X(\boldsymbol{g})= \sum_{\{a_j,b_j\in G\}^{\gamma_X}_{m=1},\{f_j\in G\}^n_{j=2}}\delta_{\prod^{\gamma_X}_{m=1}[a_m,b_m]\prod^n_{j=1}f_j^{-1} c_j f_j,1} \prod^{\gamma_X}_{m=1}\delta_{a_m,g_1a_m g_1^{-1}}\delta_{b_m,g_1b_m g_1^{-1}}\prod^n_{j=2} \delta_{f_j,g_j f_j g_1^{-1}}.
\end{equation}
Note that $\delta_{a_m,g_1a_m g_1^{-1}}\delta_{b_m,g_1b_m g_1^{-1}}$ enforces $a_m,b_m\in {\rm C}_{g_1}$, while $\delta_{f_j,g_j f_j g_1^{-1}}$ enforces $g_j \in [g_1]$ as well as $f_j^{-1}c_j f_j\in {\rm C}_{g_1}$ (provided $g_j\in {\rm C}_{c_j}$). Moreover, suppose $f_j^\star= g_j f_j^\star g_1^{-1}$ for some specific $f_j^\star$, then $f_j= g_j f_j g_1^{-1}$ iff $f_j=f_j^\star h_j$ with $h_j\in  {\rm C}_{g_1}$. Applying the regular rep technique to $\delta_{\prod^{\gamma_X}_{m=1}[a_m,b_m]\prod^n_{j=1}f_j^{-1} c_j f_j,1}$ over ${\rm C}_{g_1}$, we obtain
\begin{equation}
\Tr D_\phi^X(\boldsymbol{g})= \sum_{\beta\in {\rm Rep}({\rm C}_{g_1})} \left(\frac{|{\rm C}_{g_1}|}{d_\beta}\right)^{2\gamma_X+n-2} \chi_\beta (c_1)\chi_\beta (f_2^{\star -1}c_2f_2^\star)\cdots\chi_\beta (f_n^{\star -1}c_nf_n^\star),
\label{eq-tD}
\end{equation}
provided that $g_2,...,g_n\in [g_1]$ and $g_j\in {\rm C}_{c_j}$, but otherwise vanishes. 

To proceed, we first rewrite Eq.~(\ref{eq-tD}) into an equivalent form such that $g_1$ or $c_1$ does not appear special:
\begin{equation}
\Tr D_\phi^X(h_1gh_1^{-1},h_2gh_2^{-1},...,h_ngh_n^{-1})= \sum_{\beta\in {\rm Rep}({\rm C}_g)} \left(\frac{|{\rm C}_g|}{d_\beta}\right)^{2\gamma_X+n-2} \chi_\beta (h_1^{-1}c_1h_1)\chi_\beta (h_2^{ -1}c_2h_2)\cdots\chi_\beta (h_n^{-1}c_n h_n),
\label{eq-tD2}
\end{equation}
where $h_j$ is chosen such that $h_j gh_j^{-1} \in {\rm C}_{c_j}$. Otherwise the trace vanishes. Combining Eqs.~(\ref{eq-tD2}) and (\ref{eq-f}), we end up with
\begin{equation}
x_{\boldsymbol{\alpha}_{[\phi]}}=\sum_{[g]}\sum_{\beta\in {\rm Rep}({\rm C}_g)}  \left(\frac{|{\rm C}_g|}{d_\beta}\right)^{2\gamma_X+n-2}
\prod^n_{j=1}\left[\frac{\sum_{h_j\in G, h_j gh_j^{-1} \in {\rm C}_{c_j}}\chi_\beta(h_j^{-1}c_j h_j) \chi_{\alpha_j}(h_jgh_j^{-1})}{|{\rm C}_g||{\rm C}_{c_j}|}\right],
\label{eq-xa}
\end{equation}
where we have replace $g$ by $g^{-1}$ using the fact that $|{\rm C}_g|=|{\rm C}_{g^{-1}}|$ and $[g,c]=1\Leftrightarrow [g^{-1},c]=1$ (as well as $\sum_{[g]}=\sum_{[g^{-1}]}$). We remark that in the above product each factor 
\begin{equation}
S_{(g,\beta),(c_j,\alpha_j)}=|{\rm C}_g|^{-1}|{\rm C}_{c_j}|^{-1}\sum_{h_j\in G, h_j gh_j^{-1} \in {\rm C}_{c_j}}\chi_\beta(h_j^{-1}c_j h_j) \chi_{\alpha_j}(h_jgh_j^{-1})
\label{eq-ba}
\end{equation}
is nothing but the modular $S$ matrix for finite group $G$ \cite{costeFiniteGroupModular2000}, where $g$ should be identified if they belong to the same conjugacy class. In terms of $S$ matrix, we have
\begin{equation}
x_{\boldsymbol{\alpha}_{[\phi]}}=\sum_{[g]}\sum_{\beta\in {\rm Rep}({\rm C}_g)}  \left(\frac{|{\rm C}_g|}{d_\beta}\right)^{2\gamma_X+n-2}
\prod^n_{j=1}S_{(g,\beta),(c_j,\alpha_j)}.
\label{eq-xa2}
\end{equation}

Let us check the consistency of the total degree of freedom:
\begin{equation}
|{\rm Hom}(\pi_1(M),G)/G|=\sum_{[\phi]} \sum_{\boldsymbol{\alpha}_\phi\in{\rm Rep}(G_\phi)} x_{\boldsymbol{\alpha}_{[\phi]}}y_{\boldsymbol{\alpha}_{[\phi]}}.
\end{equation}
The lhs is well-known as $\sum_{[g]}\sum_{\beta\in{\rm Rep}({\rm C}_g)}(|{\rm C}_g|/d_\beta)^{2\gamma-2}$ \cite{ritz-zwillingPartitionFunctionKitaev2025}, as might be considered as the $n=0$ case of Eq.~(\ref{eq-xa2}). Using the realness of $x_{\boldsymbol{\alpha}_{[\phi]}},y_{\boldsymbol{\alpha}_{[\phi]}}$ and the unitarity of $S$, the rhs is obtained to be the same result.

Let us comment on the presence of modular $S$ matrix and the connection to Drinfel'd double $D(G)=\mathbb{C}^G \otimes \mathbb{C}[G]$ and rational conformal field theory (RCFT) \cite{mooreClassicalQuantumConformal1989,mooreLecturesRCFT1990}.
The appearance of Eq.~(\ref{eq-xa2}) is identical to Eq.~(A7) in \cite{mooreClassicalQuantumConformal1989}. Eq.~(A7) in \cite{mooreClassicalQuantumConformal1989} describes the dimension of Hilbert space associated on the boundary of two-dimensional RCFT.
The boundary of the RCFT is often called a puncture, where the degrees of freedom of the operator or elementary excitation are assigned.

Elementary excitations are represented by simple objects in the modular tensor category. The data of the modular tensor category contains information about the braiding of elementary excitations, from which the modular  $S$ matrix can be reconstructed.

By the way, it is known that a holographic correspondence exists between 2D RCFTs and (2+1)D TQFTs \cite{fuchsTFTConstructionRCFT2002}. Specifically, in the context of entanglement in (2+1)D TQFTs, the Li-Haldane conjecture states that the entanglement spectrum coincides with the energy spectrum of a 2D RCFT \cite{liEntanglementSpectrumGeneralization2008}.

In our case, the modular tensor category is the Drinfeld double $D(G)$ \cite{dijkgraafQUASIHOPFALGEBRAS,costeFiniteGroupModular2000}, where its simple objects are characterized by two labels: the conjugacy classes of $G$ and irreducible representations of its the centralizer group. This matches the labels of S matrix appeared in our result, meaning we have partially verified the Li-Haldane conjecture for the case where the modular tensor category is $D(G)$.

\subsubsection{Example: 2D Non-orientable Surface}
We would like to consider the case of 2D non-orientable surfaces.
The topological entanglement entropy of a minimally entangled state is again simply given by $n\ln|G| - \sum^n_{j=1}\ln (d_{\alpha_j}|[c_j]|)$.
We would like to elucidate $x_{\boldsymbol{\alpha}_{[\phi]}}$ and $y_{\boldsymbol{\alpha}_{[\phi]}}$ also for 2D non-orientable surfaces.
The majority of the derivation is shared with the case of an orientable surface; therefore, we only highlight the distinction from that scenario.

To start with, we recall that there are three cases to make $M$ non-orientable: both orientable, orientable plus non-orientable, and both non-orientable.
Currently, we are focusing solely on $X$, so we need only consider two cases: when X is orientable/non-orientable. Since the case where $X$ is orientable follows the same procedure as previously discussed, we need only consider the case where $X$ is non-orientable.

The holonomies on non-orientable surface is characterized by following relation:
\begin{equation}
    {\rm Hom}(\pi_1(X,A),G)=\left\{a_m,c_j,f_j\in G: \prod^{k_X}_{m=1}a_m^2\prod^n_{j=1}f_j^{-1} c_j f_j=1\right\}
\end{equation}

The group action is given by
\begin{equation}
\Tr D_\phi(\boldsymbol{g})= \sum_{\{a_j,b_j\in G\}^{k_X}_{m=1},\{f_j\in G\}^n_{j=2}}\delta_{\prod^{k_X}_{m=1}a_m^2\prod^n_{j=1}f_j^{-1} c_j f_j,1} \prod^{k_X}_{m=1}\delta_{a_m,g_1a_m g_1^{-1}}\prod^n_{j=2} \delta_{f_j,g_j f_j g_1^{-1}}.
\end{equation}
We find that $\delta_{a_m,g_1a_m g_1^{-1}}$ enforces $a_m\in {\rm C}_{g_1}$ (provided $g_j\in {\rm C}_{c_j}$). Applying the regular rep technique to $\delta_{\prod^{k_X}_{m=1}a_m^2\prod^n_{j=1}f_j^{-1} c_j f_j,1}$ over ${\rm C}_{g_1}$, we obtain
\begin{equation}
\Tr D_\phi(h_1gh_1^{-1},h_2gh_2^{-1},...,h_ngh_n^{-1})= \sum_{\beta\in {\rm Rep}({\rm C}_g)} (\iota^\beta)^{k_X} \left(\frac{|{\rm C}_g|}{d_\beta}\right)^{k_X+n-2} \chi_\beta (h_1^{-1}c_1h_1)\chi_\beta (h_2^{ -1}c_2h_2)\cdots\chi_\beta (h_n^{-1}c_n h_n),
\label{tD2}
\end{equation}
where $h_j$ is chosen such that $h_j gh_j^{-1} \in {\rm C}_{c_j}$. Otherwise the trace vanishes.

We end up with
\begin{equation}
  \begin{split}
  x_{\boldsymbol{\alpha}_{[\phi]}}
  &=\sum_{[g]}\sum_{\beta\in {\rm Rep}({\rm C}_g)} (\iota^\beta)^{k_X} \left(\frac{|{\rm C}_g|}{d_\beta}\right)^{k_X+n-2}
  \prod^n_{j=1}\left[\frac{\sum_{h_j\in G, h_j gh_j^{-1} \in {\rm C}_{c_j}}\chi_\beta(h_j^{-1}c_j h_j) \chi_{\alpha_j}(h_jgh_j^{-1})}{|{\rm C}_g||{\rm C}_{c_j}|}\right],\\
  &=\sum_{[g]}\sum_{\beta\in {\rm Rep}({\rm C}_g)} (\iota^\beta)^{k_X} \left(\frac{|{\rm C}_g|}{d_\beta}\right)^{k_X+n-2}
  \prod^n_{j=1}S_{(g,\beta),(c_j,\alpha_j)}.
  \end{split}
  \label{eq-nonorientable-xa2}
\end{equation}
This is consistent with the Eq.~(149) in \cite{barkeshliReflectionTimeReversal2020}.

We again check the consistency of the total degree of freedom in non-orientable surfaces.
For bipartition to two orientable surfaces, we have
\begin{equation}
\begin{split}
x_{\boldsymbol{\alpha}_{[\phi]}}
&=\sum_{[g]}\sum_{\beta\in {\rm Rep}({\rm C}_g)}  \left(\frac{|{\rm C}_g|}{d_\beta}\right)^{2\gamma_X+n-2}\prod^n_{j=1}S_{(g,\beta),(c_j,\alpha_j)}\\
y_{\boldsymbol{\alpha}_{[\phi]}}
&=\sum_{[g]}\sum_{\beta\in {\rm Rep}({\rm C}_g)}  \left(\frac{|{\rm C}_g|}{d_\beta}\right)^{2\gamma_Y+n-2}\prod^n_{j=1}S_{(g,\beta),((c_j)^{s_j},\alpha_j)}
\end{split}
\end{equation}
where $s_j$ is the label that detects if the $j$-th boundary is flipped $(s_j=-1)$ or not $(s_j=1)$.
We assume that ${}^\exists i,j,\,s_i\neq s_j$ to obtain non-orientable surface and evaluate $|{\rm Hom}(\pi_1(M),G)/G|=\sum_{[\phi]} \sum_{\boldsymbol{\alpha}_\phi\in{\rm Rep}(G_\phi)} x_{\boldsymbol{\alpha}_{[\phi]}}y_{\boldsymbol{\alpha}_{[\phi]}}$. This turns out to be
\begin{equation}
  |{\rm Hom}(\pi_1(M),G)/G|=\sum_{[g]}\sum_{\beta\in{\rm Rep}({\rm C}_g)}\left(\frac{|{\rm C}_g|}{d_\beta}\right)^{k-2} \left(\iota^\beta\right)^2
\end{equation}
by using Eq.~(\ref{eq-nonorientable-sip}). Here we defined the total crosscap number as $k=2\gamma_X+2\gamma_Y+2n-2$.
Similarly, we obtain
\begin{equation}
  |{\rm Hom}(\pi_1(M),G)/G|=\sum_{[g]}\sum_{\beta\in{\rm Rep}({\rm C}_g)}\left(\frac{|{\rm C}_g|}{d_\beta}\right)^{k-2} \left(\iota^\beta\right)^{k_Y+2}
\end{equation}
(where $k=2\gamma_X+k_Y+2n-2$) for the X orientable plus Y non-orientable case and
\begin{equation}
  |{\rm Hom}(\pi_1(M),G)/G|=\sum_{[g]}\sum_{\beta\in{\rm Rep}({\rm C}_g)}\left(\frac{|{\rm C}_g|}{d_\beta}\right)^{k-2} \left(\iota^\beta\right)^{k_X+k_Y+2}
\end{equation}
(where $k=k_X+k_Y+2n-2$) for the two non-orientable case.
Here, since the Frobenius-Schur indicator is either $0,\pm 1$, we can add any positive even number on the shoulder of the Frobenius-Schur indicator.
For two orientable cases excluding the case $\gamma_X=\gamma_Y=0$, we can modify $2\mapsto k=2\gamma_X+2\gamma_Y+2n-2$.
For other two cases, one finds that it can be modified to $k$.
Finally, we find
\begin{equation}
  |{\rm Hom}(\pi_1(M),G)/G|=\sum_{[g]}\sum_{\beta\in{\rm Rep}({\rm C}_g)}\left(\frac{|{\rm C}_g|}{d_\beta}\right)^{k-2} \left(\iota^\beta\right)^{k}
\end{equation}
holds for non-orientable surface. This result is consistent with Eq.~(\ref{eq-nonorientable-xa2}) under setting $n=0$.

\section{Entanglement Entropy for General Ground State in 2D Theories}

Let us examine this result for several 2D cases; $\Sigma_{0,1}$, $\Sigma_{0,2}$, $\Sigma_{0,3}$ and $\Sigma_{0,4}$.
We define as follows:
\begin{equation}
S'=-\sum_{[\phi]\in \mathrm{Im}/G^{\times |A|}} |[\phi]|
\sum_{\boldsymbol{\alpha}_{[\phi]}}d_{\boldsymbol{\alpha}_{[\phi]}} 
\sum_{i=1}^{x_{\boldsymbol{\alpha}_{[\phi]}}} \frac{|\psi_{\boldsymbol{\alpha}_{[\phi]},i}|^2}{\mathcal{N}}\ln \frac{|\psi_{\boldsymbol{\alpha}_{[\phi]},i}|^2}{\mathcal{N}}.
\label{eq:tee-correction}
\end{equation}
We recall that the normalization constant is given by $\mathcal{N}:=\sum_{[\phi]\in \mathrm{Im}/G^{\times |A|}} |[\phi]|\left(\sum_{\boldsymbol{\alpha}_{[\phi]}} d_{\boldsymbol{\alpha}_{[\phi]}} \Tr \Psi_{\boldsymbol{\alpha}_{[\phi]}} \Psi_{\boldsymbol{\alpha}_{[\phi]}}^\dagger\right)$.

In the case of $\Sigma_{0,1}$, boundary holonomy $\phi=c$ labels the conjugacy classes of $G$.
One of significant properties of the modular $S$ matrix, which can be used in our setup, is
\begin{equation}
    \left(\frac{|{\rm C}_g|}{d_\beta}\right)^{-1}=S_{(g,\beta),(1,1)}.
\end{equation}
Since $S$ matrix is symmetric and unitary, the following holds;
\begin{equation}
    x_{(c,\alpha)}=\sum_{[g]}\sum_{\beta\in {\rm Rep}({\rm C}_g)}
    \overline{S_{(g,\beta),(1,1)}}S_{(g,\beta),(c,\alpha)}=\delta_{c,1}\delta_{\alpha,1}.
\end{equation}
Here, $\alpha$ denotes an irrep of $\mathrm{C}_g$.
Plugging this relation into Eq.~(\ref{eq:tee-correction}), we obtain
\begin{equation}
\begin{split}
\mathcal{N}&=\sum_{[c]} |[c]|\left(\sum_\alpha d_\alpha \sum_{i=1}^{\delta_{c,1}\delta_{\alpha,1}} |\psi_{\alpha,i}|^2 \right)=|\psi_{1,1}|^2,\\
S'&=-\sum_{[c]} |[c]|
\sum_{\alpha}d_\alpha 
\delta_{c,1}\delta_{\alpha,1} 
\frac{|\psi_{\alpha,i}|^2}{|\psi_{1,1}|^2}\ln \frac{|\psi_{\alpha,i}|^2}{|\psi_{1,1}|^2}=0.
\end{split}
\end{equation}
Note that only vacuum can contribute to entanglement entropy, though they canceled out.

In the case of $\Sigma_{0,2}$, two boundary holonomies $c_1,c_2$ determines $x_{\boldsymbol{\alpha}_{[\phi]}}$ as follows;
\begin{equation}
    x_{(c_1,c_2,\alpha_1,\alpha_2)}=\sum_{(g,\beta)}S_{(c,\alpha_1),(g,\beta)}S_{(g,\beta),(c,\alpha_2)}=\delta_{c_1,c_2^{-1}}\delta_{\alpha_1,\overline{\alpha_2}}
\end{equation}
Here we used that $S$ is symmetric and $S^2$ maps $(g,\chi)\mapsto (g^{-1},\overline{\chi})$ \cite{costeFiniteGroupModular2000}.
Substituting this into Eq.~(\ref{eq:tee-correction}), we obtain
\begin{equation}
\begin{split}
\mathcal{N}&=\sum_{[c_1],[c_2]} |[c_1]||[c_2]|\left(\sum_{\alpha_1,\alpha_2} d_{\alpha_1}d_{\alpha_2} \sum_{i=1}^{\delta_{c_1,c_2^{-1}}\delta_{\alpha_1,\overline{\alpha_2}}} |\psi_{c_1,c_2,\alpha_1,\alpha_2}|^2\right)\\
&=\sum_{c,\alpha} |[c]|^2 d_\alpha^2 |\psi_{c,\alpha}|^2\\
S'&=-\sum_{c,\alpha} |[c]|^2 d_{\alpha}^2 
\frac{|\psi_{c,\alpha}|^2}{\sum_{c,\alpha} |[c]|^2 d_\alpha^2 |\psi_{c,\alpha}|^2}\ln \frac{|\psi_{c,\alpha}|^2}{\sum_{c,\alpha} |[c]|^2 d_\alpha^2 |\psi_{c,\alpha}|^2}.
\end{split}
\end{equation}
Note that only the particle-antiparticle pairs can contribute to this entanglement entropy.
Since the quantum dimension for excitation $(c,\alpha)$ is given by $|[c]| d_\alpha$, this result is consistent with the known result computed for Chern-Simons Theory \cite{dongTopologicalEntanglementEntropy2008}.

In the case of $\Sigma_{0,3}$, $x_{\boldsymbol{\alpha}_{[\phi]}}$ is determined as follows;
\begin{equation}
\begin{split}
x_{(c_1,c_2,c_3,\alpha_1,\alpha_2,\alpha_3)}&=\sum_{[g]}\sum_{\beta\in {\rm Rep}({\rm C}_g)}
\frac{|{\rm C}_g|}{d_\beta}S_{(g,\beta),(c_1,\alpha_1)}S_{(g,\beta),(c_2,\alpha_2)}S_{(g,\beta),(c_3,\alpha_3)}\\
&=\sum_{(g,\beta)}
\frac{S_{(g,\beta),(c_1,\alpha_1)}S_{(g,\beta),(c_2,\alpha_2)}S_{(g,\beta),(c_3,\alpha_3)}}{S_{(g,\beta),(1,1)}}.
\end{split}
\end{equation}
This is nothing but the Verlinde formula of fusion coefficient. We finally obtain
\begin{equation}
    x_{(c_1,c_2,c_3,\alpha_1,\alpha_2,\alpha_3)}=N^{(c_1,\alpha_1),(c_2,\alpha_2)}_{((c_3)^{-1},\overline{\alpha_3})}.
\end{equation}
The correction terms of topological entanglement entropy can be obtained by simply replacing $x_{\boldsymbol{\alpha}_{[\phi]}}$ in Eq.~(\ref{eq:tee-correction}) with the fusion coefficients.
Note that three anyons can contribute to topological entanglement entropy when they cancels out to the vacuum.

In the case of $\Sigma_{0,4}$, $x_{\boldsymbol{\alpha}_{[\phi]}}$ is determined as follows;
\begin{equation}
\begin{split}
x_{(c_1,c_2,c_3,c_4,\alpha_1,\alpha_2,\alpha_3,\alpha_4)}&=\sum_{[g]}\sum_{\beta\in {\rm Rep}({\rm C}_g)}
\left(\frac{|{\rm C}_g|}{d_\beta}\right)^2 S_{(g,\beta),(c_1,\alpha_1)}S_{(g,\beta),(c_2,\alpha_2)}S_{(g,\beta),(c_3,\alpha_3)}S_{(g,\beta),(c_4,\alpha_4)}\\
&=\sum_{(g,\beta)}\sum_{(g',\beta')}
\frac{S_{(g,\beta),(c_1,\alpha_1)}S_{(g,\beta),(c_2,\alpha_2)}\delta_{(g,\beta),(g',\beta')}S_{(g',\beta'),(c_3,\alpha_3)}S_{(g',\beta'),(c_4,\alpha_4)}}{S_{(g,\beta),(1,1)}S_{(g',\beta'),(1,1)}}\\
&=\sum_{(g,\beta)}\sum_{(g',\beta')}\sum_{(h,\alpha)}
\frac{S_{(g,\beta),(c_1,\alpha_1)}S_{(g,\beta),(c_2,\alpha_2)}S_{(g,\beta),(h,\alpha)}}{S_{(g,\beta),(1,1)}}
\frac{S_{(g',\beta'),(c_3,\alpha_3)}S_{(g',\beta'),(c_4,\alpha_4)}\overline{S_{(g',\beta'),(h,\alpha)}}}{S_{(g',\beta'),(1,1)}}\\
&=\sum_{(h,\alpha)}
N^{(c_1,\alpha_1),(c_2,\alpha_2)}_{(h^{-1},\overline{\alpha})}N^{(c_3,\alpha_3),(c_4,\alpha_4)}_{(h,\alpha)}.
\end{split}
\end{equation}
One can easily check that the result is invariant under any permutation of $\alpha_i$; the F-move is set to trivial, reflecting the fact that the theory is set to the trivial twist in $H^3(G,U(1))$.
In the twisted model, (known as the Dijkgraaf-Witten model), this F-move may not be trivial.

\begin{figure}
\begin{center}
       \includegraphics[width=10cm, clip]{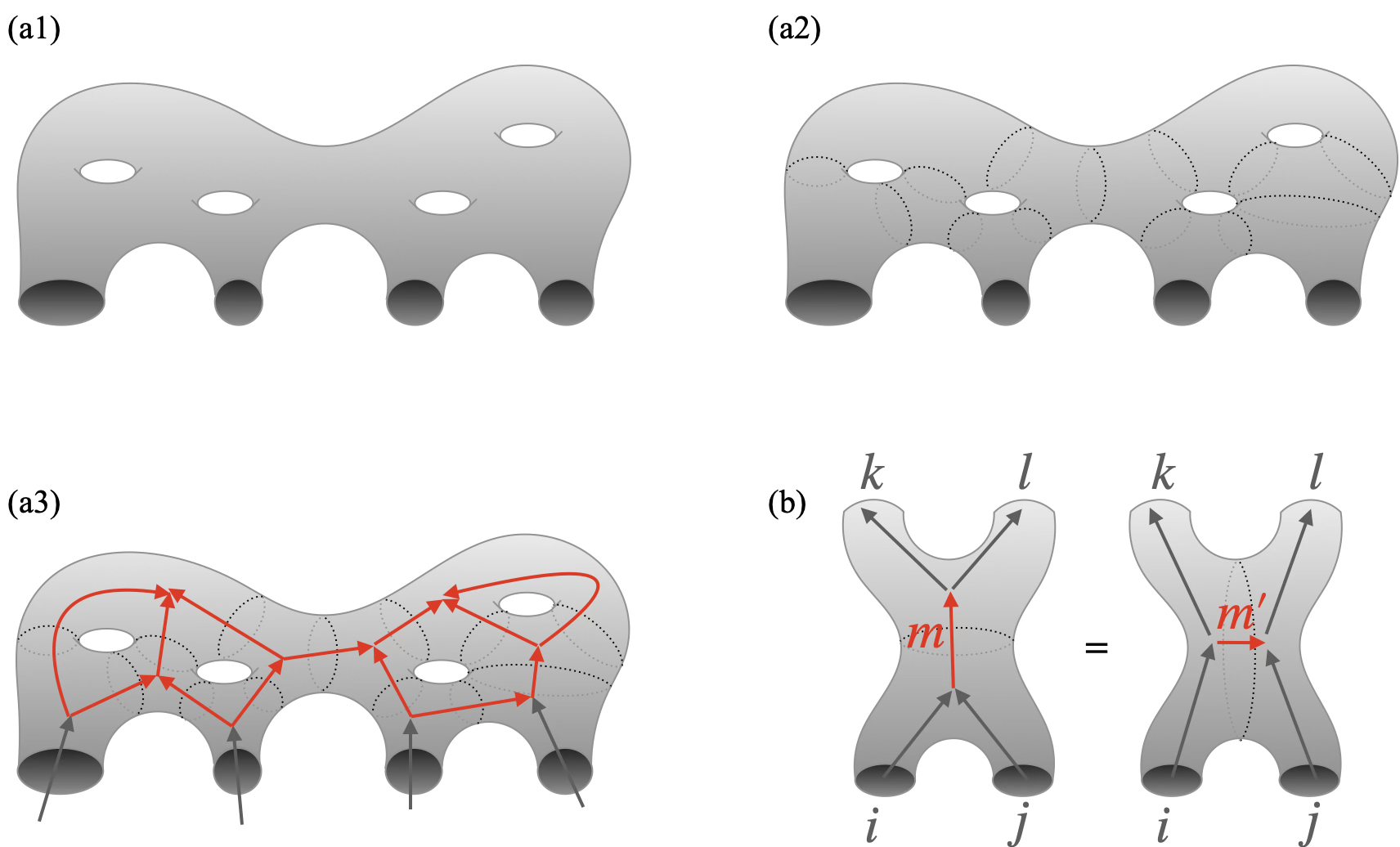}
       \end{center}
   \caption{For orientable surfaces, there are as many degree of freedom $x_{\boldsymbol{\alpha}_{[\phi]}}$ as there are fusions for which the deformation retract becomes a string diagram.
   (a1) Fix an orientable surface $X$. In this figure, we depict $\Sigma_{4,4}$ as an example.
   (a2) Disassemble $X$ into $\Sigma_{0,3}$. 
   (a3) The deformation retract is the string diagram of the fusion. For the red lines, take the sum of the fusion scores (including multiplicity defined by the $N$-symbol).
   (b) Replacing the partition corresponding to the F-move. While the partition method is not unique, any partition yields the same number of combinations.
   } 
   \label{fig:pants}
\end{figure}

We generalize this discussion in the case of $X=\Sigma_{\gamma,n}$.
By pasting together many $\Sigma_{0,3}$ and $\Sigma_{0,1}$ pieces, one can construct any $\Sigma_{\gamma,n}$.
Above, we verified that the dimension of the overall manifold can be reproduced from the dimensions of the individual manifolds when the resulting manifold is closed after gluing. However, the same calculation shows that the dimension of the overall manifold can also be reproduced from the dimensions of the individual manifolds even when the resulting manifold is not closed.
From the above facts, we understand that the block size of the $W$ matrix is equivalent to the number of ways to fuse the pieces such that the deformation retract of $\Sigma_{\gamma,n}$ becomes a string diagram (Fig.~\ref{fig:pants}).

To conclude, we have elucidated that the topological entanglement entropy for general ground state of Kitaev quantum double model is determined by the data of $D(G)$ as a modular tensor category. This is again the manifestation of the Li-Haldane correspondence.

Thus far, we have shown that the internal degrees of freedom of an orientable surface's entanglement block structure correspond to the sum of possible fusions, including their multiplicities. But for a non-orientable surface, what physical meaning can we find for the degrees of freedom per block?

Any non-orientable surface is classified by its crosscap number $k$ and number of punctures $n$, thus it can be generated by gluing many pieces of $\Sigma_{0,1},\Sigma_{0,3}$ and $N_{1,1}$.
We focus on computing $x_{\boldsymbol{\alpha}_{[\phi]}}$ for $N_{1,1}$. From Eq.~(\ref{eq-nonorientable-xa2}), we find
\begin{equation}
  \begin{split}
  x_{(c,\alpha)}
  &=\sum_{g,\beta}\iota^\beta S_{(g,\beta),(c,\alpha)}.
  \end{split}
\end{equation}
The physical implication is not clear as the orientable cases. This is equivalent to reversing the derivation of the $S$ matrix, but rewriting it by substituting the definition yields the following:
\begin{equation}
  \begin{split}
  x_{(c,\alpha)}
  &=\sum_{[g]}
  \frac{1}{|\mathrm{C}_g||\mathrm{C}_c|}\sum_{h\in G(a,b)}\overline{\chi^\alpha(h g h^{-1})} |\{t\in C_g| t^2=h^{-1} c h\}|.
  \end{split}
\end{equation}
We find that the boundary can host only the holonomy that is a square of some element. This reflects the fundamental group of the Klein bottle.

Although it is hard to proceed in the case $G$ non-Abelian, we see more feature in the case $G$ Abelian: the $S$ matrix takes a simpler form as follows:
\begin{equation}
    S_{(a,\alpha),(b,\beta)}=\frac{1}{|G|}\overline{\chi^\alpha(b)}\overline{\chi^\beta(a)},
\end{equation}
leading to a simpler form $x_{(c,\alpha)}=|\{t\in C_g| t^2=h^{-1} c h\}|\delta_{\alpha,1}$.
It is now clear that the electric excitation is not permitted for $G$ Abelian. 
For further discussion on this for Abelian modular tensor category, see \cite{Orii_2025,orii2025generalizationanomalyformulatime,Tachikawa_2017}.

\end{document}